\documentclass[notitlepage,
 aip,
 amsmath,amssymb,
reprint,%
]{revtex4-1}

\usepackage{physics}
\usepackage[version=3]{mhchem} % Formula subscripts using \ce{}
\usepackage{appendix}
\usepackage{caption}
\usepackage{subcaption}
\usepackage{graphicx}% Include figure files
\usepackage{bm}% bold math
\usepackage[section]{placeins}
\usepackage{amsmath}
\usepackage{hyperref}
\usepackage{cleveref}
\hypersetup{colorlinks,
            linkcolor=.,
            citecolor={blue!80!black},
            urlcolor=black}
\usepackage{tikz}
\usetikzlibrary{decorations.markings, shapes.misc}
\tikzset{
  mid arrow/.style={
    postaction={decorate},
    decoration={markings, mark=at position #1 with {\arrow{stealth}}}
  }
}
\tikzset{
  mid arrow rev/.style={
    postaction={decorate},
    decoration={markings, mark=at position #1 with {\arrow{stealth reversed}}}
  }
}
\usepackage[labelsep=period]{caption}
\renewcommand{\thetable}{\arabic{table}}

\begin{document}

\preprint{AIP/123-QED}

\title[]{
One-Body Properties and their Perturbative Accuracy with Aufbau Suppressed Coupled Cluster Theory
}
% Force line breaks with \\

\author{Conor Bready}
\affiliation{
Department of Chemistry, University of California, Berkeley, California 94720, USA 
\looseness=-1}

\author{Harrison Tuckman}
\affiliation{
Department of Chemistry, University of California, Berkeley, California 94720, USA 
\looseness=-1}

\author{Eric Neuscamman}
\email{eneuscamman@berkeley.edu}
\affiliation{
Department of Chemistry, University of California, Berkeley, California 94720, USA
\looseness=-1}
\affiliation{Chemical Sciences Division, Lawrence Berkeley National Laboratory, Berkeley, CA, 94720, USA
\looseness=-1}

\date{\today}% It is always \today, today,

\begin{abstract}
  We derived and implemented the calculation of the one-body reduced density matrix for Aufbau suppressed coupled cluster theory, from which excited state natural orbitals and one-body properties, like atomic populations and dipole moments, are obtained.
We utilized the natural orbitals to refine the ASCC solution
for simple valence and Rydberg systems, exploring the process
of repeatedly solving the ASCC equations in successive
natural orbital bases to achieve independence from the starting
molecular orbitals.
For dipole moments in small molecules where high-level comparison
data is available, we find that the accuracy of ASCC
essentially matches that of linear response
and equation-of-motion coupled cluster as long as
care is taken to preserve the response's perturbative completeness.
\end{abstract}

\maketitle


\section{Introduction}

The proper manipulation of electronic excited states to perform intended chemical tasks requires knowledge of both the necessary excitation energy and the characteristic properties of the desired state. Many linear response-based methods such as time dependent density functional theory (TD-DFT)\cite{rungeDensityFunctionalTheoryTimeDependent1984,burkeTimedependentDensityFunctional2005,casidaProgressTimeDependentDensityFunctional2012} and equation of motion coupled cluster theory (EOM-CC)\cite{roweEquationsofMotionMethodExtended1968,stantonEquationMotionCoupledcluster1993,krylovEquationofMotionCoupledClusterMethods2008} are able to provide both energies and properties quite accurately.\cite{loosReferenceEnergiesIntramolecular2021,knyshReferenceCC3Excitation2024,sirucekExcitedStateAbsorptionReference2025} However, these linear response methods, unlike state-specific methods, are dependent upon the proper behavior of the ground state, which has the potential to be problematic when the excited state's optimal geometry is far from that of the ground state's.\cite{wangHowAccurateAre2020,gonzalezProgressChallengesCalculation2012} Approaches such as the spin-flip method for both TD-DFT\cite{shaoSpinFlipApproach2003,bernardGeneralFormulationSpinflip2012} and EOM-CC,\cite{krylovSizeconsistentWaveFunctions,krylovSpinFlipEquationofMotionCoupledCluster2006} or double ionization potential and double electron affinity for just EOM-CC,\cite{nooijenSimilarityTransformedEquationofmotion1997,bartlettCoupledclusterRevolution2010,musialMultireferenceCoupledclusterTheory2011,kusUsingChargestabilizationTechnique2011,kusDeperturbativeCorrectionsChargestabilized2012,pereraSingletTripletSeparations2014} help ameliorate the problems that arise due to this multi-configurational character while still maintaining a single-reference formalism; however, this requires chemical knowledge that the newly chosen reference must itself remain well-behaved at the interested geometry. On the other hand, Aufbau suppressed coupled cluster (ASCC)\cite{tuckmanAufbauSuppressedCoupled2024} provides a state-specific approach, bypassing the reliance that a different potential energy surface still yields reasonable results. Furthermore, ASCC maintains all of the benefits of coupled cluster theory (CC), providing an interesting candidate for the evaluation of excited state properties. After some adjustments informed by extending the perturbative analysis performed by Tuckman et al.\cite{tuckmanImprovingAufbauSuppressed2025} to property evaluations, we demonstrate that ASCC's response can be just as accurate as other excited state CC counterparts, with the potential to be more accurate for charge transfer systems where ASCC has previously excelled.\cite{tuckmanFastAccurateCharge2025}

In addition to the benefits of state-specific methods when exploring structures outside of their optimal ground state geometries, these methods also have the benefits of orbital relaxations tailored to the excited state itself. Many approaches utilize this concept of trying to optimize the excited state orbitals, including various methods in DFT,\cite{haitExcitedStateOrbital2020,haitHighlyAccuratePrediction2020,haitOrbitalOptimizedDensity2021,yeSSCFDirectEnergytargeting2017,yeHalfProjectedSelfConsistentField2019,bagusSelfConsistentFieldWaveFunctions1965,hsuSCFMethodHole1976,navesdebritoTheoreticalStudyXray1991,besleySelfconsistentfieldCalculationsCore2009,filatovSpinrestrictedEnsemblereferencedKohn1999,kowalczykExcitationEnergiesStokes2013,kowalczykAssessmentDSCFDensity2011,zhaoDensityFunctionalExtension2020,leviVariationalDensityFunctional2020,kempfer-robertsonRoleExactExchange2022,gilbertSelfConsistentFieldCalculations2008,barcaSimpleModelsDifficult2018,carter-fenkStateTargetedEnergyProjection2020} CASSCF,\cite{tranTrackingExcitedStates2019,tranImprovingExcitedStatePotential2020,tranExploringLigandtoMetalChargeTransfer2023,hanscamApplyingGeneralizedVariational2022,roosAccurateMolecularOrbital1992,boynElucidatingMolecularOrbital2022} CI,\cite{sheaCommunicationMeanField2018,sheaGeneralizedVariationalPrinciple2020,hardikarSelfconsistentFieldFormulation2020,liuCommunicationAdjustingCharge2012,kossoskiStateSpecificConfigurationInteraction2023,kossoskiSeniorityHierarchyConfiguration2023,burtonEnergyLandscapeStateSpecific2022,tsuchimochiDoubleConfigurationInteraction2024} perturbation theory,\cite{cluneExcitationMatchedLocal2025,cluneN5ScalingExcitedStateSpecificPerturbation2020,cluneStudyingExcitedstatespecificPerturbation2023} CC,\cite{mayhallMultipleSolutionsSingleReference2010,zhengPerformanceDeltaCoupledClusterMethods2019,leeExcitedStatesCoupled2019,damourStateSpecificCoupledClusterMethods2024,kossoskiExcitedStatesStateSpecific2021,tuckmanExcitedStateSpecificPseudoprojectedCoupledCluster2023} variational Monte Carlo,\cite{pinedafloresExcitedStateSpecific2019,zhaoEfficientVariationalPrinciple2016,robinsonExcitationVarianceMatching2017,bluntChargetransferExcitedStates2017a,sheaSizeConsistentExcited2017,garnerVariationalMonteCarlo2020,otisHybridApproachExcitedstatespecific2020,shepardDoubleExcitationEnergies2022,otisOptimizationStabilityExcitedStateSpecific2023,otisPromisingIntersectionExcitedstatespecific2023,pathakExcitedStatesVariational2021,entwistleElectronicExcitedStates2023} and more. 
Likewise, ASCC was recently shown to benefit from its intrinsic orbital relaxations by demonstrating near independence from the reference provided.\cite{quadyAufbauSuppressedCoupled2025} 
Generically, utilizing state-specific orbitals often results in a more accurate description of the excited state itself and is therefore incredibly valuable when comparing to experiments. Derivatives of this accurate energy expression with respect to one-body perturbations will yield density matrices, from which key properties can be extracted. Examples of such properties include information about the locations and magnitudes of charge and spin,\cite{lowdinQuantumTheoryManyParticle1955,mcweenyDensityMatrixManyelectron1997,danilovElectronDensitybondOrder1969} electronic dipole and other higher order multipole moments,\cite{szaboModernQuantumChemistry1996,mcweenyMethodsMolecularQuantum1969} and even magnetic moments,\cite{mcweenyMethodsMolecularQuantum1969,mcweenyRecentAdvancesDensity1960,jorgensenSecondQuantizationbasedMethods1981} all of which have the potential to be compared against experimental results collected from various types of pump-probe spectroscopy.\cite{schmittStructuresDipoleMoments2018,hellwegAccuracyDipoleMoments2011,taitTransientEPRReveals2015} While the method of finite difference can be used to calculate any of these derivatives, whether it be through the direct application of an external field or just modification of the one-electron integrals in the Hamiltonian to yield the one-body reduced density matrix (1-RDM), analytic derivatives benefit from improved speed and robustness.\cite{pulayAnalyticalDerivativesForces2014,yarkonyDiabolicalConicalIntersections1996a,furcheAdiabaticTimedependentDensity2002} The true barrier to obtaining these accurate properties typically lies instead in the theoretical derivation, which becomes increasingly complex as the energy expression itself grows in intricacy.

With all of this in mind, we aim to add to ASCC the functionality of one-body property calculations that can be evaluated through the 1-RDM. This first requires a theoretical derivation of the working equations to solve for ASCC's response. As one might expect, we find that the response equations closely resemble those of ground state coupled cluster.\cite{salterAnalyticEnergyDerivatives1989,shavittManyBodyMethodsChemistry2009} However, upon completing a perturbative analysis of this response, key differences appear that require ASCC to be considerably more careful when deciding which amplitudes to include. 
We therefore test multiple amplitude inclusion approaches informed by perturbative correctness of the 1-RDM via properties.
First, we investigate whether a fully reference-independent solution can be found by iteratively using ASCC's natural orbitals (NOs) as the orbital starting point for new ASCC calculations.
We next compare excited state Mulliken and L\"owdin population analyses on various charge transfer systems, one of which where EOM-CCSD has been shown to perform poorly,\cite{cluneExcitationMatchedLocal2025,tuckmanFastAccurateCharge2025} demonstrating ASCC's reliability in more difficult systems.
Finally, we use the highly accurate QUEST database\cite{chraytehMountaineeringStrategyExcited2021,sarkarBenchmarkingTDDFTWave2021} to compare excited state dipole moments between ASCC, EOM-CC, and linear response CC (LR-CC).\cite{monkhorstCalculationPropertiesCoupledcluster1977,dalgaardAspectsTimedependentCoupledcluster1983,sekinoLinearResponseCoupledcluster1984,kochCoupledClusterResponse1990,ricoSinglereferenceTheoriesMolecular1993,kochCalculationSizeintensiveTransition1994,sneskovExcitedStateCoupled2012}

\section{Theory}

\subsection{ASCC Lagrangian}\label{sec:ASCC Lagrangian}
ASCC was designed with the goal of creating an excited-state-specific method that maintains cost-parity with single-reference ground-state coupled cluster theory while retaining all of its benefits: size consistency, size extensivity, accurate correlation energies, and systematic improvability. Additionally, when dealing with excited states, size intensivity for excitation energies was also desired. With all of these in mind, the ASCC ansatz was formulated as 
\begin{equation}\label{Ansatz}
    \ket{\Psi_{\text{ASCC}}} = e^{-\hat{S}^\dagger}e^{\hat{T}}\ket{\phi_0}
\end{equation}
where $\hat{S}^\dagger$ is a de-excitation operator, $\hat{T}$ is an excitation operator, and $\ket{\phi_0}$ is the Aufbau determinant, often coming from closed-shell Hartree-Fock.\cite{tuckmanAufbauSuppressedCoupled2024} The objective of the new exponentiated de-excitation operator is to either partially or completely cancel out the Aufbau determinant through the term $-\hat{S}^{\dagger}\hat{T}\ket{\phi_0}$ in the Taylor series expansion of \Cref{Ansatz}, which excites and then de-excites back to the original determinant, but with negative contribution.\cite{tuckmanAufbauSuppressedCoupled2024}

Utilizing this ansatz, we can proceed similar to ground state coupled cluster to formulate expressions for the excited state energy and amplitude equations. The energy equation will correspond to the expectation value of the doubly similarity transformed Hamiltonian with respect to the formal reference, $\ket{\phi_0}$, and the amplitude equations to the projections onto excited determinants $\bra{\mu}$, which can be used to optimize the $\hat{T}$ amplitudes. These can be represented concisely through the definition of an ASCC Lagrangian, where enforcing stationarity with respect to the Lagrange multipliers, $\lambda_\mu$, yields the amplitude equations.
\begin{align}\label{ASCC Lagrangian Long}
\begin{split}
    \mathcal{L}_{\text{ASCC}} &= \ev**{e^{-\hat{T}} e^{\hat{S}^\dagger} \hat{H} e^{-\hat{S}^\dagger} e^{\hat{T}}}{\phi_0} \\
    &\quad\ + \sum_{\mu} \lambda_\mu \mel**{\mu}{e^{-\hat{T}} e^{\hat{S}^\dagger} \hat{H} e^{-\hat{S}^\dagger} e^{\hat{T}}}{\phi_0}
\end{split}
\end{align}
This Lagrangian is nearly identical to that of the ground state, with the only difference coming from the similarity transform of the Hamiltonian with $\hat{S}^\dagger$. While there are other CC methods that utilize multiple similarity transforms, such as transcorrelated CC or extended CC, these other methods focus on improving the accuracy of the ground state.\cite{schraivogelTranscorrelatedCoupledCluster2021,morchenNoniterativeTriplesTranscorrelated2025,arponenExtendedCoupledclusterMethod1987} On the other hand, the purpose of the second similarity transform in ASCC is to ensure that the Aufbau contribution to the wavefunction is suppressed, thus allowing it to target the desired excited state.
% While other methods exist to performing multiple similarity transforms of the Hamiltonian within the CC framework has been is functionally very similar to transcorrelated CC (TC-CC), these two transformations serve very different purposes. In TC-CC, the inner similarity transformation with a term that depends explicitly on the distance between electrons aims to include electron correlation directly into the Hamiltonian, resulting in three-body terms and often improving convergence to the basis set limit of the ground state.\cite{schraivogelTranscorrelatedCoupledCluster2021,morchenNoniterativeTriplesTranscorrelated2025} On the other hand, by restricting our definition of $\hat{S}^\dagger$ to a linear combination of solely one-body operators, the inner similarity transform in ASCC remains two-body and has a sole purpose of ensuring that the Aufbau contribution to the wavefunction is suppressed, thus allowing it to target the excited state.} 
By continuing to follow the techniques used for the ground state Lagrangian, we can define a new $\hat{\Lambda}$ operator that closely resembles the $\hat{T}$ operator, but excites the bra instead of the ket. We use $i,j,\ldots$ to denote occupied molecular orbitals, $a,b,\ldots$ virtual molecular orbitals, and $w,x,\ldots$ any molecular orbital.
\begin{gather}\label{Lambda Def}
    \hat{\Lambda} = \hat{\Lambda}_1 + \hat{\Lambda}_2 + \cdots + \hat{\Lambda}_N \\
    \hat{\Lambda}_n = \left(\frac{1}{n}\right)^2 \sum_{ij \cdots ab\cdots}^n \lambda_{ij\cdots}^{ab\cdots} \hat{i}^\dagger \hat{j}^\dagger \cdots \hat{b} \hat{a}
\end{gather}
Putting this together with \Cref{ASCC Lagrangian Long} yields an even more concise Lagrangian.
\begin{equation}\label{ASCC Lagrangian}
    \mathcal{L}_{\text{ASCC}} = \ev**{\left(1+\hat{\Lambda}\right) e^{-\hat{T}} e^{\hat{S}^\dagger} \hat{H} e^{-\hat{S}^\dagger} e^{\hat{T}}}{\phi_0}
\end{equation}

If one was solely interested in the ASCC energy, enforcing stationarity of $\hat{\Lambda}$ would be sufficient. However, if one instead desired any properties, one would ideally want stationarity of the Lagrangian with respect to every parameter.\cite{levchenkoAnalyticGradientsSpinconserving2005} Just as in the ground state theory, enforcing stationarity with respect to $\hat{T}$ also provides values for $\hat{\Lambda}$.\cite{salterAnalyticEnergyDerivatives1989} Thus, after these stationarity conditions are met, the remaining nonzero terms in the presence of a perturbation consists of the response of $\hat{S}^\dagger$, the response of the molecular orbital coefficients from the reference of choice, and the response of the Hamiltonian itself. In this preliminary study, we restrict our excited states to those defined purely by a single configuration state function (CSF); as defined in previous works, this results in $\hat{S}^\dagger$ taking the form of 
\begin{equation}
    \hat{S}_{\text{1-CSF}}^\dagger = \frac{1}{\sqrt{2}}\left(\hat{h}_{\downarrow}^{\dagger} \hat{p}_{\downarrow} + \hat{h}_{\uparrow}^{\dagger} \hat{p}_{\uparrow} \right)
\end{equation}
with $h$ and $p$ representing the ``hole" and ``particle" orbitals, respectively. The subspace that $\hat{S}^\dagger$ acts in will be referred to as the primary subspace, as this represents the orbitals where the electronic excitation primarily occurs. Conveniently, the coefficient of $\hat{S}^\dagger$ here is constant, implying that the term containing the response of $\hat{S}^\dagger$ in the 1-CSF case will be 0. We recognize that this definition leads to an ansatz that looks very similar to that of two determinant CC (TD-CC). However, ASCC and TD-CC differ in their nonlinear terms and perturbative completeness (discussed more in \Cref{Perturbative Analysis}) due to the additional similarity transformation from $\hat{S}^\dagger$, resulting in differing energies.\cite{balkovaCoupledclusterMethodOpenshell1992,balkovaTwodeterminantCoupledclusterMethod1993,javedAufbauSuppressedCoupled2026} Furthermore, ASCC is more generalizable than TD-CC, as it can describe excited states with more than 1 CSF and it can describe doubly excited states through only minor modifications of coefficients and initial guesses.\cite{javedAufbauSuppressedCoupled2026}

In ground state CC, the response of the molecular orbitals to the perturbation is often ignored when specifically calculating one-body properties. This is because CC's own orbital relaxations on top of the reference reduce the significance of the additional molecular orbital response term.\cite{salterPropertyEvaluationOrbital1987,bartlettCoupledclusterTheoryQuantum2007} In EOM- and LR-CC, the effects of orbital relaxation are more impactful since the orbitals aren't optimized for excited states. However, including additional cluster operators mitigates these effects due to the correlation method itself performing better orbital relaxations, resulting in a smaller dependence on the original orbital's response.\cite{stantonEquationMotionCoupledcluster1993,hodeckerSimilaritiesDifferencesLagrange2019,sarkarBenchmarkingTDDFTWave2021} Conversely, as ASCC already allows for significant state-specific orbital relaxation similar to ground state CC, we expect the effects of explicitly including the orbital response to be comparable in magnitude to that of the ground state. For this reason, and because issues related to perturbative completeness are likely more impactful (see \Cref{Perturbative Analysis}), we opt to ignore this small contribution in the present study. 
We also note that, because each state gets its own orbital relaxations,
the ASCC excited state wave functions will not be exactly orthogonal
to each other or to the ground state, which is common in state-specific
theories.
So long as the ansatz is accurate for the states being studied,
the degree of this non-orthogonality should be small.

Under these approximations, enforcing stationarity with respect to $\hat{\Lambda}$ and $\hat{T}$ within the 1-CSF framework leads to the result that a perturbation with respect to any parameter $\chi$ can be represented as
\begin{align}\label{Lagrangian Derivative}
    \pdv{\mathcal{L}_{\text{ASCC}}}{\chi} = \ev**{\left(1+\hat{\Lambda}\right) e^{-\hat{T}} e^{\hat{S}^\dagger} \pdv{\hat{H}}{\chi} e^{-\hat{S}^\dagger} e^{\hat{T}}}{\phi_0}
\end{align}
which again closely parallels the ground state CC Lagrangian when also neglecting the orbital response. One can separate out the coefficients of the Hamiltonian's response projected into the molecular orbital basis from the rest of the expression to create useful definitions for the one- and two-body unrelaxed reduced density matrices (1- and 2-RDM).
\begin{gather}
    \gamma_{wx}^{\text{ASCC}} = \ev**{\left(1+\hat{\Lambda}\right) e^{-\hat{T}} e^{\hat{S}^\dagger} \hat{w}^\dagger \hat{x} e^{-\hat{S}^\dagger} e^{\hat{T}}}{\phi_0} \\
    \Gamma_{wxyz}^{\text{ASCC}} = \ev**{\left(1+\hat{\Lambda}\right) e^{-\hat{T}} e^{\hat{S}^\dagger} \hat{w}^\dagger \hat{x}^\dagger \hat{z} \hat{y} e^{-\hat{S}^\dagger} e^{\hat{T}}}{\phi_0}
\end{gather}
Additionally, it can be seen from these that the energy expression can be evaluated simply by taking the product of the Hamiltonian with the 1- and 2-RDM and summing over all indices, shown in \Cref{Energy RDMS}. Furthermore, it follows from \Cref{Lagrangian Derivative} that the response of the energy to a perturbation looks nearly identical to the regular energy expression when utilizing the density matrices, with the only difference being the use of the perturbed Hamiltonian.
\begin{gather}
    E_{\text{ASCC}} = \sum_{wx} \gamma_{wx}^{\text{ASCC}} h_{wx} + \frac{1}{4} \sum_{wxyz} \Gamma_{wxyz}^{\text{ASCC}} \expval{wx \vert\vert yz} \label{Energy RDMS} \\
    \pdv{E_{\text{ASCC}}}{\chi} = \sum_{wx} \gamma_{wx}^{\text{ASCC}} \pdv{h_{wx}}{\chi} + \frac{1}{4} \sum_{wxyz} \Gamma_{wxyz}^{\text{ASCC}} \pdv{\expval{wx \vert\vert yz}}{\chi} \label{Energy Derivative}
\end{gather}
For one-body properties, the second term in \Cref{Energy Derivative} disappears, leaving just the product of the 1-RDM with the perturbed one electron integrals. As mentioned previously though, this result is only the case when the orbital response terms are ignored and the system remains 1-CSF. For references best described by more than one CSF, response of $\hat{S}^\dagger$ would also need to be taken into consideration.

Throughout this entire framework though, the focus has been solely on the right-hand eigenvectors of the Hamiltonian and ensuring that the targeted excited state is the one described by the wavefunction. However, if one instead focuses on the left-hand side of the Lagrangian, the theory becomes less straightforward. Here, the $\hat{S}^\dagger$ no longer performs its intended function of suppressing the Aufbau contribution. Instead, due to its exponentiated form, a given $n$-CSF reference for $\hat{S}^\dagger$ can actually perform a $2n$-fold excitation within the primary subspace on the bra. 
While optimization of the left eigenvectors is still achieved through $\hat{\Lambda}$, thus making ASCC bivariational just like ground state CC,\cite{arponenVariationalPrinciplesLinkedcluster1983} the overall effect that $\hat{S}^\dagger$ has on the left-hand side of the wavefunction should still be explored in more depth, especially since it wasn't required for the calculation of energies.

\subsection{Perturbative Analysis}\label{Perturbative Analysis}
Before we complete a perturbative analysis of the left-hand side of our Lagrangian to determine how $\hat{S}^\dagger$ might alter the accuracy of ASCC's properties, we will first review the perturbative analysis of the cluster operators and Hamiltonian originally outlined by Tuckman et al.\cite{tuckmanImprovingAufbauSuppressed2025} We start by examining the ASCC ansatz. In order to achieve the desired zeroth order wavefunction, it must be the case that both $\hat{S}^\dagger$ and some parts of $\hat{T}$ be nonzero at zeroth order. As the coefficient of $\hat{S}^\dagger$ is constant, this implies that it resides exclusively at zeroth order. On the other hand, as any excitations by cluster operators outside of the primary subspace cannot be offset by the de-excitation performed by $\hat{S}^\dagger$, we can conclude that any fully non-primary cluster amplitudes and mixed amplitudes containing both primary and non-primary indices must be zero at zeroth order, as is the case for ground state CC. While the specific definition of $\hat{T}^{(0)}$ will vary depending on the number of CSFs deemed necessary to describe the excitation, the single CSF case remains quite simple.
\begin{gather}
    \label{S0 def}\hat{S}_{\text{1-CSF}}^{\dagger(0)} = \hat{S}_{\text{1-CSF}}^\dagger \\
    \label{T0 def}\hat{T}_{\text{1-CSF}}^{(0)} = \hat{S}_{\text{1-CSF}} - \frac{1}{2}\left(\hat{S}_{\text{1-CSF}}\right)^2
\end{gather}

However, this definition by itself can result in undesired consequences. Consider if one were to follow the standard perturbative partitioning of the Hamiltonian, where the zeroth order contribution arises from the block diagonal occupied-occupied and virtual-virtual parts of the Fock matrix, with everything else considered first order. From this, the off-diagonal parts of the zeroth order Hamiltonian can couple with the primary indices of $\hat{T}^{(0)}$ to generate zeroth order mixed amplitudes, which we know from our ansatz should be 0 at zeroth order. Note that, even when working in the canonical Hartree-Fock orbital basis, the Fock matrix is not diagonal, because it is constructed not from the Hartree-Fock 1-body density matrix but from that of the excited state starting point (e.g., the excited CSF).
We do expect it to be diagonally dominant, though.
Thus, to prevent mixed amplitudes from becoming zeroth order,
we instead partition our zeroth order Hamiltonian into four
block diagonal pieces,
\begin{gather}
    \label{H0 def}
    \begin{split}\hat{H}^{(0)} = \sum_{h_1h_2} f_{h_1h_2} \hat{h}_1^\dagger\hat{h}_2 + \sum_{ij} f_{ij} \hat{i}^\dagger \hat{j} \\+ \sum_{p_1p_2} f_{p_1p_2} \hat{p}_1^\dagger\hat{p}_2 + \sum_{ab} f_{ab} \hat{a}^\dagger\hat{b}
    \end{split} \\
    \label{H1 def} \hat{H}^{(1)} = \hat{H} - \hat{H}^{(0)}
\end{gather}
where $h$ and $p$ still refer to hole and particle indices, but now $i$ and $j$ refer to solely non-primary occupied indices and $a$ and $b$ to solely non-primary virtual indices.
While this new partitioning of the Hamiltonian breaks invariance to orbital rotations between the hole orbital and other occupied orbitals
and between the particle orbital and other virtual orbitals,
it retains invariance with respect to rotations within each of the four subspaces (hole, non-primary occupied, particle, and non-primary virtual) individually.\cite{tuckmanImprovingAufbauSuppressed2025}

With definitions for the perturbative partitioning of the Hamiltonian and the zeroth order portions of $\hat{T}$, the determination of which cluster amplitudes are nonzero at each order can be achieved. To mirror ground state CCSD, it was decided that the $\hat{T}$ amplitudes to be included in ASCC are those that contribute at first order to the wavefunction. For 1-CSF, those include all singles, doubles, and a slice of the triples that contain at least three primary indices.\cite{tuckmanImprovingAufbauSuppressed2025} Note that while this is more than ground state CCSD, due to the primary space scaling as $O(1)$, the worst additional amplitudes scale as $O(N^3)$ in memory and contribute no additional $O(N^6)$ cost terms, thus maintaining asymptotic cost parity. A partially linearized version of ASCC (PLASCC) also uses these same definitions, but with specific non-linear terms removed in an effort to eliminate contributions from terms that would normally be canceled out at higher orders of PT.\cite{tuckmanImprovingAufbauSuppressed2025}

Upon completion of the perturbative analysis of the Hamiltonian, $\hat{S}^\dagger$, and $\hat{T}$ for 1-CSF ASCC, we finally have the tools necessary to investigate the left-hand side of the Lagrangian. Here, we find that in order to left project with purely the desired excited state, a zeroth order $\hat{\Lambda}$ described by solely a two-body operator is required (see Appendix).
\begin{equation}\label{Lambda 0 def}
    \hat{\Lambda}_{\text{1-CSF}}^{(0)} = -\left(\hat{S}_{\text{1-CSF}}^{\dagger}\right)^2
\end{equation}
In ground state CC, the $\hat{\Lambda}_n$ amplitudes appear at the same order of perturbation theory as the corresponding $\hat{T}_n$ amplitudes. However, as can be seen from the zeroth order contributions alone, this is not true for ASCC, as there is no one-body operator in the zeroth order definition of $\hat{\Lambda}$. In the determination of the first order pieces, we find that $\hat{\Lambda}^{(1)}$ contains all the same respective parts as $\hat{T}^{(1)}$, but also includes the slice of triples containing one primary and two non-primary de-excitations and the slice of quadruples with two primary de-excitations (see Appendix for nonzero quadruples contributions). While the addition of these new slices would not increase the overall asymptotic scaling of ASCC, they would add additional $O(N^6)$ tensor contractions and additional $O(N^4)$ memory requirements beyond what
is needed for CCSD.

To test whether these additional terms are worth the effort,
we will test the effectiveness of three different choices for
which amplitudes to include in $\hat{T}$ and $\hat{\Lambda}$.
First, we allow $\hat{T}$ to contain only its amplitudes that
receive first order contributions
% in the amplitude equations, 
and we restrict $\hat{\Lambda}$ to
the same set of mirrored amplitudes (i.e. singles, doubles, and the slice of triples with at least three primary indices). 
This approach, which we refer to as ASCC$^{\left(M,1\right)}$
and PLASCC$^{\left(M,1\right)}$, leads the theory to contain
the same set of $O(N^6)$ ``bottleneck'' terms that are present
in ground state CCSD.
Next, we allow $\hat{T}$ and $\hat{\Lambda}$ to each separately contain
whichever of their amplitudes receive first order contributions
in the amplitude and response equations, respectively,
an approach we refer to as ASCC$^{\left(1,1\right)}$
and PLASCC$^{\left(1,1\right)}$.
Note that, in this case, there are more amplitudes inside
$\hat{\Lambda}$ than there are in $\hat{T}$, and so the
usual approach of defining the amplitude equations as the
derivatives of $\mathcal{L}$ with respect to
the $\hat{\Lambda}$ amplitudes yields too many equations.
Instead, we form the residual equations as usual only
for the amplitudes in $\hat{T}$ that have first order
contributions (i.e.\ those present in ASCC$^{\left(M,1\right)}$).  
We then approximate the remaining amplitudes in $\hat{T}$,
each of which would be second order, by setting them equal to zero.
The $\hat{\Lambda}$ amplitudes are then solved for in the standard way:
for each amplitude in $\hat{\Lambda}$, we get an equation
to solve by setting the corresponding $\hat{T}$-amplitude-derivative
of $\mathcal{L}$ to zero, whether or not we are allowing
that $\hat{T}$ amplitude to be nonzero.
Finally, we form the
ASCC$^{\left(1,M\right)}$ and PLASCC$^{\left(1,M\right)}$
approaches by allowing $\hat{\Lambda}$ to contain any
amplitude that receives a first order contribution and
then activating the same mirrored set of
amplitudes within $\hat{T}$.

\begin{figure}[t]
    \centering
    \includegraphics[width=\linewidth]{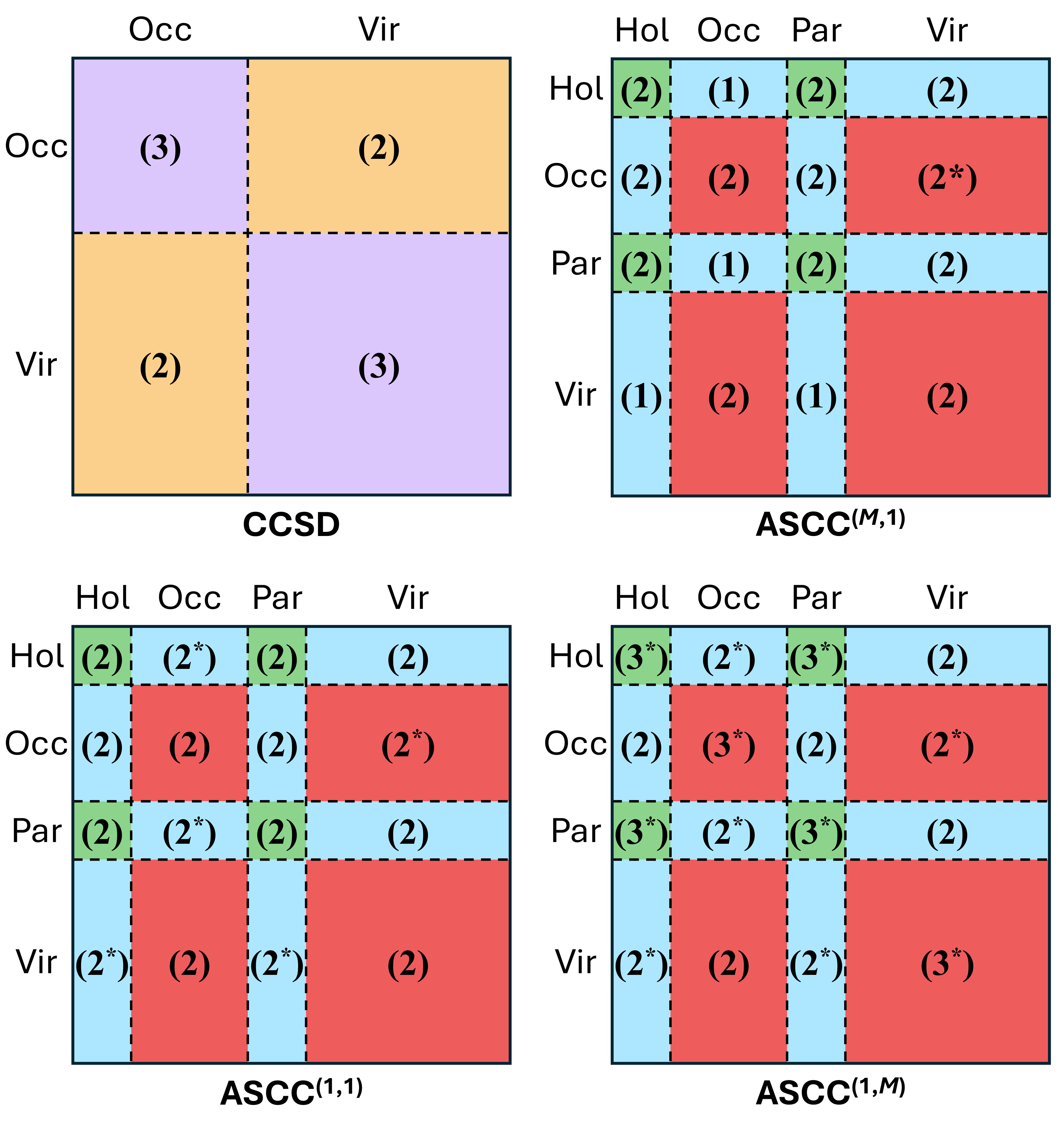}
    \caption{The CCSD and (PL)ASCC 1-RDMs with orders of perturbative correctness indicated. For CCSD, the all occupied and all virtual blocks are highlighted in purple and the occupied-virtual blocks are highlighted in orange. For (PL)ASCC, the all primary blocks are highlighted in green, the mixed blocks in blue, and the all non-primary blocks in red. The $^*$ indicates that the block is one order lower with PLASCC (e.g. \textbf{(}$\mathbf{3^*}$\textbf{)} becomes \textbf{(}$\mathbf{2}$\textbf{)} for PLASCC).}
    \label{RDMs}
\end{figure}

To get a better sense of the significance of omitting various parts of $\hat{T}$ or $\hat{\Lambda}$, we turn our attention to the perturbative correctness of the 1-RDM, shown in \Cref{RDMs}. In ground state CCSD, the 1-RDM is complete through third order in the occupied/occupied and virtual/virtual blocks, but only second order in the off-diagonal blocks. However, in ASCC$^{(M,1)}$, the four blocks along the diagonal are now only complete through second order, alongside some, but not all, of the off-diagonal sections. Interestingly, due to the lack of incorporation of the first order $\hat{\Lambda}$, the pattern of perturbative completeness in the
off-diagonal sections is not always symmetric. A more in depth analysis depicting this difference between the primary/non-primary occupied blocks ($\gamma_{hi}$ and $\gamma_{ih}$) can be found in the Appendix, \Cref{Appendix}. Upon symmetrization, the 1-RDM becomes Hermitian, with any block taking on the lesser completeness of itself and its adjoint. This results in the all primary and the all non-primary blocks remaining complete through second order, but all mixed blocks being complete through only first order. When additionally including all $\hat{\Lambda}^{(1)}$ pieces, the entirety of the ASCC$^{(1,1)}$ 1-RDM becomes complete through second order, demonstrating improvement, but still not to the level of the ground state. Finally, for ASCC$^{(1,M)}$, the perturbative completeness more closely mirrors the ground state, with the only difference being the off-diagonal all primary blocks being complete through the higher third order too. Interestingly, when including partial linearization, the perturbative completeness of the 1-RDM is equivalent for all three implementations due to the missing nonlinear $\hat{T}$ contributions. Furthermore, they are less accurate than ASCC$^{(M,1)}$.
Based off this fact alone, it seems an unnecessary expense to include
any higher order contributions when performing partial linearization. 
Indeed, we will see in \Cref{Dipole Moments} that, when using
partial linearization, including the extra amplitudes is not helpful.

\subsection{Natural Orbital Refinement}\label{NOR}

\begin{figure}[t]
    \centering
    \includegraphics[width=\linewidth]{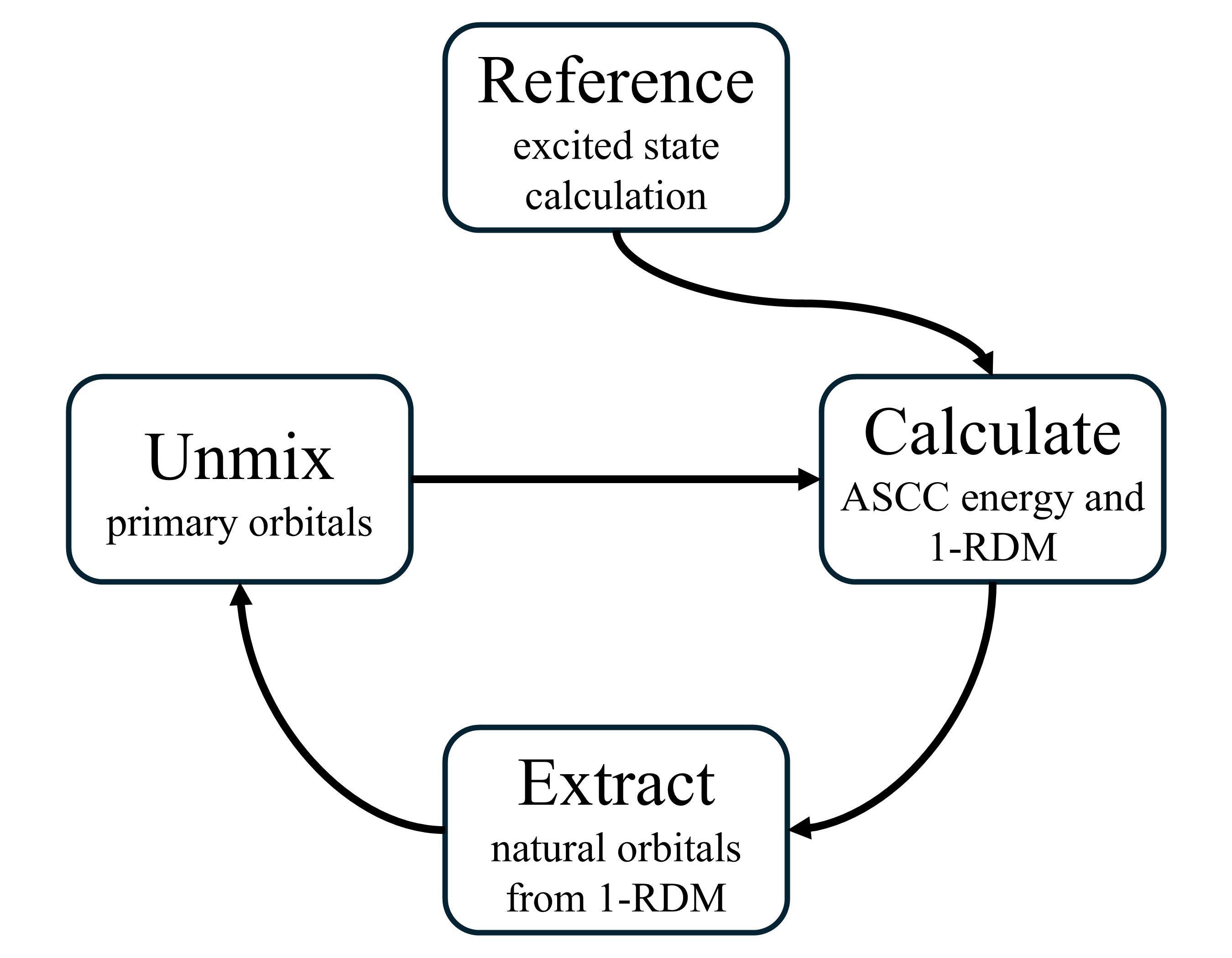}
    \caption{Schematic of the natural orbital refinement procedure. See \Cref{NOR} for details.}
    \label{NOR Method}
\end{figure}

In single-CSF ASCC, the dependence of the ansatz on the
reference wave function comes through
the starting molecular orbitals and the designation of which
are the hole and particle orbitals.
As different reference methods
(CIS, TD-DFT, ESMF, etc.)\ provide different orbital shapes,
ASCC's results are expected to
vary at least slightly from one reference to another.
Ideally, the reference orbitals would be determined by the higher accuracy ASCC method as opposed to the often more affordable (and less accurate) reference methods.
With the ability to evaluate the 1-RDM and thus ASCC's natural
orbitals (NOs), a straightforward (if a bit crude) way to explore
this concept is to iteratively feed the NOs outputted by one
ASCC calculation in as the starting molecular orbitals for another,
as depicted in \Cref{NOR Method}.
In other words, this approach allows ASCC to refine its own orbitals,
potentially eliminating any differences that might have arose from starting with CIS
vs TD-DFT vs ESMF, assuming that this SCF-like iteration
converges towards the same fixed point starting from each of
those unique references.

While this procedure sounds straightforward, an interesting problem arises very quickly. Tuckman et al.\ have previously noted that ASCC is prone to symmetry violations from a ``downward ladder effect" caused from the new zeroth order Hamiltonian being able to de-excite in the primary subspace.\cite{tuckmanImprovingAufbauSuppressed2025} Because of this symmetry violation, the off diagonal elements of the 1-RDM between orbitals of different symmetry representations do not necessarily have to be zero, meaning that, upon diagonalization, mixing between orbitals of different symmetries can occur.
As the hole and particle orbitals are typically close to half-filled
in an excited state, their diagonal elements in the 1-RDM are nearly
equal, and so even small symmetry violations in their off-diagonal
elements can result in significant particle-hole mixing in the
NOs, regardless of whether such mixing is symmetry forbidden.
To minimize this effect, we project the original particle and
hole orbitals into the two-dimensional subspace of the new
hole and particle NOs to form near-natural hole and particle
orbitals that are much more resilient to symmetry breaking.
Note that the same issue does not arise for other orbitals,
as the 1-RDM diagonal entries for
occupied, hole, particle, and unoccupied orbitals are
roughly 2.0, 1.0, 1.0, and 0.0, respectively, making the
particle-hole case the only near-degeneracy case that matters
(ASCC is invariant to mixings between doubly occupied orbitals
and mixings between unoccupied orbitals).

\subsection{Population Analysis and Dipole Moments}
In order to approximate the electronic change in populations from the ground to the excited state, both a Mulliken and a L\"owdin population analysis were employed. These analyses are two of the most common population schemes utilized due to their simplistic form and long-standing history.\cite{bachrachPopulationAnalysisElectron1994} The charge on a given atom $A$ is
\begin{equation}\label{Mulliken}
    q_A = Z_A - \sum_{\nu \in A} (\gamma_{\text{AO}} S)_{\nu \nu}
\end{equation}
for a Mulliken-style analysis and
\begin{equation}\label{Lowdin}
    q_A = Z_A - \sum_{\nu \in A} (S^{1/2} \gamma_{\text{AO}} S^{1/2})_{\nu \nu}
\end{equation}
for a L\"owdin-style analysis, where $Z_A$ is the nuclear charge of atom $A$, $\nu$ are the atomic orbitals located on atom $A$, $\gamma_{\text{AO}}$ is the 1-RDM in the atomic orbital (AO) basis, and $S$ is the atomic overlap matrix.\cite{lowdinNonOrthogonalityProblemConnected1950,szaboModernQuantumChemistry1996} The difference between the excited state and ground state atomic populations were taken to focus specifically on the charge transfer occurring. These changes were then summed over within specific moieties to gather the electronic change of an entire group instead of a specific atom. While both the Mulliken and L\"owdin population analyses are known to depend on the chosen basis set and therefore can be misleading to trust in absolutes,\cite{contrerasElectronDensitiesPopulation2017,northPopulationAnalysisEffects2023} we attempt to mitigate these uncertainties by focusing specifically on changes between the ground and excited charges, overall charges of entire fragments instead of individual atoms, and on a more qualitative analysis. Furthermore, by comparing both population schemes against each other, we lessen the chance that our results are simply a product of fortune from one specific scheme.

For excited state dipole moments, the 1-RDM in the AO basis is also needed. However, as mentioned previously, for completely accurate dipole moment calculations the response of the molecular orbitals to the changing electric field is required. These were neglected as we opted for an orbital unrelaxed 1-RDM in this study. After converting the 1-RDM into the AO basis, tracing the product of it and the atomic orbital's response to the electric field yields the electronic dipole, which can be added to the nuclear dipole for the total dipole.
\section{Results}

\subsection{Computational Details}\label{Computational Details}
Following the study by Quady et al., CIS,\cite{dreuwSingleReferenceInitioMethods2005} EOM-CCSD,\cite{roweEquationsofMotionMethodExtended1968,stantonEquationMotionCoupledcluster1993,krylovEquationofMotionCoupledClusterMethods2008} TD-DFT,\cite{rungeDensityFunctionalTheoryTimeDependent1984,burkeTimedependentDensityFunctional2005,casidaProgressTimeDependentDensityFunctional2012} and ESMF\cite{sheaCommunicationMeanField2018,sheaGeneralizedVariationalPrinciple2020,hardikarSelfconsistentFieldFormulation2020} were all used as excited state references in both the natural orbital refinement and dipole moment calculations.\cite{quadyAufbauSuppressedCoupled2025} For more details on the specifics of how initial guesses were generated from the references, we refer the reader to their paper. For the population analysis tests, ESMF was used as a reference in all systems except for the tetrafluoroethylene-ethylene charge transfer, where instead CIS was used as a reference due to convergence issues with the ESMF solution. CIS calculations were conducted with PySCF\cite{fuxOQuPyPythonPackage2024} while both EOM-CCSD and TD-DFT/$\omega$B97X-V\cite{mardirossianOB97XV10parameterRangeseparated2014} calculations were performed with Q-Chem 6.2.\cite{epifanovskySoftwareFrontiersQuantum2021} None of the calculations used the frozen core approximation, though comparisons to references that did utilize the frozen core approximation are still performed; the difference between utilizing this approximation and not is very small ($\sim$0.02 eV), so these comparisons are still valid to make.\cite{loosMountaineeringStrategyExcited2018,loosMountaineeringStrategyExcited2020} Q-Chem's excited state analysis module was used to calculate the transition density matrices and natural orbitals necessary for creating (PL)ASCC's starting guess, along with the excited state dipole moments.\cite{plasserLibwfaWavefunctionAnalysis2022,krylovOrbitalsObservablesBack2020,plasserNewToolsSystematic2014,plasserNewToolsSystematic2014a,plasserAnalysisExcitonicCharge2012} The calculation of orbital unrelaxed and relaxed LR-CCSD excited state dipoles was performed in MRCC.\cite{mesterOverviewDevelopmentsMRCC2025,kallayMRCCQuantumChemical,kallayHigherExcitationsCoupledcluster2001,kallayAnalyticFirstDerivatives2003,kallayCalculationExcitedstateProperties2004} Orbital visualizations were performed using Gabedit.\cite{alloucheGabeditGraphicalUser2011} The molecular structures used for the valence and Rydberg excitation systems of the natural orbital refinement and the dipole moment calculations can be found in the supporting information of the respective QUEST benchmark studies.\cite{loosMountaineeringStrategyExcited2018,loosMountaineeringStrategyExcited2020} The geometries for the charge transfer systems can be found in the supporting information of Kuzma et al.\cite{kozmaNewBenchmarkSet2020} and Tuckman et al.\cite{tuckmanImprovingAufbauSuppressed2025} The geometries for the water flyby test system explored using population analyses were taken from Clune et al.\cite{cluneExcitationMatchedLocal2025} For totally symmetric excitations where (PL)ASCC demonstrates two unique solutions,
\cite{tuckmanImprovingAufbauSuppressed2025}
we still average the two together to get one final energy or dipole moment value.

\subsection{ASCC Orbital Refinement}
Utilizing the framework outlined in \Cref{NOR}, we first explored ASCC natural orbital refinement on some simple valence and Rydberg excitations previously examined by Quady et al.\cite{quadyAufbauSuppressedCoupled2025} Specifically, 14 states coming from 5 different molecules were examined, with vertical excitation energy comparisons to high level theory from the QUEST database.\cite{loosMountaineeringStrategyExcited2018,loosMountaineeringStrategyExcited2020} CIS, EOM-CCSD, TD-DFT/$\omega$B97X-V, and ESMF were all used as references for both ASCC$^{(M,1)}$ and PLASCC$^{(M,1)}$. The natural orbital refinement procedure was carried out for two iterations on each state, resulting in three (PL)ASCC$^{(M,1)}$ energies for each reference: without orbital refinement, with 1 orbital refinement, and with 2 orbital refinements. The precise energy values from each calculation can be found in the Supporting Information.

\begin{figure}[t]
    \includegraphics[width=\linewidth]{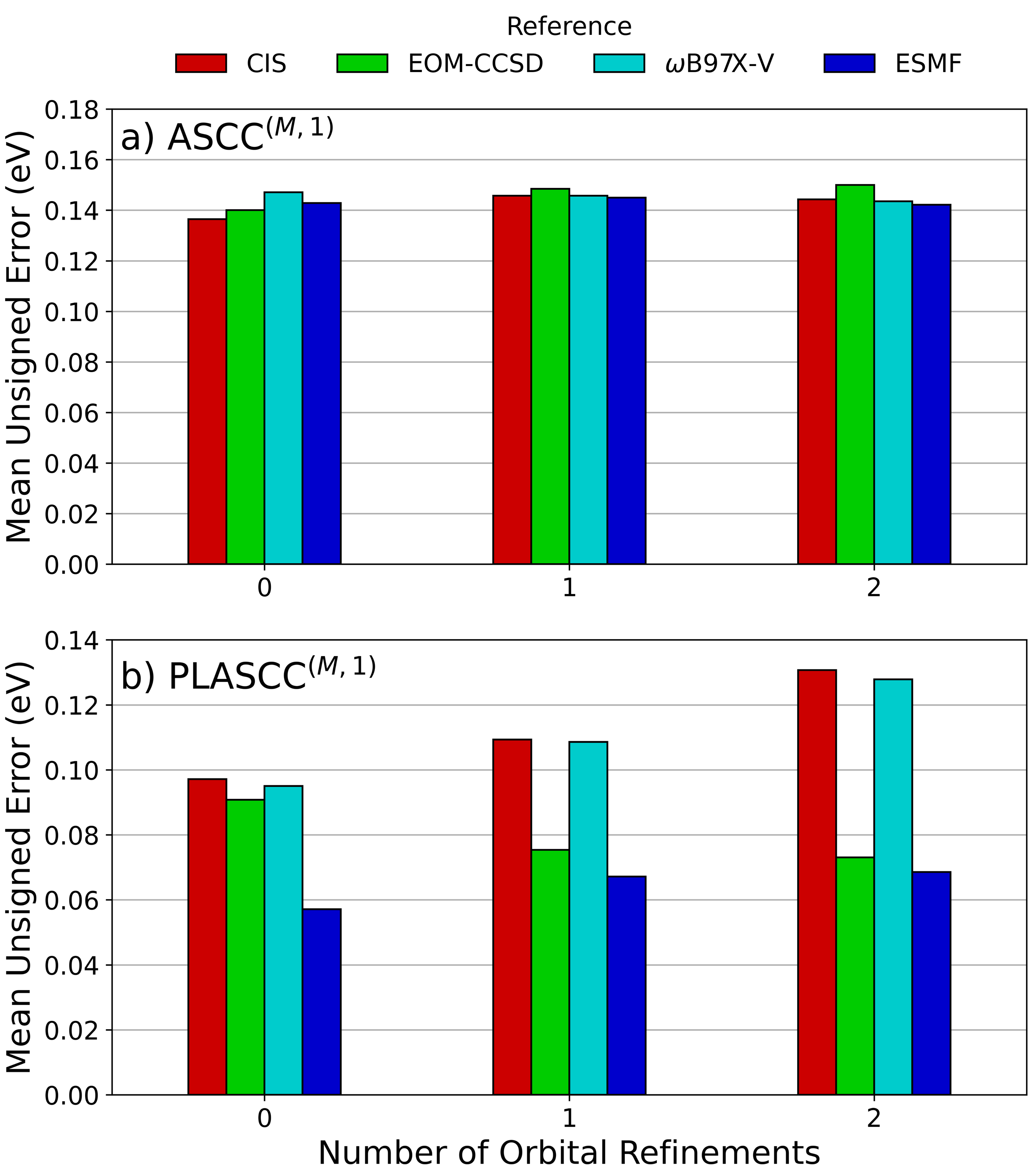}
    \caption{The mean unsigned excitation errors of 14 valence and Rydberg states from 5 molecules when starting from CIS, TD-DFT, EOM-CCSD, or ESMF starting points for a) ASCC$^{(M,1)}$ and b) PLASCC$^{(M,1)}$ after 0, 1, or 2 natural orbital refinements.}
    \label{VR NOR}
\end{figure}

\begin{figure}[t]
    \includegraphics[width=\linewidth]{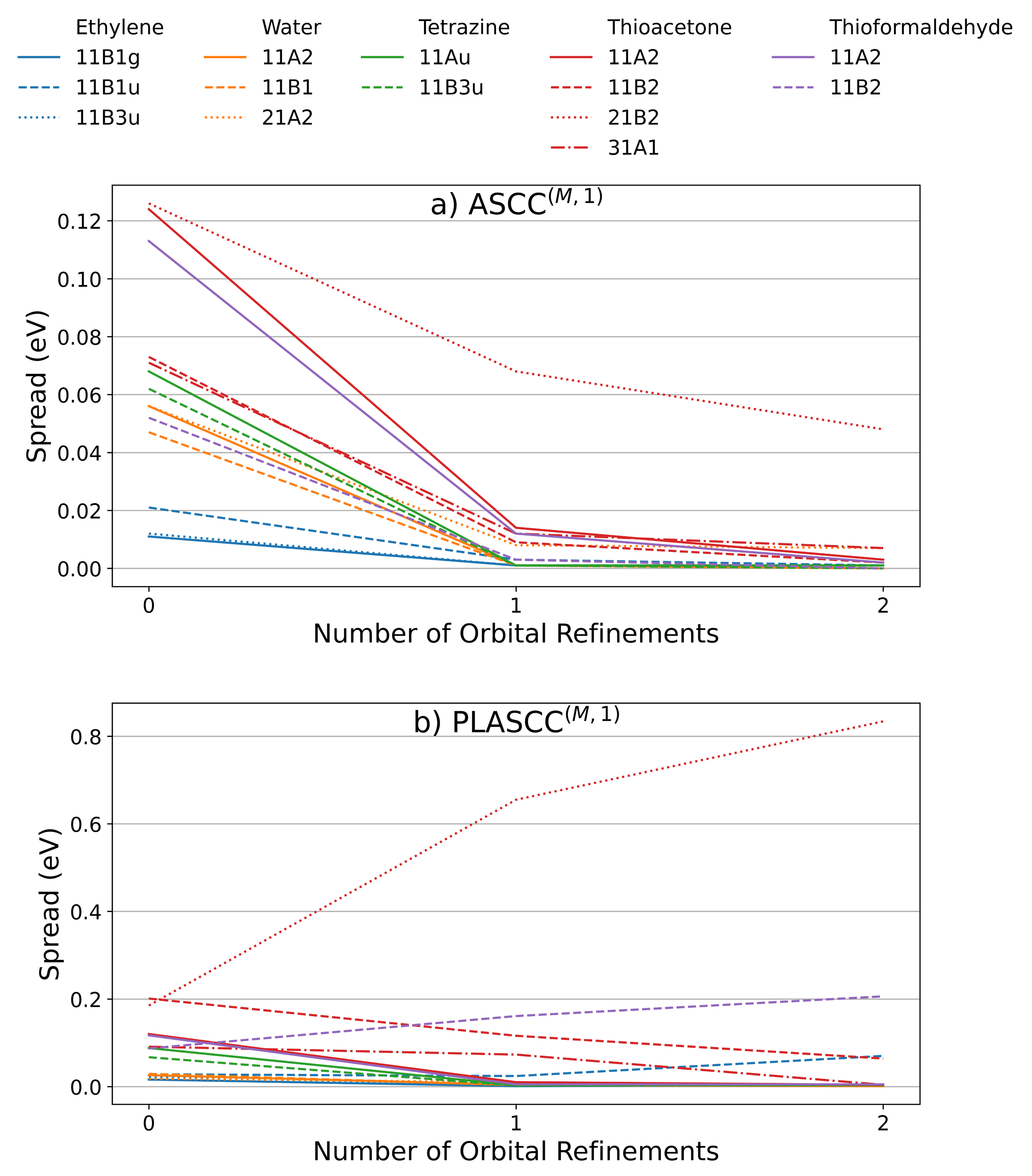}
    \caption{The spread in the excitation energies when starting from CIS, TD-DFT, ESMF, or EOM-CCSD starting points for a) ASCC$^{(M,1)}$ and b) PLASCC$^{(M,1)}$ after 0, 1, or 2 natural orbital refinements.
    The spread is measured as the difference between the maximum and minimum excitation energy predictions.}
    \label{Spread VR NOR}
\end{figure}

As seen in \Cref{VR NOR}, performing the natural orbital refinements seemed to have little effect on the overall accuracy of the vertical excitation energies in ASCC$^{(M,1)}$.
However, these refinements did reduce the methods' dependence on the
starting point, as seen in \Cref{Spread VR NOR}.
In all but one state, two iterations of natural orbital refinement brought all the starting points'
ASCC energies to within 0.01 eV of each other.
In that one state (the 2$^1B_2$ transition in thioacetone), the refinements are mitigating the
starting point dependence, but more slowly, with a roughly 0.05 eV spread left after
two cycles of refinement.
We also note that in the totally symmetric states, we witness convergence towards two unique solutions, indicating that each of the two ans{\"a}tzae is converging towards separate orbital fixed points.
With only two such states in this test set, though, we make no conclusions about whether one or the
other ansatz is more accurate.

\begin{figure*}[t]
    \centering
    \includegraphics[width=\textwidth]{Figures/Orbital_Refinement_2.png}
    \caption{The hole and particle orbitals demonstrating symmetry contamination for the acetone-difluorine charge transfer system throughout the natural orbital refinement procedure. The original reference was ESMF and ASCC$^{\left(M,1\right)}$ was utilized to perform the orbital refinements.}
    \label{Symmetry Contamination}
\end{figure*}

While PLASCC$^{(M,1)}$ typically demonstrates a lower error in vertical excitations, it did not seem to behave as well under orbital refinement as ASCC$^{(M,1)}$ did.
In fact, for 3 of the 14 states, the standard deviation of the 4 references grew after performing natural orbital refinements, as seen in \Cref{Spread VR NOR}, indicating potential divergence between the various references instead of the desired convergence. This might be caused in part by the slightly less accurate 1-RDM discussed in \Cref{Perturbative Analysis}, but as only the occupied/virtual block was of a different order, we cannot conclude that it is guaranteed to be solely from this perturbative difference. On the other hand, convergence to within 0.01 eV after two orbital refinements was still observed for 10 of the 14 states. Additionally, the $1^1B_2$ state of thioacetone showed signs of converging to this tight threshold more slowly, with an original energetic discrepancy between references of 0.20 eV shrinking to only 0.06 eV. Interestingly, the initial reference seemed to be much more significant for PLASCC. In the 3 states where divergence was observed, it seemed to be the case that EOM-CCSD and ESMF were converging towards one solution and CIS and $\omega$B97X-V converging towards another.

After seeing success for reaching a stationary point for all of the ASCC$^{(M,1)}$ calculations on simpler systems, we decided to test the more complex charge transfer (CT) systems that ASCC typically performs quite well at.\cite{tuckmanImprovingAufbauSuppressed2025} Predicting that these more difficult systems would be tougher to converge, we performed the NO refinement procedure for four iterations for each state, resulting in five (PL)ASCC$^{(M,1)}$ energies for each reference. While at times we did witness convergence to a single ASCC reference-independent solution with an error comparative to that of the original references, we also noticed issues appearing from the amplification of symmetry breaking. As stated before, we were manually attempting to unmix the new hole and particle orbitals by projecting them into the subspace of the reference hole and particle, but if the symmetry breaks a bit more each iteration, more and more of the Aufbau coefficient starts appearing when it should be 0. This resulted in some states whose references converged together, but to an incorrect solution that is not physically possible. An example of convergence to a different solution with significant symmetry contamination is depicted in \Cref{Symmetry Contamination}. Unfortunately, this wasn't always visible with just 1 or 2 orbital refinements, sometimes requiring 3 or 4 before any significant energetic issues arose. Once these problems did appear, we often then had difficulty converging the remaining calculations, indicative of the poor, symmetry contaminated starting points provided from the previous orbital refinements (see Supporting Information for calculation diagnostics). Of the 7 total CT states tested, ASCC$^{(M,1)}$ converged to a stationary point on only 3, those being the ammonia-difluorine $2^1A_1$, the pyrazine-difluorine $2^1A_2$, and the ammonia-oxygendifluorine $4^1A'$. The remaining 4 states all experienced symmetry contamination in at least 2 of the 4 references tested, and with new starting points based on these poor results, the calculations yielded nonphysical solutions if the energetic convergence threshold was even reached.

PLASCC$^{(M,1)}$ performed quite similarly on the CT systems, with the only states truly reaching convergence being ammonia-difluorine's $2^1A_1$ and tetrafluoroethylene-ethylene's $5^1B_1$. Acetone-difluorine's $3^1A''$ also seemed to reach a stationary point when using EOM-CCSD or ESMF as a reference, but not when using CIS or TD-DFT, again hinting at a potentially stronger reference dependence for PLASCC$^{(M,1)}$ than ASCC$^{(M,1)}$ for response properties. Unfortunately, the other 5 tests all either failed to converge or reached nonphysical solutions. Overall, these results demonstrate that stationary points still have the potential to be achieved, even when using the poorer 1-RDM only complete through first order and without the utilization of an acceleration scheme to improve convergence. However, this is currently only true for smaller systems where the symmetry violations are much less pronounced. To explore whether stationarity could still be achieved for larger systems, more careful consideration of how to reduce the problems arising from symmetry violations is needed.

\subsection{Population Analysis}

\begin{figure}[t]
    \centering
    \includegraphics[width=\linewidth]{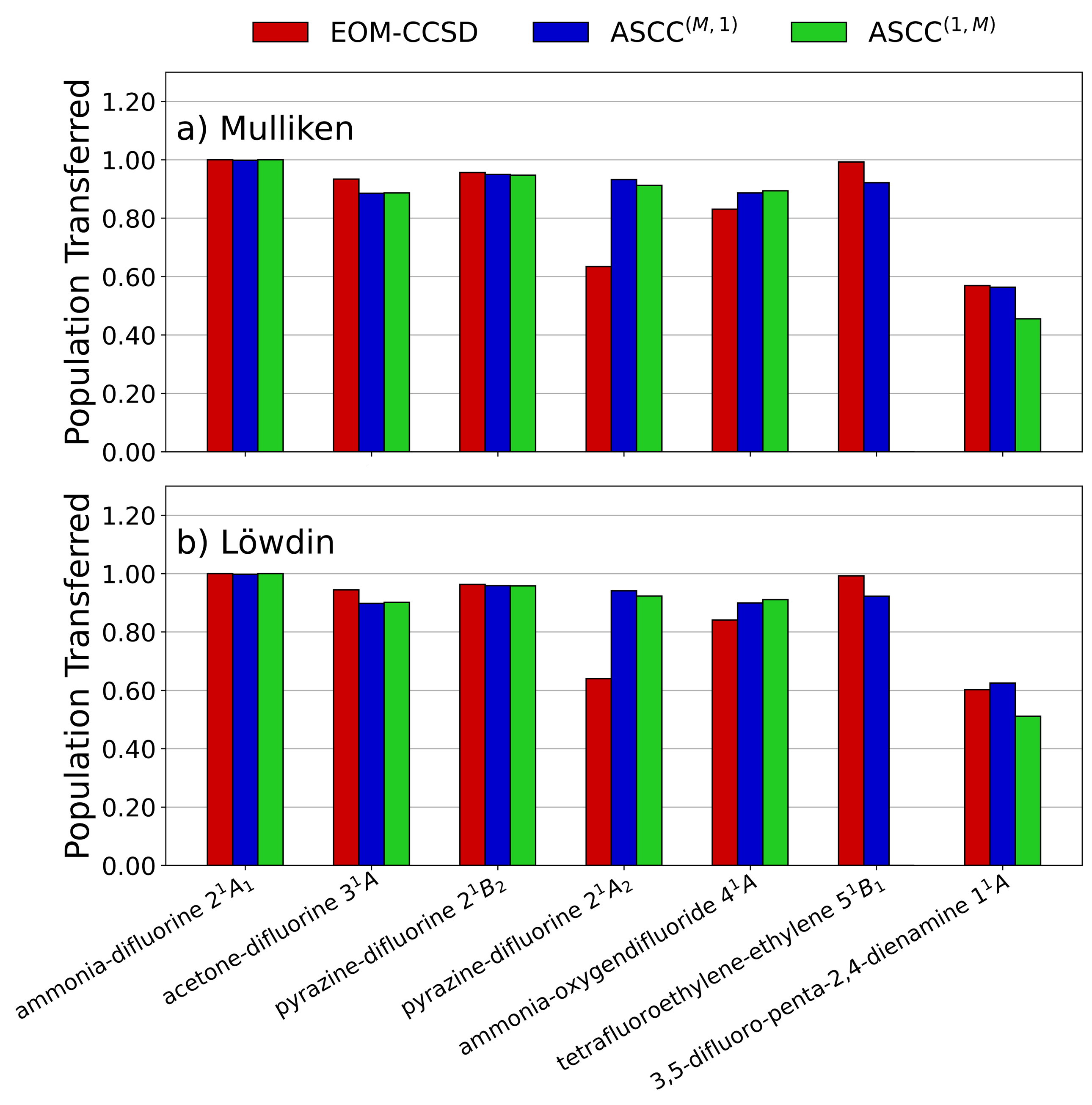}
    \caption{The a) Mulliken and b) L\"owdin populations transferred from the charge transfer excitations for both EOM-CCSD and ASCC.}
    \label{CT Populations}
\end{figure}

Despite the mixed results when attempting natural orbital refinement on the charge transfer states, the excitation energies predicted
by plain ASCC$^{\left(M,1\right)}$ for these states are more accurate than those predicted by EOM-CCSD.\cite{tuckmanImprovingAufbauSuppressed2025}
This observation begs the question:
to what degree do these methods agree or disagree about
how much charge is being transferred during these excitations?
To evaluate this,
we have carried out both Mulliken and L\"owdin population analyses.
For the intermolecular charge transfers, we add together the
populations of the atoms in the donor molecule and look at how this
total changes between the ground and excited state.
For the 3,5-difluoro-penta-2,4-dienamine intramolecular charge transfer,
we take a similar approach, with the donor moiety defined as
the primary amine and its neighboring CH$_2$ group.

As shown in \Cref{CT Populations}, EOM-CCSD and ASCC
agree about how much charge gets transferred to within 0.1 electrons
of each other for six of the seven systems tested.
We also see that the Mulliken population results are closely
matched by those of the L\"owdin approach.
The largest disagreement between EOM-CCSD and ASCC
is seen in the pyrazine-difluorine $2^1A_1$ state, where ASCC predicts
that almost a whole electron transfers, while
EOM-CCSD predicts that only a little more than 0.6 electrons transfer.
Given that this is an intermolecular charge transfer, we would expect roughly a
whole electron to transfer
(although note that this would not be our expectation for the
right-most case in the figure, as that system contains an intramolecular CT).

Without higher-level results to compare to
(none of the codes we have access to can evaluate populations for CC3
excited states),
it is difficult to draw firm conclusions about whether disagreements
between EOM-CCSD and ASCC in Figure \ref{CT Populations} imply
higher accuracy for one method or the other.
Intuition suggests that ASCC offers some advantage in the
pyrazine-difluorine $2^1A_1$ state, but overall the good agreement
in the other tests leaves us with the primary conclusion that
EOM-CCSD and ASCC tend to agree with each other on populations
in CT states.

\begin{figure}[t]
    \centering
    \includegraphics[width=\linewidth]{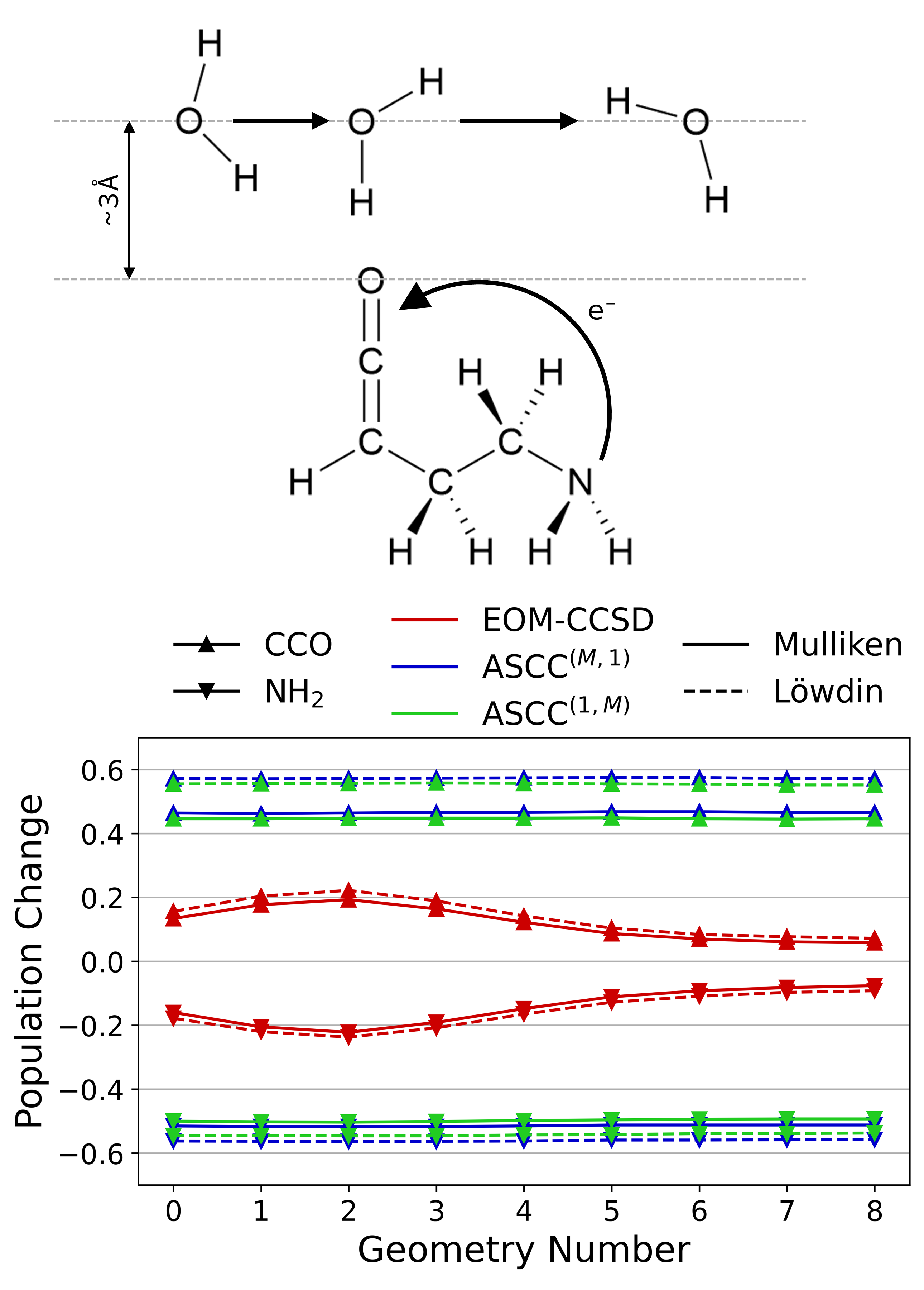}
    \caption{A schematic of the water flyby test system is shown above the EOM-CCSD and ASCC population changes of the CCO and NH$_2$ moieties upon excitation to the $^1A'$ CT state.}
    \label{Flyby}
\end{figure}

\begin{figure*}[t]
    \centering
    \includegraphics[width=\textwidth]{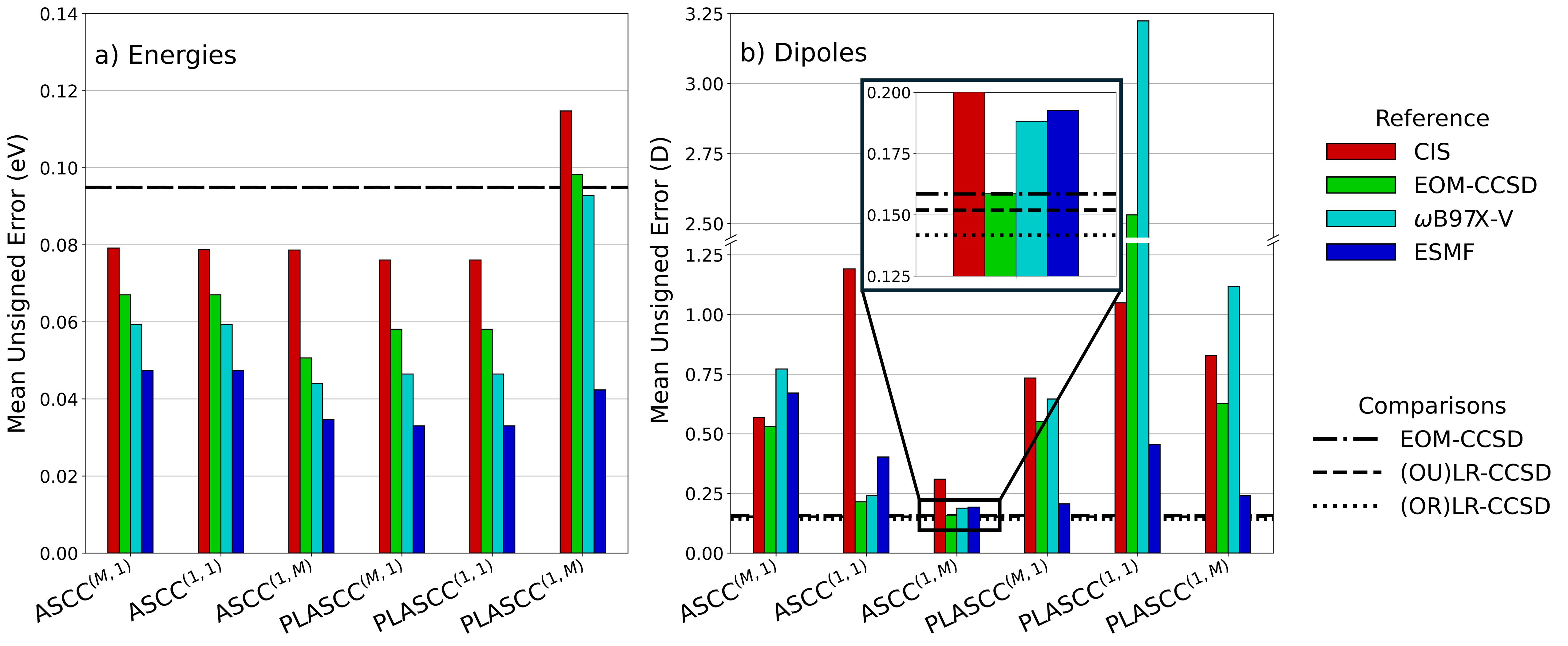}
    \caption{The mean unsigned errors for the a) vertical excitation energy (eV) and b) excited state dipole moments (D) of the various ASCC-based methods with references of CIS, EOM-CCSD, $\omega$B97X-V, and ESMF. The black lines show comparisons to EOM-CCSD, (OU)LR-CCSD, and (OR)LR-CCSD, which all produce the same energies.}
    \label{Dipoles}
\end{figure*}

Previous work has, however, made us aware of an
intramolecular CT example in which ASCC produces a qualitatively more correct
excited state than EOM-CCSD, and so it is also worth considering
their population predictions in that case.
Specifically, this test involves moving a water molecule past a donor-bridge-acceptor molecule with an intramolecular charge transfer from the nitrogen lone pair to the in-plane $\pi^*$ orbital on the ketene moiety (the $^1A'$ charge transfer state), shown in \Cref{Flyby}. Previously, it was shown that EOM-CCSD incorrectly mixes the CT state with a low-lying Rydberg excitation in a roughly 50:50 ratio.\cite{cluneExcitationMatchedLocal2025,tuckmanFastAccurateCharge2025} This implies that it should demonstrate less explicit charge transfer character, especially if any of the AOs on the primary amine contribute to the disperse MO of the Rydberg state. Additionally, interactions between the moving water molecule and the large MOs of the Rydberg excitation would alter the amount of population transfer occurring, which wouldn't be as dramatic if only the $n \rightarrow \pi^*$ charge transfer was involved.
Indeed, these are both exactly what we see in \Cref{Flyby}. The ketene moiety where the $\pi^*$ bond is localized gains a constant electronic population of roughly $0.5$ electrons as the water passes by for both ASCC$^{\left(M,1\right)}$ and ASCC$^{\left(1,M\right)}$, but EOM-CCSD predicts varying transfers ranging from $0.05$ to $0.25$ electrons, depending on the water's position. Similarly, the change on the amine group remains constant for ASCC at around $-0.5$, whereas for EOM-CCSD it again fluctuates when the water is moved. This example demonstrates that ASCC's ability to correctly
place the CT excitation below the Rydberg states and thus avoid
CT/Rydberg mixing is very significant for its population predictions,
especially when compared to the predictions of EOM-CCSD.

\subsection{Dipole Moments}\label{Dipole Moments}

Excited state dipole moment calculations were performed on 14 states across 5 different molecules. As before, CIS, EOM-CCSD, TD-DFT/$\omega$B97X-V, and ESMF were all used as references for all 3 variations of ASCC and PLASCC discussed in \Cref{Perturbative Analysis}. The overall dipole moments are compared to calculations performed by Chrayteh et al. at the LR-CCSDTQP level of theory for H$_2$O, CO, and H$_2$S and the LR-CCSDTQ level of theory for formaldehyde and thioformaldehyde.\cite{chraytehMountaineeringStrategyExcited2021} These high level LR-CC calculations all included orbital relaxation from coupled-perturbed Hartree Fock, but, as discussed previously, this effect is mitigated as more excitations are included, so it is still valid to compare to our unrelaxed ASCC values. We have also included comparisons to EOM-CCSD, orbital unrelaxed (OU) LR-CCSD, and orbital relaxed (OR) LR-CCSD to demonstrate the magnitude of errors common in these types of calculations. The results can be seen in \Cref{Dipoles}.

As expected, every variant of ASCC and nearly every variant of PLASCC performs slightly better than EOM-CCSD in terms of vertical excitation energies for the systems examined. As altering the number of left-hand amplitudes by itself has no effect on the energies, (PL)ASCC$^{(M,1)}$ and (PL)ASCC$^{(1,1)}$ both give the same results. When increasing the number of right-hand amplitudes for ASCC, the energies improve, which is rather unsurprising as the energy becomes correct through third order instead of second order. On the other hand, it seems that the addition of more amplitudes to PLASCC has the opposite effect, actually becoming worse with additional $\hat{T}$ amplitudes. This isn't especially surprising, as the inclusion of these additional amplitudes offsets the original design to balance their missing contributions. Thus, an update to this partial linearization trick is required to have the same intended effects of balancing further missing contributions. 
For this small subset of tests, ASCC with the additional $\hat{T}$ amplitudes seemed to perform roughly equal to PLASCC without them, but more testing should be conducted before concluding that this is typically observed. 

Upon investigation of dipole moments, the accuracy of the results depended heavily upon the method of use. For ASCC, we see the systematic improvability expected from the perturbative correctness of the various 1-RDMs. While the GS cost-equivalent ASCC$^{(M,1)}$ performed quite poorly, maintaining perturbative equivalence to other methods with ASCC$^{(1,M)}$ yielded errors on par with these other methods. While this test set is still quite limited, this begs the question of whether ASCC would outperform these other methods where it usually outperforms them for energies by a more significant margin, such as charge-transfer systems. Additionally, it would be interesting to explore how significantly the addition of orbital relaxation reduces the effects of the reference itself. 

Unfortunately, this same story is not true for PLASCC. As one might have expected based solely from the fact that adding in additional amplitudes does not actually improve the 1-RDM perturbative correctness for PLASCC, adding in more amplitudes also did not improve the accuracy of the dipole moments. In fact, it often actually decreased the accuracy of the calculations, pointing again quite strongly to a more careful consideration of how to perform partial linearization for both more amplitudes and for canceling potential errors on the left-hand side. Additionally, PLASCC again demonstrated a much larger dependence upon the original reference chosen for dipole moments, with ESMF references performing better than others by at least a quarter of a Debye, and often more, irrespective of the perturbative truncation scheme employed.

\section{Conclusion}

We have derived the excited state one-body reduced density matrix for single-CSF Aufbau suppressed coupled cluster theory for the evaluation of one-body properties. The 1-RDM was first used to calculate natural orbitals in an attempt to find a reference-independent stationary point for ASCC. For simple valence and Rydberg systems, we showed that this can be achieved, but the converged solution was not significantly more accurate than any of the solutions using the original references. However, with more difficult systems like charge transfer excitations, convergence was reached on only a fraction of the tests, indicating that a closer analysis of the effects and potential prevention of symmetry violation in both the right- and left-hand amplitudes should occur. For this reason, we currently do not recommend performing natural orbital refinement within ASCC as a general strategy, but we have hope that an improved approach to reach the reference-independent stationary point can be devised. Despite this setback, we still showed that ASCC's response can be useful where other methods struggle; this was demonstrated through the successful predictions of minimal change to atomic populations on a charge transfer despite a small modification to the surroundings, whereas EOM-CCSD incorrectly predicted less charge transfer characteristics and a larger dependence upon the surroundings. Through a perturbative analysis, we found that when only including the first order excitation amplitudes (similar to ground state CCSD), the 1-RDM is significantly less accurate than its ground state counterpart. However, upon the inclusion of some additional amplitudes that don't change the asymptotic scaling, we can complete the 1-RDM through the same orders of PT as CCSD. While doing so had minimal effect on the accuracy of vertical excitation energies, for which regular ASCC already performed well in the systems tested, the response-informed variant ASCC$^{(1,M)}$ demonstrated a significantly higher accuracy in predicting excited state dipole moments, roughly on par with orbital unrelaxed LR-CCSD. Thus, while ASCC$^{(1,M)}$ mostly reproduces the accuracy of EOM-CCSD for excited state populations and dipoles,
it should be emphasized that it does so while simultaneously
delivering improved excitation energies in CT systems.

The completion of analytic first derivatives would allow for multiple new areas to explore in the future. Likely one of the most exciting next steps is the addition of the 2-RDM in order to calculate nuclear gradients, which can in turn be used for excited state geometry optimizations. To be completely accurate, one would also want to include the response of the initial reference's orbital basis so that orbital relaxations from the perturbation can be taken into account, which are especially important when nuclei are moved. Additionally, if one had the response of the reference method, we could extend the results past the 1-CSF case, allowing for the calculation of properties even if the state is multireference in character. With analytic first derivatives completed, a reformulation of PLASCC, or devising an alternative to PLASCC, to better accommodate response accuracy should also be explored. Focusing instead on speed, one could extend the Lagrangian formalism to calculate properties of the excited state using a second-order Aufbau suppressed perturbation theory, which is non-iterative $O(N^5)$ scaling, as originally proposed by Tuckman and Neuscamman.\cite{tuckmanFastAccurateCharge2025} This would also mitigate the effect of adding in the additional amplitudes required to bring ASCC's response accuracy on par with that of the ground state. Combining these approaches, one could also potentially nest the CC response equations within the perturbation code, providing CC level accuracy at the PT level of cost.
\section{Appendix}

\subsection{Perturbative Analysis of the Response Amplitudes and 1-RDM}\label{Appendix}

While the Lagrange formalism is a useful way to represent the ASCC equations concisely, it can alternatively be useful to work within the framework of fully connected diagrams, in which the residual equations for the lambda amplitudes become 
\begin{equation}\label{Lambda Residual}
\begin{split}
    &0 = \mel{\phi_0}{\left(1+\hat{\Lambda}\right) \left(\left(e^{\hat{S}^\dagger}\hat{H}_N\right)_C e^{\hat{T}}\right)_C }{\phi_{ij\ldots}^{ab\ldots}}_C \\
    &+ \sum_{\substack{k<l<\cdots\\c<d<\ldots}} \mel{\phi_0}{\left(e^{\hat{S}^\dagger}\hat{H}_N\right)_C e^{\hat{T}}}{\phi_{kl\ldots}^{cd\ldots}}_C \mel{\phi_{kl\ldots}^{cd\ldots}}{\hat{\Lambda}}{\phi_{ij\ldots}^{ab\ldots}}
\end{split}
\end{equation}
with $\hat{H}_N$ representing the normal ordered Hamiltonian and the subscript $C$ designating only fully connected diagrams.\cite{shavittManyBodyMethodsChemistry2009,salterAnalyticEnergyDerivatives1989} This representation will allow us to specify which term specifically leads to the odd behavior observed in the following perturbative analysis. However, for uncovering the zeroth order $\hat{\Lambda}$ amplitudes, it is arguably simpler to focus on just the wavefunction itself. As discussed in \Cref{sec:ASCC Lagrangian} and demonstrated by Tuckman et al.,\cite{tuckmanImprovingAufbauSuppressed2025} evaluating the Taylor series expansion of \Cref{Ansatz} at zeroth order using definitions from \Cref{S0 def,T0 def} leads to the result that the zeroth order wavefunction is simply $\hat{S}_{\text{1-CSF}}\ket{\phi_0}$. From here on out, we will drop the 1-CSF label, as this analysis is confined to this specific type of system. As we similarly desire the zeroth order bra to be $\bra{\phi_0}\hat{S}^\dagger$, then we must ensure that the left projection onto our Hamiltonian in the Lagrangian simplifies to this, providing an expression relating $\hat{\Lambda}^{(0)}$ to the zeroth order wavefunction.
\begin{align}
    &\bra{\phi_0}\hat{S}^\dagger = \bra{\phi_0}\left(1+\hat{\Lambda}^{(0)}\right)e^{-\hat{T}^{(0)}}e^{\hat{S}^\dagger} \\&
    = \bra{\phi_0}\left(1+\hat{\Lambda}^{(0)}\right)e^{-\hat{S} + \frac{1}{2}\hat{S}^2}e^{\hat{S}^\dagger} \\
    \label{app:expanded}&= \bra{\phi_0} \left(1+\hat{\Lambda}^{(0)}\right)\left(1-\hat{S} + \hat{S}^2\right)\left(1+\hat{S}^\dagger + \frac{1}{2}(\hat{S}^\dagger)^2\right)
\end{align}
From \Cref{app:expanded}, we find that $\hat{\Lambda}^{(0)}$ must reside solely within the primary subspace just like $\hat{T}^{(0)}$, as any extension outside of this subspace could not be balanced using just $\hat{S}$ and its adjoint. Equivalently, this implies that $\hat{\Lambda}^{(0)}=a\hat{S}^\dagger + b(\hat{S}^\dagger)^2$. By right projecting with 
$\ket{\phi_0}$ and $\hat{S}^2\ket{\phi_0}$ onto both sides of \Cref{app:expanded}, we get a two-by-two system of equations. Solving this, we find that $a=b+1$ from the first equation and $a=-(b+1)$ in the second, implying that $a=0$ and that $b=-1$. Thus, we are left with the result from \Cref{Lambda 0 def}, that $\hat{\Lambda}^{(0)}=-(\hat{S}^\dagger)^2$. A perturbative analysis to zeroth order of \Cref{Lambda Residual} yields an equivalent result, further confirming that this Taylor series approach is valid.

With this definition of $\hat{\Lambda}^{(0)}$ and \Cref{Lambda Residual}, one can verify that the additional slices of the triples and quadruples discussed in \Cref{Perturbative Analysis} are indeed nonzero at first order. To demonstrate this, we will show that there is a nonzero contribution to the slice of quadruples at first order, as this term is the most unusual. By working in the untruncated limit, consider the right projection with the slice of the quadruples containing four primary indices in \Cref{Lambda Residual}. To first order, the first term only provides nonzero contributions containing $\hat{\Lambda}_4^{(1)}$ itself. If the second term were zero, like in the ground state, then this would guarantee $\hat{\Lambda}_4^{(1)}=0$. However, the second term can provide contribution from the product of the double de-excitation of $\hat{H}^{(1)}$ in the solely non-primary space and the double de-excitation of $\hat{\Lambda}^{(0)}$ in the all primary space. While this term alone is enough to demonstrate that the quadruples are not all zero, there is an additional contribution from the single primary de-excitation resulting from the similarity transformation of $\hat{H}^{(0)}$ with $\hat{S}^\dagger$ paired with the first order $\hat{\Lambda}_3^{(1)}$ with two primary indices and a primary de-excitation. These contributions arise solely due to the $\hat{\Lambda}$ amplitudes containing contributions from disconnected terms, which is unlike the $\hat{T}$ amplitudes.

Completing a full perturbative analysis of both the $\hat{T}$ and $\hat{\Lambda}$ operators at each order, it is likely unsurprising that more differences appear at higher orders. We will focus our attention on ASCC$^{(M,1)}$, where we can show that these differences lead to the asymmetric perturbative accuracy of the 1-RDM shown in \Cref{RDMs}. All of the $\hat{T}$ amplitudes that are included in ASCC$^{(M,1)}$ are complete through second order, but there are no pieces complete through third order. On the other hand, since $\hat{\Lambda}$ is missing some parts of its first order amplitudes, the only amplitudes complete through second order are restricted to the all primary singles and doubles and the all non-primary singles. For example, the mixed single de-excitation of $\hat{\Lambda}_1^{(2)}$ has contribution from the missing $\hat{\Lambda}_{3}^{(1)}$ slice with two primary indices and one primary de-excitation when contracted with the slice of $\hat{T}_{3}^{(1)}$ with three primary indices and one primary excitation, demonstrated by the diagrammatic sketch of \Cref{Lambda 2}. Here, the think black bar on top designates $\hat{\Lambda}$, whereas the thinner black bar on top designates $\hat{S}^\dagger$.
\begin{equation}\label{Lambda 2}
    (\lambda_{h}^{a})^{(2)} \leftarrow 
    \vcenter{\hbox{\begin{tikzpicture}
        \node (gl) at (0,0) {};
        \node[draw, cross out, thick, minimum size=5pt,inner sep=0pt,label={[label distance=-1pt]right:(0)}] (gr) at (0.67,0) {};
        \draw[dashed,thick] (gl.center) -- (gr);
        \draw[thick,mid arrow rev=0.5] (gl.center) -- (-0.25,0.5) node[pos=1] (s) {} node[pos=0.6,label={[label distance=-3pt]right:$h$}] {};
        \draw[very thick] (s.center) -- ++(-0.2,0);
        \draw[very thick] (s.center) -- ++(0.2,0) node[pos=1,label={[label distance=-2pt]right:(0)}] {};
        \draw[thick, mid arrow=0.6] (gl.center) -- (-0.2,-0.4) node[pos=0.6,label={[label distance=-3pt]right:$h$}] {};
        \draw[thick,mid arrow rev=0.5] (s.center) -- ++(-0.4,-1) coordinate (tright) node[pos=0.25,label={[label distance=-4pt]left:$p$}] {};
        \draw[very thick] (tright.center) -- ++(0.2,0);
        \draw[very thick] (tright.center) -- ++(-1.4,0) node[pos=1] (tmid) {};
        \draw[very thick] (tmid.center) -- ++(-1,0) node[pos=1] (tleft) {};
        \draw[very thick] (tleft.center) -- ++(-0.2,0) node[pos=1,label={[label distance=-5pt]left:(1)}] {};
        \draw[thick, mid arrow rev=0.5] (tright.center) -- ++(-0.4,1) node[pos=1] (lright) {} node[pos=0.3,label={[label distance=-4pt]left:$c$}] {};
        \draw[line width=3pt] (lright.center) -- ++(0.2,0);
        \draw[line width=3pt] (lright.center) -- ++(-1,0) node[pos=1] (lmid) {};
        \draw[line width=3pt] (lmid.center) -- ++(-1,0) node[pos=1] (lleft) {};
        \draw[line width=3pt] (lleft.center) -- ++(-0.2,0) node[pos=1,label={[label distance=-7pt]left:(1)}] {};
        \draw[thick, mid arrow rev=0.75] (lright.center) -- ++(-0.2,-0.4) node[pos=0.75,label={[label distance=-4pt]left:$a$}] {};
        \draw[thick, mid arrow=0.5375] (tmid.center) to[out=60,in=-60] node[midway,label={[label distance=-4pt]right:$b$}] {} (lmid.center);
        \draw[thick, mid arrow rev=0.525] (tmid.center) to[out=120,in=-120] node[midway,label={[label distance=-4pt]left:$i$}] {} (lmid.center);
        \draw[thick, mid arrow=0.5375] (tleft.center) to[out=60,in=-60] node[midway,label={[label distance=-4pt]right:$p$}] {} (lleft.center);
        \draw[thick, mid arrow rev=0.525] (tleft.center) to[out=120,in=-120] node[midway,label={[label distance=-4pt]left:$h$}] {} (lleft.center);
    \end{tikzpicture}}}
\end{equation}
With this in mind, consider the $\gamma_{ih}$ and $\gamma_{hi}$ elements of the 1-RDM. For $\gamma_{ih}$, the non-primary creation operator must be balanced by a $\hat{T}$ operator, implying that $\hat{T}$ must be at least first order, restricting $\hat{\Lambda}$ to first or zeroth order. This does not cause any issues, as we have all $\hat{T}^{(1)}$ and the corresponding $\hat{\Lambda}^{(1)}$ needed to balance them out. Alternatively, if we wanted to use a $\hat{T}^{(2)}$, then we are restricted to $\hat{\Lambda}^{(0)}$, forcing us to use a slice of $\hat{T}_2^{(2)}$ with three primary indices, which we also have. Thus, $\gamma_{ih}$ is complete through second order. However, for $\gamma_{hi}$, these restrictions are reversed. Thus, a valid term appears containing the all-primary single excitation of $\hat{T}_1^{(0)}$ and the mixed single de-excitation of $\hat{\Lambda}_{1}^{(2)}$, of which we don't have because of contributions like those in \Cref{Lambda 2}. This type of interaction leads to the imbalances of the perturbative accuracy in the 1-RDM witnessed in \Cref{RDMs}. Similarly, one can show that the removal of terms from the partial linearlization of PLASCC will reduce which parts of $\hat{T}$ and $\hat{\Lambda}$ are complete through second order and higher, resulting in the lower perturbative accuracy of PLASCC in these same types of situations.

\section{Acknowledgements}

This work was supported by the National Science Foundation,
Award Number 2320936.
Calculations were performed 
using the Savio computational cluster resource provided by the Berkeley Research Computing program at the University of California, Berkeley.
C.B. acknowledges that this 
material is based upon work supported by the U.S. Department of Energy, Office of Science, Office of Advanced Scientific Computing Research, Department of Energy Computational Science Graduate Fellowship under Award Number DE-SC0025528.
H.T. acknowledges that this material is based upon work supported by the National Science Foundation Graduate Research Fellowship Program under Grant No. DGE 2146752.
Any opinions, findings, and conclusions or recommendations expressed in this material are those of the authors and do not necessarily reflect the views of the Department of Energy or the National Science Foundation.

\section{References}
\bibliographystyle{achemso}
\bibliography{main}

@article{alloucheGabeditGraphicalUser2011,
  title = {Gabedit---{{A}} Graphical User Interface for Computational Chemistry Softwares},
  author = {Allouche, Abdul-Rahman},
  year = 2011,
  journal = {J. Comput. Chem.},
  volume = {32},
  number = {1},
  pages = {174--182},
  issn = {1096-987X},
  doi = {10.1002/jcc.21600},
  urldate = {2026-02-12},
  copyright = {Copyright \copyright{} 2010 Wiley Periodicals, Inc.},
  langid = {english}
}

@article{arponenExtendedCoupledclusterMethod1987,
  title = {Extended Coupled-Cluster Method. {{I}}. {{Generalized}} Coherent Bosonization as a Mapping of Quantum Theory into Classical {{Hamiltonian}} Mechanics},
  author = {Arponen, J. S. and Bishop, R. F. and Pajanne, E.},
  year = 1987,
  month = sep,
  journal = {Phys. Rev. A},
  volume = {36},
  number = {6},
  pages = {2519--2538},
  issn = {0556-2791},
  doi = {10.1103/PhysRevA.36.2519},
  urldate = {2026-02-12},
  copyright = {http://link.aps.org/licenses/aps-default-license},
  langid = {english}
}

@article{arponenVariationalPrinciplesLinkedcluster1983,
  title = {Variational Principles and Linked-Cluster Exp {{{\emph{S}}}} Expansions for Static and Dynamic Many-Body Problems},
  author = {Arponen, Jouko},
  year = 1983,
  month = dec,
  journal = {Annals of Physics},
  volume = {151},
  number = {2},
  pages = {311--382},
  issn = {0003-4916},
  doi = {10.1016/0003-4916(83)90284-1},
  urldate = {2026-02-11}
}

@incollection{bachrachPopulationAnalysisElectron1994,
  title = {Population {{Analysis}} and {{Electron Densities}} from {{Quantum Mechanics}}},
  booktitle = {Reviews in {{Computational Chemistry}}},
  author = {Bachrach, Steven M.},
  year = 1994,
  pages = {171--228},
  publisher = {John Wiley \& Sons, Ltd},
  doi = {10.1002/9780470125823.ch3},
  urldate = {2026-02-02},
  isbn = {978-0-470-12582-3},
  langid = {english}
}

@article{bagusSelfConsistentFieldWaveFunctions1965,
  title = {Self-{{Consistent-Field Wave Functions}} for {{Hole States}} of {{Some Ne-Like}} and {{Ar-Like Ions}}},
  author = {Bagus, P. S.},
  year = 1965,
  month = aug,
  journal = {Phys. Rev.},
  volume = {139},
  number = {3A},
  pages = {A619-A634},
  issn = {0031-899X},
  doi = {10.1103/PhysRev.139.A619},
  urldate = {2025-10-20},
  copyright = {http://link.aps.org/licenses/aps-default-license},
  langid = {english}
}

@article{balkovaCoupledclusterMethodOpenshell1992,
  title = {Coupled-Cluster Method for Open-Shell Singlet States},
  author = {Balkov{\'a}, Anna and Bartlett, Rodney J.},
  year = 1992,
  month = jun,
  journal = {Chemical Physics Letters},
  volume = {193},
  number = {5},
  pages = {364--372},
  issn = {00092614},
  doi = {10.1016/0009-2614(92)85644-P},
  urldate = {2026-01-27},
  copyright = {https://www.elsevier.com/tdm/userlicense/1.0/},
  langid = {english}
}

@article{balkovaTwodeterminantCoupledclusterMethod1993,
  title = {The Two-Determinant Coupled-Cluster Method for Electric Properties of Excited Electronic States: {{The}} Lowest 1 {{{\emph{B}}}} 1 and 3 {{{\emph{B}}}} 1 States of the Water Molecule},
  shorttitle = {The Two-Determinant Coupled-Cluster Method for Electric Properties of Excited Electronic States},
  author = {Balkov{\'a}, Anna and Bartlett, Rodney J.},
  year = 1993,
  month = nov,
  journal = {J. Chem. Phys.},
  volume = {99},
  number = {10},
  pages = {7907--7915},
  issn = {0021-9606, 1089-7690},
  doi = {10.1063/1.465668},
  urldate = {2026-01-27},
  langid = {english}
}

@article{barcaSimpleModelsDifficult2018,
  title = {Simple {{Models}} for {{Difficult Electronic Excitations}}},
  author = {Barca, Giuseppe M. J. and Gilbert, Andrew T. B. and Gill, Peter M. W.},
  year = 2018,
  month = mar,
  journal = {J. Chem. Theory Comput.},
  volume = {14},
  number = {3},
  pages = {1501--1509},
  publisher = {American Chemical Society},
  issn = {1549-9618},
  doi = {10.1021/acs.jctc.7b00994},
  urldate = {2025-10-20}
}

@article{bartlettCoupledclusterRevolution2010,
  title = {The Coupled-Cluster Revolution},
  author = {Bartlett, Rodney J.},
  year = 2010,
  month = nov,
  journal = {Mol. Phys.},
  volume = {108},
  number = {21-23},
  pages = {2905--2920},
  publisher = {Taylor \& Francis},
  issn = {0026-8976},
  doi = {10.1080/00268976.2010.531773},
  urldate = {2026-01-29}
}

@article{bartlettCoupledclusterTheoryQuantum2007,
  title = {Coupled-Cluster Theory in Quantum Chemistry},
  author = {Bartlett, Rodney J. and Musia{\l}, Monika},
  year = 2007,
  month = feb,
  journal = {Rev. Mod. Phys.},
  volume = {79},
  number = {1},
  pages = {291--352},
  issn = {0034-6861, 1539-0756},
  doi = {10.1103/RevModPhys.79.291},
  urldate = {2026-02-02},
  copyright = {http://link.aps.org/licenses/aps-default-license},
  langid = {english}
}

@article{bernardGeneralFormulationSpinflip2012,
  title = {General Formulation of Spin-Flip Time-Dependent Density Functional Theory Using Non-Collinear Kernels: {{Theory}}, Implementation, and Benchmarks},
  shorttitle = {General Formulation of Spin-Flip Time-Dependent Density Functional Theory Using Non-Collinear Kernels},
  author = {Bernard, Yves A. and Shao, Yihan and Krylov, Anna I.},
  year = 2012,
  month = may,
  journal = {J. Chem. Phys.},
  volume = {136},
  number = {20},
  pages = {204103},
  issn = {0021-9606, 1089-7690},
  doi = {10.1063/1.4714499},
  urldate = {2026-01-29},
  langid = {english}
}

@article{besleySelfconsistentfieldCalculationsCore2009,
  title = {Self-Consistent-Field Calculations of Core Excited States},
  author = {Besley, Nicholas A. and Gilbert, Andrew T. B. and Gill, Peter M. W.},
  year = 2009,
  month = mar,
  journal = {J. Chem. Phys.},
  volume = {130},
  number = {12},
  pages = {124308},
  issn = {0021-9606},
  doi = {10.1063/1.3092928},
  urldate = {2025-10-20}
}

@article{bluntChargetransferExcitedStates2017a,
  title = {Charge-Transfer Excited States: {{Seeking}} a Balanced and Efficient Wave Function Ansatz in Variational {{Monte Carlo}}},
  shorttitle = {Charge-Transfer Excited States},
  author = {Blunt, N. S. and Neuscamman, Eric},
  year = 2017,
  month = nov,
  journal = {J. Chem. Phys.},
  volume = {147},
  number = {19},
  pages = {194101},
  issn = {0021-9606},
  doi = {10.1063/1.4998197},
  urldate = {2025-10-20}
}

@article{boynElucidatingMolecularOrbital2022,
  title = {Elucidating the Molecular Orbital Dependence of the Total Electronic Energy in Multireference Problems},
  author = {Boyn, Jan-Niklas and Mazziotti, David A.},
  year = 2022,
  month = may,
  journal = {J. Chem. Phys.},
  volume = {156},
  number = {19},
  pages = {194104},
  issn = {0021-9606},
  doi = {10.1063/5.0090342},
  urldate = {2025-10-20}
}

@article{burkeTimedependentDensityFunctional2005,
  title = {Time-Dependent Density Functional Theory: {{Past}}, Present, and Future},
  shorttitle = {Time-Dependent Density Functional Theory},
  author = {Burke, Kieron and Werschnik, Jan and Gross, E. K. U.},
  year = 2005,
  month = aug,
  journal = {J. Chem. Phys.},
  volume = {123},
  number = {6},
  pages = {062206},
  issn = {0021-9606, 1089-7690},
  doi = {10.1063/1.1904586},
  urldate = {2025-09-11},
  langid = {english}
}

@article{burtonEnergyLandscapeStateSpecific2022,
  title = {Energy {{Landscape}} of {{State-Specific Electronic Structure Theory}}},
  author = {Burton, Hugh G. A.},
  year = 2022,
  month = mar,
  journal = {J. Chem. Theory Comput.},
  volume = {18},
  number = {3},
  pages = {1512--1526},
  publisher = {American Chemical Society},
  issn = {1549-9618},
  doi = {10.1021/acs.jctc.1c01089},
  urldate = {2025-10-20}
}

@article{carter-fenkStateTargetedEnergyProjection2020,
  title = {State-{{Targeted Energy Projection}}: {{A Simple}} and {{Robust Approach}} to {{Orbital Relaxation}} of {{Non-Aufbau Self-Consistent Field Solutions}}},
  shorttitle = {State-{{Targeted Energy Projection}}},
  author = {{Carter-Fenk}, Kevin and Herbert, John M.},
  year = 2020,
  month = aug,
  journal = {J. Chem. Theory Comput.},
  volume = {16},
  number = {8},
  pages = {5067--5082},
  publisher = {American Chemical Society},
  issn = {1549-9618},
  doi = {10.1021/acs.jctc.0c00502},
  urldate = {2025-10-20}
}

@article{casidaProgressTimeDependentDensityFunctional2012,
  title = {Progress in {{Time-Dependent Density-Functional Theory}}},
  author = {Casida, M. E. and {Huix-Rotllant}, M.},
  year = 2012,
  month = may,
  journal = {Annu. Rev. Phys. Chem.},
  volume = {63},
  number = {Volume 63, 2012},
  pages = {287--323},
  publisher = {Annual Reviews},
  issn = {0066-426X, 1545-1593},
  doi = {10.1146/annurev-physchem-032511-143803},
  urldate = {2025-09-11},
  langid = {english}
}

@article{chraytehMountaineeringStrategyExcited2021,
  title = {Mountaineering {{Strategy}} to {{Excited States}}: {{Highly Accurate Oscillator Strengths}} and {{Dipole Moments}} of {{Small Molecules}}},
  shorttitle = {Mountaineering {{Strategy}} to {{Excited States}}},
  author = {Chrayteh, Amara and Blondel, Aymeric and Loos, Pierre-Fran{\c c}ois and Jacquemin, Denis},
  year = 2021,
  month = jan,
  journal = {J. Chem. Theory Comput.},
  volume = {17},
  number = {1},
  pages = {416--438},
  publisher = {American Chemical Society},
  issn = {1549-9618},
  doi = {10.1021/acs.jctc.0c01111},
  urldate = {2025-08-07}
}

@article{cluneExcitationMatchedLocal2025,
  title = {An Excitation Matched Local Correlation Approach to Excited State Specific Perturbation Theory},
  author = {Clune, Rachel and Neuscamman, Eric},
  year = 2025,
  month = sep,
  journal = {J. Chem. Phys.},
  volume = {163},
  number = {9},
  pages = {094109},
  issn = {0021-9606},
  doi = {10.1063/5.0280479},
  urldate = {2025-10-20}
}

@article{cluneN5ScalingExcitedStateSpecificPerturbation2020,
  title = {N5-{{Scaling Excited-State-Specific Perturbation Theory}}},
  author = {Clune, Rachel and Shea, Jacqueline A. R. and Neuscamman, Eric},
  year = 2020,
  month = oct,
  journal = {J. Chem. Theory Comput.},
  volume = {16},
  number = {10},
  pages = {6132--6141},
  publisher = {American Chemical Society},
  issn = {1549-9618},
  doi = {10.1021/acs.jctc.0c00308},
  urldate = {2025-10-20}
}

@article{cluneStudyingExcitedstatespecificPerturbation2023,
  title = {Studying Excited-State-Specific Perturbation Theory on the {{Thiel}} Set},
  author = {Clune, Rachel and Shea, Jacqueline A. R. and Hardikar, Tarini S. and Tuckman, Harrison and Neuscamman, Eric},
  year = 2023,
  month = jun,
  journal = {J. Chem. Phys.},
  volume = {158},
  number = {22},
  pages = {224113},
  issn = {0021-9606, 1089-7690},
  doi = {10.1063/5.0146975},
  urldate = {2025-09-12},
  langid = {english}
}

@incollection{contrerasElectronDensitiesPopulation2017,
  title = {Electron {{Densities}}: {{Population Analysis}} and {{Beyond}}},
  shorttitle = {Electron {{Densities}}},
  booktitle = {Encyclopedia of {{Physical Organic Chemistry}}},
  author = {Contreras, Renato and Domingo, Luis R. and Silvi, Bernard},
  year = 2017,
  pages = {1--114},
  publisher = {John Wiley \& Sons, Ltd},
  doi = {10.1002/9781118468586.epoc4003},
  urldate = {2026-02-03},
  copyright = {Copyright \copyright{} 2017 John Wiley \& Sons, Inc.},
  isbn = {978-1-118-46858-6},
  langid = {english}
}

@article{dalgaardAspectsTimedependentCoupledcluster1983,
  title = {Some Aspects of the Time-Dependent Coupled-Cluster Approach to Dynamic Response Functions},
  author = {Dalgaard, Esper and Monkhorst, Hendrik J.},
  year = 1983,
  month = sep,
  journal = {Phys. Rev. A},
  volume = {28},
  number = {3},
  pages = {1217--1222},
  issn = {0556-2791},
  doi = {10.1103/PhysRevA.28.1217},
  urldate = {2025-09-11},
  copyright = {http://link.aps.org/licenses/aps-default-license},
  langid = {english}
}

@article{damourStateSpecificCoupledClusterMethods2024,
  title = {State-{{Specific Coupled-Cluster Methods}} for {{Excited States}}},
  author = {Damour, Yann and Scemama, Anthony and Jacquemin, Denis and Kossoski, F{\'a}bris and Loos, Pierre-Fran{\c c}ois},
  year = 2024,
  month = may,
  journal = {J. Chem. Theory Comput.},
  volume = {20},
  number = {10},
  pages = {4129--4145},
  publisher = {American Chemical Society},
  issn = {1549-9618},
  doi = {10.1021/acs.jctc.4c00034},
  urldate = {2025-10-20}
}

@article{danilovElectronDensitybondOrder1969,
  title = {The Electron Density-Bond Order Matrix and the Spin Density in the Restricted {{CI}} Method},
  author = {Danilov, V. I. and Kruglyak, Yuri A. and Pechenaya, V. I.},
  year = 1969,
  month = jan,
  journal = {Theoret. Chim. Acta},
  volume = {13},
  number = {4},
  pages = {288--296},
  issn = {1432-2234},
  doi = {10.1007/BF00529020},
  urldate = {2025-09-26},
  langid = {english}
}

@article{dreuwSingleReferenceInitioMethods2005,
  title = {Single-{{Reference}} Ab {{Initio Methods}} for the {{Calculation}} of {{Excited States}} of {{Large Molecules}}},
  author = {Dreuw, Andreas and {Head-Gordon}, Martin},
  year = 2005,
  month = nov,
  journal = {Chem. Rev.},
  volume = {105},
  number = {11},
  pages = {4009--4037},
  publisher = {American Chemical Society},
  issn = {0009-2665},
  doi = {10.1021/cr0505627},
  urldate = {2025-06-09}
}

@article{entwistleElectronicExcitedStates2023,
  title = {Electronic Excited States in Deep Variational {{Monte Carlo}}},
  author = {Entwistle, M. T. and Sch{\"a}tzle, Z. and Erdman, P. A. and Hermann, J. and No{\'e}, F.},
  year = 2023,
  month = jan,
  journal = {Nat Commun},
  volume = {14},
  number = {1},
  pages = {274},
  publisher = {Nature Publishing Group},
  issn = {2041-1723},
  doi = {10.1038/s41467-022-35534-5},
  urldate = {2025-10-20},
  copyright = {2023 The Author(s)},
  langid = {english}
}

@article{epifanovskySoftwareFrontiersQuantum2021,
  title = {Software for the Frontiers of Quantum Chemistry: {{An}} Overview of Developments in the {{Q-Chem}} 5 Package},
  shorttitle = {Software for the Frontiers of Quantum Chemistry},
  author = {Epifanovsky, Evgeny and Gilbert, Andrew T. B. and Feng, Xintian and Lee, Joonho and Mao, Yuezhi and Mardirossian, Narbe and Pokhilko, Pavel and White, Alec F. and Coons, Marc P. and Dempwolff, Adrian L. and Gan, Zhengting and Hait, Diptarka and Horn, Paul R. and Jacobson, Leif D. and Kaliman, Ilya and Kussmann, J{\"o}rg and Lange, Adrian W. and Lao, Ka Un and Levine, Daniel S. and Liu, Jie and McKenzie, Simon C. and Morrison, Adrian F. and Nanda, Kaushik D. and Plasser, Felix and Rehn, Dirk R. and Vidal, Marta L. and You, Zhi-Qiang and Zhu, Ying and Alam, Bushra and Albrecht, Benjamin J. and Aldossary, Abdulrahman and Alguire, Ethan and Andersen, Josefine H. and Athavale, Vishikh and Barton, Dennis and Begam, Khadiza and Behn, Andrew and Bellonzi, Nicole and Bernard, Yves A. and Berquist, Eric J. and Burton, Hugh G. A. and Carreras, Abel and {Carter-Fenk}, Kevin and Chakraborty, Romit and Chien, Alan D. and Closser, Kristina D. and {Cofer-Shabica}, Vale and Dasgupta, Saswata and De Wergifosse, Marc and Deng, Jia and Diedenhofen, Michael and Do, Hainam and Ehlert, Sebastian and Fang, Po-Tung and Fatehi, Shervin and Feng, Qingguo and Friedhoff, Triet and Gayvert, James and Ge, Qinghui and Gidofalvi, Gergely and Goldey, Matthew and Gomes, Joe and {Gonz{\'a}lez-Espinoza}, Cristina E. and Gulania, Sahil and Gunina, Anastasia O. and {Hanson-Heine}, Magnus W. D. and Harbach, Phillip H. P. and Hauser, Andreas and Herbst, Michael F. and Hern{\'a}ndez Vera, Mario and Hodecker, Manuel and Holden, Zachary C. and Houck, Shannon and Huang, Xunkun and Hui, Kerwin and Huynh, Bang C. and Ivanov, Maxim and J{\'a}sz, {\'A}d{\'a}m and Ji, Hyunjun and Jiang, Hanjie and Kaduk, Benjamin and K{\"a}hler, Sven and Khistyaev, Kirill and Kim, Jaehoon and Kis, Gergely and Klunzinger, Phil and {Koczor-Benda}, Zsuzsanna and Koh, Joong Hoon and Kosenkov, Dimitri and Koulias, Laura and Kowalczyk, Tim and Krauter, Caroline M. and Kue, Karl and Kunitsa, Alexander and Kus, Thomas and Ladj{\'a}nszki, Istv{\'a}n and Landau, Arie and Lawler, Keith V. and Lefrancois, Daniel and Lehtola, Susi and Li, Run R. and Li, Yi-Pei and Liang, Jiashu and Liebenthal, Marcus and Lin, Hung-Hsuan and Lin, You-Sheng and Liu, Fenglai and Liu, Kuan-Yu and Loipersberger, Matthias and Luenser, Arne and Manjanath, Aaditya and Manohar, Prashant and Mansoor, Erum and Manzer, Sam F. and Mao, Shan-Ping and Marenich, Aleksandr V. and Markovich, Thomas and Mason, Stephen and Maurer, Simon A. and McLaughlin, Peter F. and Menger, Maximilian F. S. J. and Mewes, Jan-Michael and Mewes, Stefanie A. and Morgante, Pierpaolo and Mullinax, J. Wayne and Oosterbaan, Katherine J. and Paran, Garrette and Paul, Alexander C. and Paul, Suranjan K. and Pavo{\v s}evi{\'c}, Fabijan and Pei, Zheng and Prager, Stefan and Proynov, Emil I. and R{\'a}k, {\'A}d{\'a}m and {Ramos-Cordoba}, Eloy and Rana, Bhaskar and Rask, Alan E. and Rettig, Adam and Richard, Ryan M. and Rob, Fazle and Rossomme, Elliot and Scheele, Tarek and Scheurer, Maximilian and Schneider, Matthias and Sergueev, Nickolai and Sharada, Shaama M. and Skomorowski, Wojciech and Small, David W. and Stein, Christopher J. and Su, Yu-Chuan and Sundstrom, Eric J. and Tao, Zhen and Thirman, Jonathan and Tornai, G{\'a}bor J. and Tsuchimochi, Takashi and Tubman, Norm M. and Veccham, Srimukh Prasad and Vydrov, Oleg and Wenzel, Jan and Witte, Jon and Yamada, Atsushi and Yao, Kun and Yeganeh, Sina and Yost, Shane R. and Zech, Alexander and Zhang, Igor Ying and Zhang, Xing and Zhang, Yu and Zuev, Dmitry and {Aspuru-Guzik}, Al{\'a}n and Bell, Alexis T. and Besley, Nicholas A. and Bravaya, Ksenia B. and Brooks, Bernard R. and Casanova, David and Chai, Jeng-Da and Coriani, Sonia and Cramer, Christopher J. and Cserey, Gy{\"o}rgy and DePrince, A. Eugene and DiStasio, Robert A. and Dreuw, Andreas and Dunietz, Barry D. and Furlani, Thomas R. and Goddard, William A. and {Hammes-Schiffer}, Sharon and {Head-Gordon}, Teresa and Hehre, Warren J. and Hsu, Chao-Ping and Jagau, Thomas-C. and Jung, Yousung and Klamt, Andreas and Kong, Jing and Lambrecht, Daniel S. and Liang, WanZhen and Mayhall, Nicholas J. and McCurdy, C. William and Neaton, Jeffrey B. and Ochsenfeld, Christian and Parkhill, John A. and Peverati, Roberto and Rassolov, Vitaly A. and Shao, Yihan and Slipchenko, Lyudmila V. and Stauch, Tim and Steele, Ryan P. and Subotnik, Joseph E. and Thom, Alex J. W. and Tkatchenko, Alexandre and Truhlar, Donald G. and Van Voorhis, Troy and Wesolowski, Tomasz A. and Whaley, K. Birgitta and Woodcock, H. Lee and Zimmerman, Paul M. and Faraji, Shirin and Gill, Peter M. W. and {Head-Gordon}, Martin and Herbert, John M. and Krylov, Anna I.},
  year = 2021,
  month = aug,
  journal = {J. Chem. Phys.},
  volume = {155},
  number = {8},
  pages = {084801},
  issn = {0021-9606, 1089-7690},
  doi = {10.1063/5.0055522},
  urldate = {2025-08-12},
  langid = {english}
}

@article{filatovSpinrestrictedEnsemblereferencedKohn1999,
  title = {A Spin-Restricted Ensemble-Referenced {{Kohn}}--{{Sham}} Method and Its Application to Diradicaloid Situations},
  author = {Filatov, Michael and Shaik, Sason},
  year = 1999,
  month = may,
  journal = {Chemical Physics Letters},
  volume = {304},
  number = {5},
  pages = {429--437},
  issn = {0009-2614},
  doi = {10.1016/S0009-2614(99)00336-X},
  urldate = {2025-10-20}
}

@article{furcheAdiabaticTimedependentDensity2002,
  title = {Adiabatic Time-Dependent Density Functional Methods for Excited State Properties},
  author = {Furche, Filipp and Ahlrichs, Reinhart},
  year = 2002,
  month = oct,
  journal = {J. Chem. Phys.},
  volume = {117},
  number = {16},
  pages = {7433--7447},
  issn = {0021-9606, 1089-7690},
  doi = {10.1063/1.1508368},
  urldate = {2026-01-27},
  langid = {english}
}

@article{fuxOQuPyPythonPackage2024,
  title = {{{OQuPy}}: {{A Python}} Package to Efficiently Simulate Non-{{Markovian}} Open Quantum Systems with Process Tensors},
  shorttitle = {{{OQuPy}}},
  author = {Fux, Gerald E. and {Fowler-Wright}, Piper and Beckles, Joel and Butler, Eoin P. and Eastham, Paul R. and Gribben, Dominic and Keeling, Jonathan and Kilda, Dainius and Kirton, Peter and Lawrence, Ewen D. C. and Lovett, Brendon W. and O'Neill, Eoin and Strathearn, Aidan and De Wit, Roosmarijn},
  year = 2024,
  month = sep,
  journal = {J. Chem. Phys.},
  volume = {161},
  number = {12},
  pages = {124108},
  issn = {0021-9606, 1089-7690},
  doi = {10.1063/5.0225367},
  urldate = {2025-08-12},
  langid = {english}
}

@article{garnerVariationalMonteCarlo2020,
  title = {A Variational {{Monte Carlo}} Approach for Core Excitations},
  author = {Garner, Scott M. and Neuscamman, Eric},
  year = 2020,
  month = oct,
  journal = {J. Chem. Phys.},
  volume = {153},
  number = {14},
  pages = {144108},
  issn = {0021-9606},
  doi = {10.1063/5.0020310},
  urldate = {2025-10-20}
}

@article{gilbertSelfConsistentFieldCalculations2008,
  title = {Self-{{Consistent Field Calculations}} of {{Excited States Using}} the {{Maximum Overlap Method}} ({{MOM}})},
  author = {Gilbert, Andrew T. B. and Besley, Nicholas A. and Gill, Peter M. W.},
  year = 2008,
  month = dec,
  journal = {J. Phys. Chem. A},
  volume = {112},
  number = {50},
  pages = {13164--13171},
  publisher = {American Chemical Society},
  issn = {1089-5639},
  doi = {10.1021/jp801738f},
  urldate = {2025-10-20}
}

@article{gonzalezProgressChallengesCalculation2012,
  title = {Progress and {{Challenges}} in the {{Calculation}} of {{Electronic Excited States}}},
  author = {Gonz{\'a}lez, Leticia and Escudero, Daniel and {Serrano-Andr{\'e}s}, Luis},
  year = 2012,
  journal = {ChemPhysChem},
  volume = {13},
  number = {1},
  pages = {28--51},
  issn = {1439-7641},
  doi = {10.1002/cphc.201100200},
  urldate = {2025-09-26},
  copyright = {Copyright \copyright{} 2012 WILEY-VCH Verlag GmbH \& Co. KGaA, Weinheim},
  langid = {english}
}

@article{haitExcitedStateOrbital2020,
  title = {Excited {{State Orbital Optimization}} via {{Minimizing}} the {{Square}} of the {{Gradient}}: {{General Approach}} and {{Application}} to {{Singly}} and {{Doubly Excited States}} via {{Density Functional Theory}}},
  shorttitle = {Excited {{State Orbital Optimization}} via {{Minimizing}} the {{Square}} of the {{Gradient}}},
  author = {Hait, Diptarka and {Head-Gordon}, Martin},
  year = 2020,
  month = mar,
  journal = {J. Chem. Theory Comput.},
  volume = {16},
  number = {3},
  pages = {1699--1710},
  publisher = {American Chemical Society},
  issn = {1549-9618},
  doi = {10.1021/acs.jctc.9b01127},
  urldate = {2025-10-20}
}

@article{haitHighlyAccuratePrediction2020,
  title = {Highly {{Accurate Prediction}} of {{Core Spectra}} of {{Molecules}} at {{Density Functional Theory Cost}}: {{Attaining Sub-electronvolt Error}} from a {{Restricted Open-Shell Kohn}}--{{Sham Approach}}},
  shorttitle = {Highly {{Accurate Prediction}} of {{Core Spectra}} of {{Molecules}} at {{Density Functional Theory Cost}}},
  author = {Hait, Diptarka and {Head-Gordon}, Martin},
  year = 2020,
  month = feb,
  journal = {J. Phys. Chem. Lett.},
  volume = {11},
  number = {3},
  pages = {775--786},
  publisher = {American Chemical Society},
  doi = {10.1021/acs.jpclett.9b03661},
  urldate = {2025-10-20}
}

@article{haitOrbitalOptimizedDensity2021,
  title = {Orbital {{Optimized Density Functional Theory}} for {{Electronic Excited States}}},
  author = {Hait, Diptarka and {Head-Gordon}, Martin},
  year = 2021,
  month = may,
  journal = {J. Phys. Chem. Lett.},
  volume = {12},
  number = {19},
  pages = {4517--4529},
  publisher = {American Chemical Society},
  doi = {10.1021/acs.jpclett.1c00744},
  urldate = {2025-10-20}
}

@article{hanscamApplyingGeneralizedVariational2022,
  title = {Applying {{Generalized Variational Principles}} to {{Excited-State-Specific Complete Active Space Self-consistent Field Theory}}},
  author = {Hanscam, Rebecca and Neuscamman, Eric},
  year = 2022,
  month = nov,
  journal = {J. Chem. Theory Comput.},
  volume = {18},
  number = {11},
  pages = {6608--6621},
  publisher = {American Chemical Society},
  issn = {1549-9618},
  doi = {10.1021/acs.jctc.2c00639},
  urldate = {2025-10-20}
}

@article{hardikarSelfconsistentFieldFormulation2020,
  title = {A Self-Consistent Field Formulation of Excited State Mean Field Theory},
  author = {Hardikar, Tarini S. and Neuscamman, Eric},
  year = 2020,
  month = oct,
  journal = {J. Chem. Phys.},
  volume = {153},
  number = {16},
  pages = {164108},
  issn = {0021-9606, 1089-7690},
  doi = {10.1063/5.0019557},
  urldate = {2025-09-12},
  langid = {english}
}

@article{hellwegAccuracyDipoleMoments2011,
  title = {The Accuracy of Dipole Moments from Spin-Component Scaled {{CC2}} in Ground and Electronically Excited States},
  author = {Hellweg, Arnim},
  year = 2011,
  month = feb,
  journal = {J. Chem. Phys.},
  volume = {134},
  number = {6},
  pages = {064103},
  issn = {0021-9606},
  doi = {10.1063/1.3549818},
  urldate = {2025-09-26}
}

@article{hodeckerSimilaritiesDifferencesLagrange2019,
  title = {Similarities and Differences of the {{Lagrange}} Formalism and the Intermediate State Representation in the Treatment of Molecular Properties},
  author = {Hodecker, Manuel and Rehn, Dirk R. and Dreuw, Andreas and H{\"o}fener, Sebastian},
  year = 2019,
  month = apr,
  journal = {J. Chem. Phys.},
  volume = {150},
  number = {16},
  pages = {164125},
  issn = {0021-9606},
  doi = {10.1063/1.5093606},
  urldate = {2025-08-19}
}

@article{hsuSCFMethodHole1976,
  title = {An {{SCF}} Method for Hole States},
  author = {Hsu, Hsiang-lin and Davidson, Ernest R. and Pitzer, Russell M.},
  year = 1976,
  month = jul,
  journal = {J. Chem. Phys.},
  volume = {65},
  number = {2},
  pages = {609--613},
  issn = {0021-9606, 1089-7690},
  doi = {10.1063/1.433118},
  urldate = {2025-10-20},
  langid = {english}
}

@article{javedAufbauSuppressedCoupled2026,
  title = {Aufbau {{Suppressed Coupled Cluster Theory}} for {{Doubly Excited States}}},
  author = {Javed, Qasim and Tuckman, Harrison and Neuscamman, Eric},
  year = 2026,
  journal = {arXiv.org},
  pages = {2601.20089},
  doi = {10.48550/arXiv.2601.20089},
  urldate = {2026-02-11},
}

@book{jorgensenSecondQuantizationbasedMethods1981,
  title = {Second Quantization-Based Methods in Quantum Chemistry},
  author = {J{\o}rgensen, Poul and Simons, Jack},
  year = 1981,
  publisher = {Academic Press},
  address = {New York},
  isbn = {978-0-12-390220-7 978-1-299-36306-9},
  langid = {english}
}

@article{kallayAnalyticFirstDerivatives2003,
  title = {Analytic First Derivatives for General Coupled-Cluster and Configuration Interaction Models},
  author = {K{\'a}llay, Mih{\'a}ly and Gauss, J{\"u}rgen and Szalay, P{\'e}ter G.},
  year = 2003,
  month = aug,
  journal = {J. Chem. Phys.},
  volume = {119},
  number = {6},
  pages = {2991--3004},
  issn = {0021-9606, 1089-7690},
  doi = {10.1063/1.1589003},
  urldate = {2025-09-24},
  langid = {english}
}

@article{kallayCalculationExcitedstateProperties2004,
  title = {Calculation of Excited-State Properties Using General Coupled-Cluster and Configuration-Interaction Models},
  author = {K{\'a}llay, Mih{\'a}ly and Gauss, J{\"u}rgen},
  year = 2004,
  month = nov,
  journal = {J. Chem. Phys.},
  volume = {121},
  number = {19},
  pages = {9257--9269},
  issn = {0021-9606, 1089-7690},
  doi = {10.1063/1.1805494},
  urldate = {2025-09-24},
  langid = {english}
}

@article{kallayHigherExcitationsCoupledcluster2001,
  title = {Higher Excitations in Coupled-Cluster Theory},
  author = {K{\'a}llay, Mih{\'a}ly and Surj{\'a}n, P{\'e}ter R.},
  year = 2001,
  month = aug,
  journal = {J. Chem. Phys.},
  volume = {115},
  number = {7},
  pages = {2945--2954},
  issn = {0021-9606, 1089-7690},
  doi = {10.1063/1.1383290},
  urldate = {2025-09-24},
  langid = {english}
}

@misc{kallayMRCCQuantumChemical,
  title = {{{MRCC}}, a Quantum Chemical Program Suite},
  author = {K{\'a}llay, M. and Nagy, P. R. and Mester, D. and {Gyevi-Nagy}, L. and Cs{\'o}ka, J. and Szab{\'o}, P. B. and Rolik, Z. and Samu, G. and H{\'e}gely, B. and Lad{\'o}czki, B. and Petrov, K. and Csontos, J. and Ganyecz, {\'A}. and Ladj{\'a}nszki, I. and Szegedy, L. and Farkas, M. and Mezei, P. D. and Horv{\'a}th, R. A. and L{\H o}rincz, B. D.},
  howpublished = {www.mrcc.hu}
}

@article{kempfer-robertsonRoleExactExchange2022,
  title = {Role of {{Exact Exchange}} in {{Difference Projected Double-Hybrid Density Functional Theory}} for {{Treatment}} of {{Local}}, {{Charge Transfer}}, and {{Rydberg Excitations}}},
  author = {{Kempfer-Robertson}, Emily M. and Haase, Meagan N. and Bersson, Jonathan S. and Avdic, Irma and Thompson, Lee M.},
  year = 2022,
  month = nov,
  journal = {J. Phys. Chem. A},
  volume = {126},
  number = {43},
  pages = {8058--8069},
  publisher = {American Chemical Society},
  issn = {1089-5639},
  doi = {10.1021/acs.jpca.2c04338},
  urldate = {2025-10-20}
}

@article{knyshReferenceCC3Excitation2024,
  title = {Reference {{CC3 Excitation Energies}} for {{Organic Chromophores}}: {{Benchmarking TD-DFT}}, {{BSE}}/{{GW}}, and {{Wave Function Methods}}},
  shorttitle = {Reference {{CC3 Excitation Energies}} for {{Organic Chromophores}}},
  author = {Knysh, Iryna and Lipparini, Filippo and Blondel, Aymeric and Duchemin, Ivan and Blase, Xavier and Loos, Pierre-Fran{\c c}ois and Jacquemin, Denis},
  year = 2024,
  month = sep,
  journal = {J. Chem. Theory Comput.},
  volume = {20},
  number = {18},
  pages = {8152--8174},
  publisher = {American Chemical Society},
  issn = {1549-9618},
  doi = {10.1021/acs.jctc.4c00906},
  urldate = {2025-12-05}
}

@article{kochCalculationSizeintensiveTransition1994,
  title = {Calculation of Size-Intensive Transition Moments from the Coupled Cluster Singles and Doubles Linear Response Function},
  author = {Koch, Henrik and Kobayashi, Rika and Sanchez De Mer{\'a}s, Alfredo and Jo/rgensen, Poul},
  year = 1994,
  month = mar,
  journal = {J. Chem. Phys.},
  volume = {100},
  number = {6},
  pages = {4393--4400},
  issn = {0021-9606, 1089-7690},
  doi = {10.1063/1.466321},
  urldate = {2025-09-11},
  langid = {english}
}

@article{kochCoupledClusterResponse1990,
  title = {Coupled Cluster Response Functions},
  author = {Koch, Henrik and Jo/rgensen, Poul},
  year = 1990,
  month = sep,
  journal = {J. Chem. Phys.},
  volume = {93},
  number = {5},
  pages = {3333--3344},
  issn = {0021-9606, 1089-7690},
  doi = {10.1063/1.458814},
  urldate = {2025-09-11},
  langid = {english}
}

@article{kossoskiExcitedStatesStateSpecific2021,
  title = {Excited {{States}} from {{State-Specific Orbital-Optimized Pair Coupled Cluster}}},
  author = {Kossoski, F{\'a}bris and Marie, Antoine and Scemama, Anthony and Caffarel, Michel and Loos, Pierre-Fran{\c c}ois},
  year = 2021,
  month = aug,
  journal = {J. Chem. Theory Comput.},
  volume = {17},
  number = {8},
  pages = {4756--4768},
  publisher = {American Chemical Society},
  issn = {1549-9618},
  doi = {10.1021/acs.jctc.1c00348},
  urldate = {2025-10-20}
}

@article{kossoskiSeniorityHierarchyConfiguration2023,
  title = {Seniority and {{Hierarchy Configuration Interaction}} for {{Radicals}} and {{Excited States}}},
  author = {Kossoski, F{\'a}bris and Loos, Pierre-Fran{\c c}ois},
  year = 2023,
  month = dec,
  journal = {J. Chem. Theory Comput.},
  volume = {19},
  number = {23},
  pages = {8654--8670},
  publisher = {American Chemical Society},
  issn = {1549-9618},
  doi = {10.1021/acs.jctc.3c00946},
  urldate = {2025-10-20}
}

@article{kossoskiStateSpecificConfigurationInteraction2023,
  title = {State-{{Specific Configuration Interaction}} for {{Excited States}}},
  author = {Kossoski, F{\'a}bris and Loos, Pierre-Fran{\c c}ois},
  year = 2023,
  month = apr,
  journal = {J. Chem. Theory Comput.},
  volume = {19},
  number = {8},
  pages = {2258--2269},
  publisher = {American Chemical Society},
  issn = {1549-9618},
  doi = {10.1021/acs.jctc.3c00057},
  urldate = {2025-10-20}
}

@article{kowalczykAssessmentDSCFDensity2011,
  title = {Assessment of the {{$\Delta$SCF}} Density Functional Theory Approach for Electronic Excitations in Organic Dyes},
  author = {Kowalczyk, Tim and Yost, Shane R. and Voorhis, Troy Van},
  year = 2011,
  month = feb,
  journal = {J. Chem. Phys.},
  volume = {134},
  number = {5},
  pages = {054128},
  issn = {0021-9606},
  doi = {10.1063/1.3530801},
  urldate = {2025-10-20}
}

@article{kowalczykExcitationEnergiesStokes2013,
  title = {Excitation Energies and {{Stokes}} Shifts from a Restricted Open-Shell {{Kohn-Sham}} Approach},
  author = {Kowalczyk, Tim and Tsuchimochi, Takashi and Chen, Po-Ta and Top, Laken and Van Voorhis, Troy},
  year = 2013,
  month = apr,
  journal = {J. Chem. Phys.},
  volume = {138},
  number = {16},
  pages = {164101},
  issn = {0021-9606},
  doi = {10.1063/1.4801790},
  urldate = {2025-10-20}
}

@article{kozmaNewBenchmarkSet2020,
  title = {A {{New Benchmark Set}} for {{Excitation Energy}} of {{Charge Transfer States}}: {{Systematic Investigation}} of {{Coupled Cluster Type Methods}}},
  shorttitle = {A {{New Benchmark Set}} for {{Excitation Energy}} of {{Charge Transfer States}}},
  author = {Kozma, Bal{\'a}zs and Tajti, Attila and Demoulin, Baptiste and Izs{\'a}k, R{\'o}bert and Nooijen, Marcel and Szalay, P{\'e}ter G.},
  year = 2020,
  month = jul,
  journal = {J. Chem. Theory Comput.},
  volume = {16},
  number = {7},
  pages = {4213--4225},
  issn = {1549-9618, 1549-9626},
  doi = {10.1021/acs.jctc.0c00154},
  urldate = {2025-10-16},
  copyright = {http://pubs.acs.org/page/policy/authorchoice\_ccby\_termsofuse.html},
  langid = {english}
}

@article{krylovEquationofMotionCoupledClusterMethods2008,
  title = {Equation-of-{{Motion Coupled-Cluster Methods}} for {{Open-Shell}} and {{Electronically Excited Species}}: {{The Hitchhiker}}'s {{Guide}} to {{Fock Space}}},
  shorttitle = {Equation-of-{{Motion Coupled-Cluster Methods}} for {{Open-Shell}} and {{Electronically Excited Species}}},
  author = {Krylov, Anna I.},
  year = 2008,
  month = may,
  journal = {Annu. Rev. Phys. Chem.},
  volume = {59},
  number = {Volume 59, 2008},
  pages = {433--462},
  publisher = {Annual Reviews},
  issn = {0066-426X, 1545-1593},
  doi = {10.1146/annurev.physchem.59.032607.093602},
  urldate = {2025-09-11},
  langid = {english}
}

@article{krylovOrbitalsObservablesBack2020,
  title = {From Orbitals to Observables and Back},
  author = {Krylov, Anna I.},
  year = 2020,
  month = aug,
  journal = {J. Chem. Phys.},
  volume = {153},
  number = {8},
  pages = {080901},
  issn = {0021-9606, 1089-7690},
  doi = {10.1063/5.0018597},
  urldate = {2026-02-03},
  langid = {english}
}

@article{krylovSizeconsistentWaveFunctions,
  title = {Size-Consistent Wave Functions for Bond-Breaking: The Equation-of-Motion Spin-\textasciimacron ip Model},
  author = {Krylov, Anna I},
  langid = {english}
}

@article{krylovSpinFlipEquationofMotionCoupledCluster2006,
  title = {Spin-{{Flip Equation-of-Motion Coupled-Cluster Electronic Structure Method}} for a {{Description}} of {{Excited States}}, {{Bond Breaking}}, {{Diradicals}}, and {{Triradicals}}},
  author = {Krylov, Anna I.},
  year = 2006,
  month = feb,
  journal = {Acc. Chem. Res.},
  volume = {39},
  number = {2},
  pages = {83--91},
  issn = {0001-4842, 1520-4898},
  doi = {10.1021/ar0402006},
  urldate = {2026-01-29},
  langid = {english}
}

@article{kusDeperturbativeCorrectionsChargestabilized2012,
  title = {De-Perturbative Corrections for Charge-Stabilized Double Ionization Potential Equation-of-Motion Coupled-Cluster Method},
  author = {Ku{\'s}, Tomasz and Krylov, Anna I.},
  year = 2012,
  month = jun,
  journal = {J. Chem. Phys.},
  volume = {136},
  number = {24},
  pages = {244109},
  issn = {0021-9606, 1089-7690},
  doi = {10.1063/1.4730296},
  urldate = {2026-01-29},
  langid = {english}
}

@article{kusUsingChargestabilizationTechnique2011,
  title = {Using the Charge-Stabilization Technique in the Double Ionization Potential Equation-of-Motion Calculations with Dianion References},
  author = {Ku{\'s}, Tomasz and Krylov, Anna I.},
  year = 2011,
  month = aug,
  journal = {J. Chem. Phys.},
  volume = {135},
  number = {8},
  pages = {084109},
  issn = {0021-9606, 1089-7690},
  doi = {10.1063/1.3626149},
  urldate = {2026-01-29},
  langid = {english}
}

@article{leeExcitedStatesCoupled2019,
  title = {Excited States via Coupled Cluster Theory without Equation-of-Motion Methods: {{Seeking}} Higher Roots with Application to Doubly Excited States and Double Core Hole States},
  shorttitle = {Excited States via Coupled Cluster Theory without Equation-of-Motion Methods},
  author = {Lee, Joonho and Small, David W. and {Head-Gordon}, Martin},
  year = 2019,
  month = dec,
  journal = {J. Chem. Phys.},
  volume = {151},
  number = {21},
  pages = {214103},
  issn = {0021-9606},
  doi = {10.1063/1.5128795},
  urldate = {2025-10-20}
}

@article{levchenkoAnalyticGradientsSpinconserving2005,
  title = {Analytic Gradients for the Spin-Conserving and Spin-Flipping Equation-of-Motion Coupled-Cluster Models with Single and Double Substitutions},
  author = {Levchenko, Sergey V. and Wang, Tao and Krylov, Anna I.},
  year = 2005,
  month = jun,
  journal = {J. Chem. Phys.},
  volume = {122},
  number = {22},
  pages = {224106},
  issn = {0021-9606},
  doi = {10.1063/1.1877072},
  urldate = {2025-08-19}
}

@article{leviVariationalDensityFunctional2020,
  title = {Variational {{Density Functional Calculations}} of {{Excited States}} via {{Direct Optimization}}},
  author = {Levi, Gianluca and Ivanov, Aleksei V. and J{\'o}nsson, Hannes},
  year = 2020,
  month = nov,
  journal = {J. Chem. Theory Comput.},
  volume = {16},
  number = {11},
  pages = {6968--6982},
  publisher = {American Chemical Society},
  issn = {1549-9618},
  doi = {10.1021/acs.jctc.0c00597},
  urldate = {2025-10-20}
}

@article{liuCommunicationAdjustingCharge2012,
  title = {Communication: {{Adjusting}} Charge Transfer State Energies for Configuration Interaction Singles: {{Without}} Any Parameterization and with Minimal Cost},
  shorttitle = {Communication},
  author = {Liu, Xinle and Fatehi, Shervin and Shao, Yihan and Veldkamp, Brad S. and Subotnik, Joseph E.},
  year = 2012,
  month = apr,
  journal = {J. Chem. Phys.},
  volume = {136},
  number = {16},
  pages = {161101},
  issn = {0021-9606},
  doi = {10.1063/1.4705757},
  urldate = {2025-10-20}
}

@article{loosMountaineeringStrategyExcited2018,
  title = {A {{Mountaineering Strategy}} to {{Excited States}}: {{Highly Accurate Reference Energies}} and {{Benchmarks}}},
  shorttitle = {A {{Mountaineering Strategy}} to {{Excited States}}},
  author = {Loos, Pierre-Fran{\c c}ois and Scemama, Anthony and Blondel, Aymeric and Garniron, Yann and Caffarel, Michel and Jacquemin, Denis},
  year = 2018,
  month = aug,
  journal = {J. Chem. Theory Comput.},
  volume = {14},
  number = {8},
  pages = {4360--4379},
  issn = {1549-9618, 1549-9626},
  doi = {10.1021/acs.jctc.8b00406},
  urldate = {2025-09-11},
  langid = {english}
}

@article{loosMountaineeringStrategyExcited2020,
  title = {A {{Mountaineering Strategy}} to {{Excited States}}: {{Highly Accurate Energies}} and {{Benchmarks}} for {{Medium Sized Molecules}}},
  shorttitle = {A {{Mountaineering Strategy}} to {{Excited States}}},
  author = {Loos, Pierre-Fran{\c c}ois and Lipparini, Filippo and {Boggio-Pasqua}, Martial and Scemama, Anthony and Jacquemin, Denis},
  year = 2020,
  month = mar,
  journal = {J. Chem. Theory Comput.},
  volume = {16},
  number = {3},
  pages = {1711--1741},
  issn = {1549-9618, 1549-9626},
  doi = {10.1021/acs.jctc.9b01216},
  urldate = {2025-09-11},
  copyright = {https://doi.org/10.15223/policy-029},
  langid = {english}
}

@article{loosReferenceEnergiesIntramolecular2021,
  title = {Reference {{Energies}} for {{Intramolecular Charge-Transfer Excitations}}},
  author = {Loos, Pierre-Fran{\c c}ois and Comin, Massimiliano and Blase, Xavier and Jacquemin, Denis},
  year = 2021,
  month = jun,
  journal = {J. Chem. Theory Comput.},
  volume = {17},
  number = {6},
  pages = {3666--3686},
  publisher = {American Chemical Society},
  issn = {1549-9618},
  doi = {10.1021/acs.jctc.1c00226},
  urldate = {2025-12-05}
}

@article{lowdinNonOrthogonalityProblemConnected1950,
  title = {On the {{Non-Orthogonality Problem Connected}} with the {{Use}} of {{Atomic Wave Functions}} in the {{Theory}} of {{Molecules}} and {{Crystals}}},
  author = {L{\"o}wdin, Per-Olov},
  year = 1950,
  month = mar,
  journal = {J. Chem. Phys.},
  volume = {18},
  number = {3},
  pages = {365--375},
  issn = {0021-9606, 1089-7690},
  doi = {10.1063/1.1747632},
  urldate = {2025-09-24},
  langid = {english}
}

@article{lowdinQuantumTheoryManyParticle1955,
  title = {Quantum {{Theory}} of {{Many-Particle Systems}}. {{I}}. {{Physical Interpretations}} by {{Means}} of {{Density Matrices}}, {{Natural Spin-Orbitals}}, and {{Convergence Problems}} in the {{Method}} of {{Configurational Interaction}}},
  author = {L{\"o}wdin, Per-Olov},
  year = 1955,
  month = mar,
  journal = {Phys. Rev.},
  volume = {97},
  number = {6},
  pages = {1474--1489},
  issn = {0031-899X},
  doi = {10.1103/PhysRev.97.1474},
  urldate = {2025-09-26},
  copyright = {http://link.aps.org/licenses/aps-default-license},
  langid = {english}
}

@article{mardirossianOB97XV10parameterRangeseparated2014,
  title = {{{$\omega$B97X-V}}: {{A}} 10-Parameter, Range-Separated Hybrid, Generalized Gradient Approximation Density Functional with Nonlocal Correlation, Designed by a Survival-of-the-Fittest Strategy},
  shorttitle = {{{$\omega$B97X-V}}},
  author = {Mardirossian, Narbe and {Head-Gordon}, Martin},
  year = 2014,
  journal = {Phys. Chem. Chem. Phys.},
  volume = {16},
  number = {21},
  pages = {9904},
  issn = {1463-9076, 1463-9084},
  doi = {10.1039/c3cp54374a},
  urldate = {2025-08-12},
  langid = {english}
}

@article{mayhallMultipleSolutionsSingleReference2010,
  title = {Multiple {{Solutions}} to the {{Single-Reference CCSD Equations}} for {{NiH}}},
  author = {Mayhall, Nicholas J. and Raghavachari, Krishnan},
  year = 2010,
  month = sep,
  journal = {J. Chem. Theory Comput.},
  volume = {6},
  number = {9},
  pages = {2714--2720},
  publisher = {American Chemical Society},
  issn = {1549-9618},
  doi = {10.1021/ct100321k},
  urldate = {2025-10-20}
}

@article{mcweenyDensityMatrixManyelectron1997,
  title = {The Density Matrix in Many-Electron Quantum Mechanics {{I}}. {{Generalized}} Product Functions. {{Factorization}} and Physical Interpretation of the Density Matrices},
  author = {McWeeny, R.},
  year = 1997,
  month = jan,
  journal = {Proc. R. Soc. Lond. Ser. Math. Phys. Sci.},
  volume = {253},
  number = {1273},
  pages = {242--259},
  publisher = {Royal Society},
  doi = {10.1098/rspa.1959.0191},
  urldate = {2025-09-26}
}

@book{mcweenyMethodsMolecularQuantum1969,
  title = {Methods of {{Molecular Quantum Mechanics}}},
  author = {McWeeny, R. and Sutcliffe, Brian T.},
  year = 1969,
  publisher = {Academic Press},
  googlebooks = {D\_Ph20OZu0YC},
  isbn = {978-0-12-486550-1},
  langid = {english}
}

@article{mcweenyRecentAdvancesDensity1960,
  title = {Some {{Recent Advances}} in {{Density Matrix Theory}}},
  author = {McWeeny, R.},
  year = 1960,
  month = apr,
  journal = {Rev. Mod. Phys.},
  volume = {32},
  number = {2},
  pages = {335--369},
  issn = {0034-6861},
  doi = {10.1103/RevModPhys.32.335},
  urldate = {2025-09-26},
  copyright = {http://link.aps.org/licenses/aps-default-license},
  langid = {english}
}

@article{mesterOverviewDevelopmentsMRCC2025,
  title = {Overview of {{Developments}} in the {{MRCC Program System}}},
  author = {Mester, D{\'a}vid and Nagy, P{\'e}ter R. and Cs{\'o}ka, J{\'o}zsef and {Gyevi-Nagy}, L{\'a}szl{\'o} and Szab{\'o}, P. Bern{\'a}t and Horv{\'a}th, R{\'e}ka A. and Petrov, Kl{\'a}ra and H{\'e}gely, Bence and Lad{\'o}czki, Bence and Samu, Gyula and L{\H o}rincz, Bal{\'a}zs D. and K{\'a}llay, Mih{\'a}ly},
  year = 2025,
  month = feb,
  journal = {J. Phys. Chem. A},
  volume = {129},
  number = {8},
  pages = {2086--2107},
  issn = {1089-5639, 1520-5215},
  doi = {10.1021/acs.jpca.4c07807},
  urldate = {2025-09-24},
  copyright = {https://creativecommons.org/licenses/by/4.0/},
  langid = {english}
}

@article{monkhorstCalculationPropertiesCoupledcluster1977,
  title = {{Calculation of properties with the coupled-cluster method}},
  author = {Monkhorst, Hendrik J.},
  year = 1977,
  journal = {Int. J. Quantum Chem.},
  volume = {12},
  number = {S11},
  pages = {421--432},
  issn = {1097-461X},
  doi = {10.1002/qua.560120850},
  urldate = {2025-09-11},
  langid = {ngerman}
}

@article{morchenNoniterativeTriplesTranscorrelated2025,
  title = {Non-Iterative {{Triples}} for {{Transcorrelated Coupled Cluster Theory}}},
  author = {M{\"o}rchen, Maximilian and Baiardi, Alberto and Lesiuk, Micha{\l} and Reiher, Markus},
  year = 2025,
  month = feb,
  journal = {J. Chem. Theory Comput.},
  volume = {21},
  number = {4},
  pages = {1588--1601},
  issn = {1549-9618, 1549-9626},
  doi = {10.1021/acs.jctc.4c01062},
  urldate = {2026-01-29},
  copyright = {https://creativecommons.org/licenses/by/4.0/},
  langid = {english}
}

@article{musialMultireferenceCoupledclusterTheory2011,
  title = {Multireference Coupled-Cluster Theory: {{The}} Easy Way},
  shorttitle = {Multireference Coupled-Cluster Theory},
  author = {Musia{\l}, Monika and Perera, Ajith and Bartlett, Rodney J.},
  year = 2011,
  month = mar,
  journal = {J. Chem. Phys.},
  volume = {134},
  number = {11},
  pages = {114108},
  issn = {0021-9606, 1089-7690},
  doi = {10.1063/1.3567115},
  urldate = {2026-01-29},
  langid = {english}
}

@article{navesdebritoTheoreticalStudyXray1991,
  title = {A Theoretical Study of X-ray Photoelectron Spectra of Model Molecules for Polymethylmethacrylate},
  author = {{Naves de Brito}, A. and Correia, N. and Svensson, S. and {\AA}gren, H.},
  year = 1991,
  month = aug,
  journal = {J. Chem. Phys.},
  volume = {95},
  number = {4},
  pages = {2965--2974},
  issn = {0021-9606},
  doi = {10.1063/1.460898},
  urldate = {2025-10-20}
}

@article{nooijenSimilarityTransformedEquationofmotion1997,
  title = {Similarity Transformed Equation-of-Motion Coupled-Cluster Theory: {{Details}}, Examples, and Comparisons},
  shorttitle = {Similarity Transformed Equation-of-Motion Coupled-Cluster Theory},
  author = {Nooijen, Marcel and Bartlett, Rodney J.},
  year = 1997,
  month = nov,
  journal = {J. Chem. Phys.},
  volume = {107},
  number = {17},
  pages = {6812--6830},
  issn = {0021-9606, 1089-7690},
  doi = {10.1063/1.474922},
  urldate = {2026-01-29},
  langid = {english}
}

@article{northPopulationAnalysisEffects2023,
  title = {Population Analysis and the Effects of {{Gaussian}} Basis Set Quality and Quantum Mechanical Approach: Main Group through Heavy Element Species},
  shorttitle = {Population Analysis and the Effects of {{Gaussian}} Basis Set Quality and Quantum Mechanical Approach},
  author = {North, Sasha C. and Jorgensen, Kameron R. and Pricetolstoy, Jason and Wilson, Angela K.},
  year = 2023,
  month = apr,
  journal = {Front Chem},
  volume = {11},
  pages = {1152500},
  issn = {2296-2646},
  doi = {10.3389/fchem.2023.1152500},
  urldate = {2026-02-02},
  pmcid = {PMC10154537},
  pmid = {37153525}
}

@article{otisHybridApproachExcitedstatespecific2020,
  title = {A Hybrid Approach to Excited-State-Specific Variational {{Monte Carlo}} and Doubly Excited States},
  author = {Otis, Leon and Craig, Isaac M. and Neuscamman, Eric},
  year = 2020,
  month = dec,
  journal = {J. Chem. Phys.},
  volume = {153},
  number = {23},
  pages = {234105},
  issn = {0021-9606},
  doi = {10.1063/5.0024572},
  urldate = {2025-10-20}
}

@article{otisOptimizationStabilityExcitedStateSpecific2023,
  title = {Optimization {{Stability}} in {{Excited-State-Specific Variational Monte Carlo}}},
  author = {Otis, Leon and Neuscamman, Eric},
  year = 2023,
  month = feb,
  journal = {J. Chem. Theory Comput.},
  volume = {19},
  number = {3},
  pages = {767--782},
  publisher = {American Chemical Society},
  issn = {1549-9618},
  doi = {10.1021/acs.jctc.2c00642},
  urldate = {2025-10-20}
}

@article{otisPromisingIntersectionExcitedstatespecific2023,
  title = {A Promising Intersection of Excited-State-Specific Methods from Quantum Chemistry and Quantum {{Monte Carlo}}},
  author = {Otis, Leon and Neuscamman, Eric},
  year = 2023,
  journal = {WIREs Comput. Mol. Sci.},
  volume = {13},
  number = {5},
  pages = {e1659},
  issn = {1759-0884},
  doi = {10.1002/wcms.1659},
  urldate = {2025-10-20},
  copyright = {\copyright{} 2023 Wiley Periodicals LLC.},
  langid = {english}
}

@article{pathakExcitedStatesVariational2021,
  title = {Excited States in Variational {{Monte Carlo}} Using a Penalty Method},
  author = {Pathak, Shivesh and Busemeyer, Brian and Rodrigues, Jo{\~a}o N. B. and Wagner, Lucas K.},
  year = 2021,
  month = jan,
  journal = {J. Chem. Phys.},
  volume = {154},
  number = {3},
  pages = {034101},
  issn = {0021-9606},
  doi = {10.1063/5.0030949},
  urldate = {2025-10-20}
}

@article{pereraSingletTripletSeparations2014,
  title = {Singlet--Triplet Separations of Di-Radicals Treated by the {{DEA}}/{{DIP-EOM-CCSD}} Methods},
  author = {Perera, Ajith and Molt, Robert W. and Lotrich, Victor F. and Bartlett, Rodney J.},
  year = 2014,
  month = jun,
  journal = {Theor Chem Acc},
  volume = {133},
  number = {8},
  pages = {1514},
  issn = {1432-2234},
  doi = {10.1007/s00214-014-1514-5},
  urldate = {2026-01-29},
  langid = {english}
}

@article{pinedafloresExcitedStateSpecific2019,
  title = {Excited {{State Specific Multi-Slater Jastrow Wave Functions}}},
  author = {Pineda Flores, Sergio D. and Neuscamman, Eric},
  year = 2019,
  month = feb,
  journal = {J. Phys. Chem. A},
  volume = {123},
  number = {8},
  pages = {1487--1497},
  publisher = {American Chemical Society},
  issn = {1089-5639},
  doi = {10.1021/acs.jpca.8b10671},
  urldate = {2025-10-20}
}

@article{plasserAnalysisExcitonicCharge2012,
  title = {Analysis of {{Excitonic}} and {{Charge Transfer Interactions}} from {{Quantum Chemical Calculations}}},
  author = {Plasser, Felix and Lischka, Hans},
  year = 2012,
  month = aug,
  journal = {J. Chem. Theory Comput.},
  volume = {8},
  number = {8},
  pages = {2777--2789},
  publisher = {American Chemical Society},
  issn = {1549-9618},
  doi = {10.1021/ct300307c},
  urldate = {2026-02-03}
}

@article{plasserLibwfaWavefunctionAnalysis2022,
  title = {Libwfa: {{Wavefunction}} Analysis Tools for Excited and Open-Shell Electronic States},
  shorttitle = {Libwfa},
  author = {Plasser, Felix and Krylov, Anna I. and Dreuw, Andreas},
  year = 2022,
  journal = {WIREs Comput. Mol. Sci.},
  volume = {12},
  number = {4},
  pages = {e1595},
  issn = {1759-0884},
  doi = {10.1002/wcms.1595},
  urldate = {2026-02-03},
  langid = {english}
}

@article{plasserNewToolsSystematic2014,
  title = {New Tools for the Systematic Analysis and Visualization of Electronic Excitations. {{I}}. {{Formalism}}},
  author = {Plasser, Felix and Wormit, Michael and Dreuw, Andreas},
  year = 2014,
  month = jul,
  journal = {J. Chem. Phys.},
  volume = {141},
  number = {2},
  pages = {024106},
  issn = {0021-9606, 1089-7690},
  doi = {10.1063/1.4885819},
  urldate = {2026-02-03},
  langid = {english}
}

@article{plasserNewToolsSystematic2014a,
  title = {New Tools for the Systematic Analysis and Visualization of Electronic Excitations. {{II}}. {{Applications}}},
  author = {Plasser, Felix and B{\"a}ppler, Stefanie A. and Wormit, Michael and Dreuw, Andreas},
  year = 2014,
  month = jul,
  journal = {J. Chem. Phys.},
  volume = {141},
  number = {2},
  pages = {024107},
  issn = {0021-9606},
  doi = {10.1063/1.4885820},
  urldate = {2026-02-03}
}

@article{pulayAnalyticalDerivativesForces2014,
  title = {Analytical Derivatives, Forces, Force Constants, Molecular Geometries, and Related Response Properties in Electronic Structure Theory},
  author = {Pulay, Peter},
  year = 2014,
  journal = {WIREs Comput. Mol. Sci.},
  volume = {4},
  number = {3},
  pages = {169--181},
  issn = {1759-0884},
  doi = {10.1002/wcms.1171},
  urldate = {2025-10-20},
  copyright = {\copyright{} 2013 John Wiley \& Sons, Ltd.}
}

@article{quadyAufbauSuppressedCoupled2025,
  title = {Aufbau {{Suppressed Coupled Cluster As}} a {{Post-Linear-Response Method}}},
  author = {Quady, Trine Kay and Tuckman, Harrison and Neuscamman, Eric},
  year = 2025,
  month = sep,
  journal = {J. Chem. Theory Comput.},
  volume = {21},
  number = {18},
  pages = {8843--8852},
  publisher = {American Chemical Society},
  issn = {1549-9618},
  doi = {10.1021/acs.jctc.5c01027},
  urldate = {2025-10-20}
}

@article{ricoSinglereferenceTheoriesMolecular1993,
  title = {Single-Reference Theories of Molecular Excited States with Single and Double Substitutions},
  author = {Rico, Rudolph J. and {Head-Gordon}, Martin},
  year = 1993,
  month = oct,
  journal = {Chemical Physics Letters},
  volume = {213},
  number = {3-4},
  pages = {224--232},
  issn = {00092614},
  doi = {10.1016/0009-2614(93)85124-7},
  urldate = {2025-09-11},
  langid = {english}
}

@article{robinsonExcitationVarianceMatching2017,
  title = {Excitation Variance Matching with Limited Configuration Interaction Expansions in Variational {{Monte Carlo}}},
  author = {Robinson, Paul J. and Pineda Flores, Sergio D. and Neuscamman, Eric},
  year = 2017,
  month = oct,
  journal = {J. Chem. Phys.},
  volume = {147},
  number = {16},
  pages = {164114},
  issn = {0021-9606},
  doi = {10.1063/1.5008743},
  urldate = {2025-10-20}
}

@article{roosAccurateMolecularOrbital1992,
  title = {Towards an Accurate Molecular Orbital Theory for Excited States: The Benzene Molecule},
  shorttitle = {Towards an Accurate Molecular Orbital Theory for Excited States},
  author = {Roos, Bj{\"o}rn O. and Andersson, Kerstin and F{\"u}lscher, Markus P.},
  year = 1992,
  month = apr,
  journal = {Chemical Physics Letters},
  volume = {192},
  number = {1},
  pages = {5--13},
  issn = {0009-2614},
  doi = {10.1016/0009-2614(92)85419-B},
  urldate = {2025-10-20}
}

@article{roweEquationsofMotionMethodExtended1968,
  title = {Equations-of-{{Motion Method}} and the {{Extended Shell Model}}},
  author = {Rowe, D. J.},
  year = 1968,
  month = jan,
  journal = {Rev. Mod. Phys.},
  volume = {40},
  number = {1},
  pages = {153--166},
  issn = {0034-6861},
  doi = {10.1103/RevModPhys.40.153},
  urldate = {2025-09-11},
  copyright = {http://link.aps.org/licenses/aps-default-license},
  langid = {english}
}

@article{rungeDensityFunctionalTheoryTimeDependent1984,
  title = {Density-{{Functional Theory}} for {{Time-Dependent Systems}}},
  author = {Runge, Erich and Gross, E. K. U.},
  year = 1984,
  month = mar,
  journal = {Phys. Rev. Lett.},
  volume = {52},
  number = {12},
  pages = {997--1000},
  issn = {0031-9007},
  doi = {10.1103/PhysRevLett.52.997},
  urldate = {2025-09-11},
  copyright = {http://link.aps.org/licenses/aps-default-license},
  langid = {english}
}

@article{salterAnalyticEnergyDerivatives1989,
  title = {Analytic Energy Derivatives in Many-Body Methods. {{I}}. {{First}} Derivatives},
  author = {Salter, E. A. and Trucks, Gary W. and Bartlett, Rodney J.},
  year = 1989,
  month = feb,
  journal = {J. Chem. Phys.},
  volume = {90},
  number = {3},
  pages = {1752--1766},
  issn = {0021-9606, 1089-7690},
  doi = {10.1063/1.456069},
  urldate = {2025-08-11},
  langid = {english}
}

@article{salterPropertyEvaluationOrbital1987,
  title = {Property Evaluation and Orbital Relaxation in Coupled Cluster Methods},
  author = {Salter, E. A. and Sekino, Hideo and Bartlett, Rodney J.},
  year = 1987,
  month = jul,
  journal = {J. Chem. Phys.},
  volume = {87},
  number = {1},
  pages = {502--509},
  issn = {0021-9606},
  doi = {10.1063/1.453596},
  urldate = {2025-08-19}
}

@article{sarkarBenchmarkingTDDFTWave2021,
  title = {Benchmarking {{TD-DFT}} and {{Wave Function Methods}} for {{Oscillator Strengths}} and {{Excited-State Dipole Moments}}},
  author = {Sarkar, Rudraditya and {Boggio-Pasqua}, Martial and Loos, Pierre-Fran{\c c}ois and Jacquemin, Denis},
  year = 2021,
  month = feb,
  journal = {J. Chem. Theory Comput.},
  volume = {17},
  number = {2},
  pages = {1117--1132},
  publisher = {American Chemical Society},
  issn = {1549-9618},
  doi = {10.1021/acs.jctc.0c01228},
  urldate = {2025-08-07}
}

@incollection{schmittStructuresDipoleMoments2018,
  title = {Structures and {{Dipole Moments}} of {{Molecules}} in {{Their Electronically Excited States}}},
  booktitle = {Frontiers and {{Advances}} in {{Molecular Spectroscopy}}},
  author = {Schmitt, Michael and Meerts, Leo},
  year = 2018,
  pages = {143--193},
  publisher = {Elsevier},
  doi = {10.1016/B978-0-12-811220-5.00005-8},
  urldate = {2025-09-26},
  copyright = {https://www.elsevier.com/tdm/userlicense/1.0/},
  isbn = {978-0-12-811220-5},
  langid = {english}
}

@article{schraivogelTranscorrelatedCoupledCluster2021,
  title = {Transcorrelated Coupled Cluster Methods},
  author = {Schraivogel, Thomas and Cohen, Aron J. and Alavi, Ali and Kats, Daniel},
  year = 2021,
  month = nov,
  journal = {J. Chem. Phys.},
  volume = {155},
  number = {19},
  pages = {191101},
  issn = {0021-9606, 1089-7690},
  doi = {10.1063/5.0072495},
  urldate = {2026-01-27},
  langid = {english}
}

@article{sekinoLinearResponseCoupledcluster1984,
  title = {A Linear Response, Coupled-Cluster Theory for Excitation Energy},
  author = {Sekino, Hideo and Bartlett, Rodney J.},
  year = 1984,
  journal = {Int. J. Quantum Chem.},
  volume = {26},
  number = {S18},
  pages = {255--265},
  issn = {1097-461X},
  doi = {10.1002/qua.560260826},
  urldate = {2025-09-11},
  langid = {english}
}

@article{shaoSpinFlipApproach2003,
  title = {The Spin--Flip Approach within Time-Dependent Density Functional Theory: {{Theory}} and Applications to Diradicals},
  shorttitle = {The Spin--Flip Approach within Time-Dependent Density Functional Theory},
  author = {Shao, Yihan and {Head-Gordon}, Martin and Krylov, Anna I.},
  year = 2003,
  month = mar,
  journal = {J. Chem. Phys.},
  volume = {118},
  number = {11},
  pages = {4807--4818},
  issn = {0021-9606, 1089-7690},
  doi = {10.1063/1.1545679},
  urldate = {2026-01-29},
  langid = {english}
}

@book{shavittManyBodyMethodsChemistry2009,
  title = {Many-{{Body Methods}} in {{Chemistry}} and {{Physics}}: {{MBPT}} and {{Coupled-Cluster Theory}}},
  shorttitle = {Many-{{Body Methods}} in {{Chemistry}} and {{Physics}}},
  author = {Shavitt, Isaiah and Bartlett, Rodney J.},
  year = 2009,
  series = {Cambridge {{Molecular Science}}},
  publisher = {Cambridge University Press},
  address = {Cambridge},
  doi = {10.1017/CBO9780511596834},
  urldate = {2025-09-26},
  isbn = {978-0-521-81832-2}
}

@article{sheaCommunicationMeanField2018,
  title = {Communication: {{A}} Mean Field Platform for Excited State Quantum Chemistry},
  shorttitle = {Communication},
  author = {Shea, Jacqueline A. R. and Neuscamman, Eric},
  year = 2018,
  month = aug,
  journal = {J. Chem. Phys.},
  volume = {149},
  number = {8},
  pages = {081101},
  issn = {0021-9606, 1089-7690},
  doi = {10.1063/1.5045056},
  urldate = {2025-09-26},
  langid = {english}
}

@article{sheaGeneralizedVariationalPrinciple2020,
  title = {A {{Generalized Variational Principle}} with {{Applications}} to {{Excited State Mean Field Theory}}},
  author = {Shea, Jacqueline A. R. and Gwin, Elise and Neuscamman, Eric},
  year = 2020,
  month = mar,
  journal = {J. Chem. Theory Comput.},
  volume = {16},
  number = {3},
  pages = {1526--1540},
  issn = {1549-9618, 1549-9626},
  doi = {10.1021/acs.jctc.9b01105},
  urldate = {2025-09-26},
  copyright = {http://pubs.acs.org/page/policy/authorchoice\_termsofuse.html},
  langid = {english}
}

@article{sheaSizeConsistentExcited2017,
  title = {Size {{Consistent Excited States}} via {{Algorithmic Transformations}} between {{Variational Principles}}},
  author = {Shea, Jacqueline A. R. and Neuscamman, Eric},
  year = 2017,
  month = dec,
  journal = {J. Chem. Theory Comput.},
  volume = {13},
  number = {12},
  pages = {6078--6088},
  publisher = {American Chemical Society},
  issn = {1549-9618},
  doi = {10.1021/acs.jctc.7b00923},
  urldate = {2025-10-20}
}

@article{shepardDoubleExcitationEnergies2022,
  title = {Double {{Excitation Energies}} from {{Quantum Monte Carlo Using State-Specific Energy Optimization}}},
  author = {Shepard, Stuart and {Panad{\'e}s-Barrueta}, Ram{\'o}n L. and Moroni, Saverio and Scemama, Anthony and Filippi, Claudia},
  year = 2022,
  month = nov,
  journal = {J. Chem. Theory Comput.},
  volume = {18},
  number = {11},
  pages = {6722--6731},
  publisher = {American Chemical Society},
  issn = {1549-9618},
  doi = {10.1021/acs.jctc.2c00769},
  urldate = {2025-10-20}
}

@article{sirucekExcitedStateAbsorptionReference2025,
  title = {Excited-{{State Absorption}}: {{Reference Oscillator Strengths}}, {{Wave Function}}, and {{TDDFT Benchmarks}}},
  shorttitle = {Excited-{{State Absorption}}},
  author = {{\v S}ir{\r u}{\v c}ek, Jakub and Le Guennic, Boris and Damour, Yann and Loos, Pierre-Fran{\c c}ois and Jacquemin, Denis},
  year = 2025,
  month = may,
  journal = {J. Chem. Theory Comput.},
  volume = {21},
  number = {9},
  pages = {4688--4703},
  publisher = {American Chemical Society},
  issn = {1549-9618},
  doi = {10.1021/acs.jctc.5c00159},
  urldate = {2025-12-05}
}

@article{sneskovExcitedStateCoupled2012,
  title = {Excited State Coupled Cluster Methods},
  author = {Sneskov, Kristian and Christiansen, Ove},
  year = 2012,
  journal = {WIREs Comput. Mol. Sci.},
  volume = {2},
  number = {4},
  pages = {566--584},
  issn = {1759-0884},
  doi = {10.1002/wcms.99},
  urldate = {2025-09-11},
  langid = {english}
}

@article{stantonEquationMotionCoupledcluster1993,
  title = {The Equation of Motion Coupled-Cluster Method. {{A}} Systematic Biorthogonal Approach to Molecular Excitation Energies, Transition Probabilities, and Excited State Properties},
  author = {Stanton, John F. and Bartlett, Rodney J.},
  year = 1993,
  month = may,
  journal = {J. Chem. Phys.},
  volume = {98},
  number = {9},
  pages = {7029--7039},
  issn = {0021-9606, 1089-7690},
  doi = {10.1063/1.464746},
  urldate = {2025-09-11},
  langid = {english}
}

@book{szaboModernQuantumChemistry1996,
  title = {Modern {{Quantum Chemistry}}},
  author = {Szabo, Attila and Ostlund, Neil},
  year = 1996,
  publisher = {Dover},
  urldate = {2025-09-24},
  langid = {english}
}

@article{taitTransientEPRReveals2015,
  title = {Transient {{EPR Reveals Triplet State Delocalization}} in a {{Series}} of {{Cyclic}} and {{Linear}} {$\pi$}-{{Conjugated Porphyrin Oligomers}}},
  author = {Tait, Claudia E. and Neuhaus, Patrik and Peeks, Martin D. and Anderson, Harry L. and Timmel, Christiane R.},
  year = 2015,
  month = jul,
  journal = {J. Am. Chem. Soc.},
  volume = {137},
  number = {25},
  pages = {8284--8293},
  publisher = {American Chemical Society},
  issn = {0002-7863},
  doi = {10.1021/jacs.5b04511},
  urldate = {2025-09-26}
}

@article{tranExploringLigandtoMetalChargeTransfer2023,
  title = {Exploring {{Ligand-to-Metal Charge-Transfer States}} in the {{Photo-Ferrioxalate System Using Excited-State Specific Optimization}}},
  author = {Tran, Lan Nguyen and Neuscamman, Eric},
  year = 2023,
  month = aug,
  journal = {J. Phys. Chem. Lett.},
  volume = {14},
  number = {33},
  pages = {7454--7460},
  publisher = {American Chemical Society},
  doi = {10.1021/acs.jpclett.3c01308},
  urldate = {2025-10-20}
}

@article{tranImprovingExcitedStatePotential2020,
  title = {Improving {{Excited-State Potential Energy Surfaces}} via {{Optimal Orbital Shapes}}},
  author = {Tran, Lan Nguyen and Neuscamman, Eric},
  year = 2020,
  month = oct,
  journal = {J. Phys. Chem. A},
  volume = {124},
  number = {40},
  pages = {8273--8279},
  publisher = {American Chemical Society},
  issn = {1089-5639},
  doi = {10.1021/acs.jpca.0c07593},
  urldate = {2025-10-20}
}

@article{tranTrackingExcitedStates2019,
  title = {Tracking {{Excited States}} in {{Wave Function Optimization Using Density Matrices}} and {{Variational Principles}}},
  author = {Tran, Lan Nguyen and Shea, Jacqueline A. R. and Neuscamman, Eric},
  year = 2019,
  month = sep,
  journal = {J. Chem. Theory Comput.},
  volume = {15},
  number = {9},
  pages = {4790--4803},
  publisher = {American Chemical Society},
  issn = {1549-9618},
  doi = {10.1021/acs.jctc.9b00351},
  urldate = {2025-10-20}
}

@article{tsuchimochiDoubleConfigurationInteraction2024,
  title = {Double Configuration Interaction Singles: {{Scalable}} and Size-Intensive Approach for Orbital Relaxation in Excited States and Bond-Dissociation},
  shorttitle = {Double Configuration Interaction Singles},
  author = {Tsuchimochi, Takashi},
  year = 2024,
  month = dec,
  journal = {J. Chem. Phys.},
  volume = {161},
  number = {24},
  pages = {241102},
  issn = {0021-9606},
  doi = {10.1063/5.0243710},
  urldate = {2025-10-20}
}

@article{tuckmanAufbauSuppressedCoupled2024,
  title = {Aufbau {{Suppressed Coupled Cluster Theory}} for {{Electronically Excited States}}},
  author = {Tuckman, Harrison and Neuscamman, Eric},
  year = 2024,
  month = apr,
  journal = {J. Chem. Theory Comput.},
  volume = {20},
  number = {7},
  pages = {2761--2773},
  publisher = {American Chemical Society},
  issn = {1549-9618},
  doi = {10.1021/acs.jctc.3c01285},
  urldate = {2025-08-11}
}

@article{tuckmanExcitedStateSpecificPseudoprojectedCoupledCluster2023,
  title = {Excited-{{State-Specific Pseudoprojected Coupled-Cluster Theory}}},
  author = {Tuckman, Harrison and Neuscamman, Eric},
  year = 2023,
  month = sep,
  journal = {J. Chem. Theory Comput.},
  volume = {19},
  number = {18},
  pages = {6160--6171},
  publisher = {American Chemical Society},
  issn = {1549-9618},
  doi = {10.1021/acs.jctc.3c00194},
  urldate = {2025-10-20}
}

@article{tuckmanFastAccurateCharge2025,
  title = {Fast and {{Accurate Charge Transfer Excitations}} via {{Nested Aufbau Suppressed Coupled Cluster}}},
  author = {Tuckman, Harrison and Neuscamman, Eric},
  year = 2025,
  month = aug,
  journal = {J. Phys. Chem. Lett.},
  volume = {16},
  number = {31},
  pages = {7889--7897},
  issn = {1948-7185, 1948-7185},
  doi = {10.1021/acs.jpclett.5c01576},
  urldate = {2025-09-11},
  copyright = {https://doi.org/10.15223/policy-029},
  langid = {english}
}

@article{tuckmanImprovingAufbauSuppressed2025,
  title = {Improving {{Aufbau Suppressed Coupled Cluster}} through {{Perturbative Analysis}}},
  author = {Tuckman, Harrison and Ma, Ziheng and Neuscamman, Eric},
  year = 2025,
  month = apr,
  journal = {J. Chem. Theory Comput.},
  volume = {21},
  number = {8},
  pages = {3993--4005},
  publisher = {American Chemical Society},
  issn = {1549-9618},
  doi = {10.1021/acs.jctc.5c00096},
  urldate = {2025-08-11}
}

@article{wangHowAccurateAre2020,
  title = {How Accurate Are {{TD-DFT}} Excited-State Geometries Compared to {{DFT}} Ground-State Geometries?},
  author = {Wang, Jun and Durbeej, Bo},
  year = 2020,
  journal = {J. Comput. Chem.},
  volume = {41},
  number = {18},
  pages = {1718--1729},
  issn = {1096-987X},
  doi = {10.1002/jcc.26213},
  urldate = {2025-09-26},
  langid = {english}
}

@article{yarkonyDiabolicalConicalIntersections1996a,
  title = {Diabolical Conical Intersections},
  author = {Yarkony, David R.},
  year = 1996,
  month = oct,
  journal = {Rev. Mod. Phys.},
  volume = {68},
  number = {4},
  pages = {985--1013},
  issn = {0034-6861, 1539-0756},
  doi = {10.1103/RevModPhys.68.985},
  urldate = {2026-01-27},
  copyright = {http://link.aps.org/licenses/aps-default-license},
  langid = {english}
}

@article{yeHalfProjectedSelfConsistentField2019,
  title = {Half-{{Projected}} {$\sigma$} {{Self-Consistent Field For Electronic Excited States}}},
  author = {Ye, Hong-Zhou and Van Voorhis, Troy},
  year = 2019,
  month = may,
  journal = {J. Chem. Theory Comput.},
  volume = {15},
  number = {5},
  pages = {2954--2965},
  publisher = {American Chemical Society},
  issn = {1549-9618},
  doi = {10.1021/acs.jctc.8b01224},
  urldate = {2025-10-20}
}

@article{yeSSCFDirectEnergytargeting2017,
  title = {{$\sigma$}-{{SCF}}: {{A}} Direct Energy-Targeting Method to Mean-Field Excited States},
  shorttitle = {{$\sigma$}-{{SCF}}},
  author = {Ye, Hong-Zhou and Welborn, Matthew and Ricke, Nathan D. and Van Voorhis, Troy},
  year = 2017,
  month = dec,
  journal = {J. Chem. Phys.},
  volume = {147},
  number = {21},
  pages = {214104},
  issn = {0021-9606},
  doi = {10.1063/1.5001262},
  urldate = {2025-10-20}
}

@article{zhaoDensityFunctionalExtension2020,
  title = {Density {{Functional Extension}} to {{Excited-State Mean-Field Theory}}},
  author = {Zhao, Luning and Neuscamman, Eric},
  year = 2020,
  month = jan,
  journal = {J. Chem. Theory Comput.},
  volume = {16},
  number = {1},
  pages = {164--178},
  publisher = {American Chemical Society},
  issn = {1549-9618},
  doi = {10.1021/acs.jctc.9b00530},
  urldate = {2025-10-20}
}

@article{zhaoEfficientVariationalPrinciple2016,
  title = {An {{Efficient Variational Principle}} for the {{Direct Optimization}} of {{Excited States}}},
  author = {Zhao, Luning and Neuscamman, Eric},
  year = 2016,
  month = aug,
  journal = {J. Chem. Theory Comput.},
  volume = {12},
  number = {8},
  pages = {3436--3440},
  publisher = {American Chemical Society},
  issn = {1549-9618},
  doi = {10.1021/acs.jctc.6b00508},
  urldate = {2025-10-20}
}

@article{zhengPerformanceDeltaCoupledClusterMethods2019,
  title = {Performance of {{Delta-Coupled-Cluster Methods}} for {{Calculations}} of {{Core-Ionization Energies}} of {{First-Row Elements}}},
  author = {Zheng, Xuechen and Cheng, Lan},
  year = 2019,
  month = sep,
  journal = {J. Chem. Theory Comput.},
  volume = {15},
  number = {9},
  pages = {4945--4955},
  publisher = {American Chemical Society},
  issn = {1549-9618},
  doi = {10.1021/acs.jctc.9b00568},
  urldate = {2025-10-20}
}

\clearpage
\onecolumngrid

\section{Supporting Information}
\renewcommand{\thesection}{S\arabic{section}}
\renewcommand{\theequation}{S\arabic{equation}}
\renewcommand{\thefigure}{S\arabic{figure}}
\renewcommand{\thetable}{S\arabic{table}}
\setcounter{section}{0}
\setcounter{figure}{0}
\setcounter{equation}{0}
\setcounter{table}{0}
%This will usually read something like: ``Experimental procedures and
%characterization data for all new compounds. The class will
%automatically add a sentence pointing to the information on-line:
\section{Raw Data}

The attached .csv file contains all of the raw data from the various calculations used throughout this study with topics separated by worksheet. The first worksheet contains excitation energies (eV) for every iteration of the natural orbital refinement procedure using ASCC$^{(M,1)}$ and PLASCC$^{(M,1)}$ on the four different references: CIS, EOM-CCSD, TD-DFT/$\omega$B97X-V, and ESMF. For totally symmetric states, the ``flipped" refers to the second (PL)ASCC solution. Values of N/A indicate that convergence was not achieved for the provided calculation. The reference values for all valence and Rydberg states are calculated at the exFCI level of theory except for thioformaldehyde's 1$^1\text{A}_2$, which is calculated at EOM-CCSDTQ. The reference values for all charge transfer states are calculated using EOM-CCSDT except for 3,5-difluoro-penta-2,3-dienamine, which is calculated using LR-CC3. The second worksheet contains the Mulliken and L\"owdin population analyses results for EOM-CCSD, ASCC$^{(M,1)}$, and ASCC$^{(1,M)}$ on each of the charge transfer excitations. The third worksheet contains the Mulliken and L\"owdin population analyses results for EOM-CCSD, ASCC$^{(M,1)}$, and ASCC$^{(1,M)}$ at each geometry in the water flyby system. The final worksheet contains the excitation energies (eV) and dipole moments (D) of the various versions of (PL)ASCC using each of the references alongside the values calculated with EOM-CCSD, (OU)LR-CCSD, and (OR)LR-CCSD. Reference excitation energies are at the exFCI level of theory, again except for thioformaldehyde's 1$^1\text{A}_2$, which is calculated at EOM-CCSDTQ. Reference dipole moments are calculated at (OR)LR-CCSDTQP for H$_2$O, H$_2$S, and CO and (OR)LR-CCSDTQ for formaldehyde and thioformaldehyde.

\newpage
\section{Charge Transfer Natural Orbital Refinement Summary}

\begin{table}[ht]
    \centering
    \caption{Diagnostics when performing natural orbital refinement on the ammonia-difluorine $2^1A_1$ charge transfer excitation. N/A indicates that convergence could not be reached for the specific calculation or the previous calculation(s).}
    \label{CT_Diagnostics_1}
    \begin{tabular}{lllccccc}
        & & & \multicolumn{5}{c}{Orbital Refinements} \\
        ASCC Variant & Reference & Diagnostic & 0 & 1 & 2 & 3 & 4 \\
        \hline
        ASCC$^{\left(M,1\right)}$ & CIS & Energy & 9.13 & 9.17 & 9.17 & 9.17 & 9.17 \\
        & & Max $\abs{T-T^{(0)}}$ & 0.09 & 0.03 & 0.11 & 0.03 & 0.10 \\
        & & Max $\abs{\Lambda-\Lambda^{(0)}}$ & 0.11 & 0.08 & 0.08 & 0.08 & \\
        ASCC$^{\left(M,1\right)}$ & EOM-CCSD & Energy & 9.20 & 9.18 & 9.17 & 9.18 & 9.17 \\
        & & Max $\abs{T-T^{(0)}}$ & 0.06 & 0.07 & 0.23 & 0.08 & 0.16 \\
        & & Max $\abs{\Lambda-\Lambda^{(0)}}$ & 0.10 & 0.09 & 0.10 & 0.09 & \\
        ASCC$^{\left(M,1\right)}$ & $\omega$B97X-V & Energy & 9.14 & 9.17 & 9.17 & 9.17 & 9.17 \\
        & & Max $\abs{T-T^{(0)}}$ & 0.09 & 0.03 & 0.13 & 0.03 & 0.11 \\
        & & Max $\abs{\Lambda-\Lambda^{(0)}}$ & 0.11 & 0.08 & 0.08 & 0.08 & \\
        ASCC$^{\left(M,1\right)}$ & ESMF & Energy & 9.17 & 9.17 & 9.17 & 9.17 & 9.17 \\
        & & Max $\abs{T-T^{(0)}}$ & 0.03 & 0.20 & 0.07 & 0.16 & 0.03 \\
        & & Max $\abs{\Lambda-\Lambda^{(0)}}$ & 0.08 & 0.09 & 0.09 & 0.08 & \\   
        PLASCC$^{\left(M,1\right)}$ & CIS & Energy & 9.28 & 9.38 & 9.38 & 9.38 & 9.38 \\
        & & Max $\abs{T-T^{(0)}}$ & 0.09 & 0.03 & 0.03 & 0.03 & 0.03 \\
        & & Max $\abs{\Lambda-\Lambda^{(0)}}$ & 0.09 & 0.07 & 0.07 & 0.07 & \\
        PLASCC$^{\left(M,1\right)}$ & EOM-CCSD & Energy & 9.34 & 9.39 & 9.38 & 9.38 & 9.38 \\
        & & Max $\abs{T-T^{(0)}}$ & 0.06 & 0.03 & 0.03 & 0.03 & 0.03 \\
        & & Max $\abs{\Lambda-\Lambda^{(0)}}$ & 0.09 & 0.07 & 0.07 & 0.07 & \\
        PLASCC$^{\left(M,1\right)}$ & $\omega$B97X-V & Energy & 9.32 & 9.39 & 9.39 & 9.39 & 9.39 \\
        & & Max $\abs{T-T^{(0)}}$ & 0.09 & 0.03 & 0.03 & 0.03 & 0.03 \\
        & & Max $\abs{\Lambda-\Lambda^{(0)}}$ & 0.09 & 0.07 & 0.07 & 0.07 & \\
        PLASCC$^{\left(M,1\right)}$ & ESMF & Energy & 9.37 & 9.39 & 9.39 & 9.39 & 9.39 \\
        & & Max $\abs{T-T^{(0)}}$ & 0.03 & 0.03 & 0.03 & 0.03 & 0.03 \\
        & & Max $\abs{\Lambda-\Lambda^{(0)}}$ & 0.07 & 0.07 & 0.07 & 0.07 & \\
    \end{tabular}
\end{table}

\begin{table}[ht]
    \centering
    \caption{Diagnostics when performing natural orbital refinement on the acetone-difluorine $3^1A$ charge transfer excitation. N/A indicates that convergence could not be reached for the specific calculation or the previous calculation(s).}
    \label{CT_Diagnostics_2}
    \begin{tabular}{lllccccc}
        & & & \multicolumn{5}{c}{Orbital Refinements} \\
        ASCC Variant & Reference & Diagnostic & 0 & 1 & 2 & 3 & 4 \\
        \hline
        ASCC$^{\left(M,1\right)}$ & CIS & Energy & 5.77 & 5.83 & 5.80 & 5.75 & 5.21 \\
        & & Max $\abs{T-T^{(0)}}$ & 0.14 & 0.07 & 0.15 & 0.41 & 0.34 \\
        & & Max $\abs{\Lambda-\Lambda^{(0)}}$ & 0.20 & 0.11 & 0.13 & 0.16 &  \\
        ASCC$^{\left(M,1\right)}$ & EOM-CCSD & Energy & 5.84 & 5.50 & 4.66 & 4.44 & 4.40 \\
        & & Max $\abs{T-T^{(0)}}$ & 0.13 & 0.43 & 0.17 & 0.19 & 0.19 \\
        & & Max $\abs{\Lambda-\Lambda^{(0)}}$ & 0.28 & 0.39 & 0.19 & 0.17 &  \\
        ASCC$^{\left(M,1\right)}$ & $\omega$B97X-V & Energy & 5.74 & 5.82 & 5.80 & 5.80 & 5.74 \\
        & & Max $\abs{T-T^{(0)}}$ & 0.16 & 0.05 & 0.12 & 0.22 & 0.37 \\
        & & Max $\abs{\Lambda-\Lambda^{(0)}}$ & 0.22 & 0.09 & 0.11 & 0.07 &  \\
        ASCC$^{\left(M,1\right)}$ & ESMF & Energy & 5.74 & 5.80 & 5.79 & 5.71 & 4.97 \\
        & & Max $\abs{T-T^{(0)}}$ & 0.09 & 0.10 & 0.17 & 0.34 & 0.26 \\
        & & Max $\abs{\Lambda-\Lambda^{(0)}}$ & 0.15 & 0.11 & 0.09 & 0.22 &  \\
        PLASCC$^{\left(M,1\right)}$ & CIS & Energy & 5.79 & 7.09 & 6.08 & N/A & N/A \\
        & & Max $\abs{T-T^{(0)}}$ & 0.16 & 0.80 & 0.40 & N/A & N/A \\
        & & Max $\abs{\Lambda-\Lambda^{(0)}}$ & 0.32 & 0.44 & N/A & N/A & N/A \\
        PLASCC$^{\left(M,1\right)}$ & EOM-CCSD & Energy & 5.87 & 5.94 & 5.94 & 5.93 & 5.92 \\
        & & Max $\abs{T-T^{(0)}}$ & 0.08 & 0.08 & 0.04 & 0.07 & 0.09 \\
        & & Max $\abs{\Lambda-\Lambda^{(0)}}$ & 0.14 & 0.11 & 0.11 & 0.11 &  \\
        PLASCC$^{\left(M,1\right)}$ & $\omega$B97X-V & Energy & 5.81 & 7.04 & N/A & N/A & N/A \\
        & & Max $\abs{T-T^{(0)}}$ & 0.16 & 0.68 & N/A & N/A & N/A \\
        & & Max $\abs{\Lambda-\Lambda^{(0)}}$ & 0.47 & 0.35 & N/A & N/A &  \\
        PLASCC$^{\left(M,1\right)}$ & ESMF & Energy & 5.85 & 5.96 & 5.96 & 5.96 & 5.96 \\
        & & Max $\abs{T-T^{(0)}}$ & 0.10 & 0.04 & 0.04 & 0.03 & 0.04 \\
        & & Max $\abs{\Lambda-\Lambda^{(0)}}$ & 0.12 & 0.08 & 0.08 & 0.08 &  \\
    \end{tabular}
\end{table}

\begin{table}[ht]
    \centering
    \caption{Diagnostics when performing natural orbital refinement on the pyrazine-difluorine $2^1B_2$ charge transfer excitation. N/A indicates that convergence could not be reached for the specific calculation or the previous calculation(s).}
    \label{CT_Diagnostics_3}
    \begin{tabular}{lllccccc}
        & & & \multicolumn{5}{c}{Orbital Refinements} \\
        ASCC Variant & Reference& Diagnostic & 0 & 1 & 2 & 3 & 4 \\
        \hline
        ASCC$^{\left(M,1\right)}$ & CIS & Energy & 6.42 & 6.56 & 6.55 & 6.59 & N/A \\
        & & Max $\abs{T-T^{(0)}}$ & 0.09 & 0.12 & 0.09 & 0.49 & N/A \\
        & & Max $\abs{\Lambda-\Lambda^{(0)}}$ & 0.19 & 0.12 & 0.12 & 0.24 &  \\
        ASCC$^{\left(M,1\right)}$ & EOM-CCSD & Energy & 6.56 & 6.57 & 6.46 & N/A & N/A \\
        & & Max $\abs{T-T^{(0)}}$ & 0.09 & 0.45 & 0.60 & N/A & N/A \\
        & & Max $\abs{\Lambda-\Lambda^{(0)}}$ & 0.14 & 0.16 & 0.30 & N/A &  \\
        ASCC$^{\left(M,1\right)}$ & $\omega$B97X-V & Energy & 6.48 & 6.56 & 6.55 & 6.61 & 6.29 \\
        & & Max $\abs{T-T^{(0)}}$ & 0.09 & 0.14 & 0.12 & 0.49 & 0.98 \\
        & & Max $\abs{\Lambda-\Lambda^{(0)}}$ & 0.18 & 0.12 & 0.12 & 0.26 &  \\
        ASCC$^{\left(M,1\right)}$ & ESMF & Energy & 6.51 & 6.56 & 6.56 & 6.56 & 6.53 \\
        & & Max $\abs{T-T^{(0)}}$ & 0.09 & 0.09 & 0.09 & 0.17 & 0.33 \\
        & & Max $\abs{\Lambda-\Lambda^{(0)}}$ & 0.13 & 0.12 & 0.12 & 0.14 &  \\
        PLASCC$^{\left(M,1\right)}$ & CIS & Energy & 6.33 & 6.61 & 6.72 & N/A & N/A \\
        & & Max $\abs{T-T^{(0)}}$ & 0.09 & 0.19 & 0.19 & N/A & N/A \\
        & & Max $\abs{\Lambda-\Lambda^{(0)}}$ & 0.16 & 0.24 & 0.27 & N/A &  \\
        PLASCC$^{\left(M,1\right)}$ & EOM-CCSD & Energy & 6.48 & 6.69 & N/A & N/A & N/A \\
        & & Max $\abs{T-T^{(0)}}$ & 0.09 & 0.16 & N/A & N/A & N/A \\
        & & Max $\abs{\Lambda-\Lambda^{(0)}}$ & 0.16 & 0.27 & N/A & N/A &  \\
        PLASCC$^{\left(M,1\right)}$ & $\omega$B97X-V & Energy & 6.43 & 6.55 & 6.64 & N/A & N/A \\
        & & Max $\abs{T-T^{(0)}}$ & 0.09 & 0.09 & 0.21 & N/A & N/A \\
        & & Max $\abs{\Lambda-\Lambda^{(0)}}$ & 0.15 & 0.13 & 0.28 & N/A &  \\
        PLASCC$^{\left(M,1\right)}$ & ESMF & Energy & 6.47 & 6.56 & 6.74 & N/A & N/A \\
        & & Max $\abs{T-T^{(0)}}$ & 0.09 & 0.11 & 0.22 & N/A & N/A \\
        & & Max $\abs{\Lambda-\Lambda^{(0)}}$ & 0.12 & 0.15 & 0.34 & N/A &  \\
    \end{tabular}
\end{table}

\begin{table}[ht]
    \centering
    \caption{Diagnostics when performing natural orbital refinement on the pyrazine-difluorine $2^1A_2$ charge transfer excitation. N/A indicates that convergence could not be reached for the specific calculation or the previous calculation(s).}
    \label{CT_Diagnostics_4}
    \begin{tabular}{lllccccc}
        & & & \multicolumn{5}{c}{Orbital Refinements} \\
        ASCC Variant & Reference & Diagnostic & 0 & 1 & 2 & 3 & 4 \\
        \hline
        ASCC$^{\left(M,1\right)}$ & CIS & Energy & 6.36 & 6.48 & 6.47 & 6.47 & 6.47 \\
        & & Max $\abs{T-T^{(0)}}$ & 0.10 & 0.09 & 0.09 & 0.09 & 0.09 \\
        & & Max $\abs{\Lambda-\Lambda^{(0)}}$ & 0.17 & 0.13 & 0.13 & 0.13 &  \\
        ASCC$^{\left(M,1\right)}$ & EOM-CCSD & Energy & 6.72 & 6.60 & 6.52 & 6.49 & 6.48 \\
        & & Max $\abs{T-T^{(0)}}$ & 0.12 & 0.09 & 0.09 & 0.09 & 0.09 \\
        & & Max $\abs{\Lambda-\Lambda^{(0)}}$ & 0.14 & 0.13 & 0.12 & 0.12 &  \\
        ASCC$^{\left(M,1\right)}$ & $\omega$B97X-V & Energy & 6.37 & 6.47 & 6.46 & 6.47 & 6.47 \\
        & & Max $\abs{T-T^{(0)}}$ & 0.12 & 0.09 & 0.09 & 0.09 & 0.09 \\
        & & Max $\abs{\Lambda-\Lambda^{(0)}}$ & 0.19 & 0.13 & 0.13 & 0.13 &  \\
        ASCC$^{\left(M,1\right)}$ & ESMF & Energy & 6.41 & 6.47 & 6.46 & 6.47 & 6.47 \\
        & & Max $\abs{T-T^{(0)}}$ & 0.09 & 0.09 & 0.09 & 0.09 & 0.09 \\
        & & Max $\abs{\Lambda-\Lambda^{(0)}}$ & 0.15 & 0.13 & 0.13 & 0.13 &  \\
        PLASCC$^{\left(M,1\right)}$ & CIS & Energy & 6.36 & 6.73 & 6.60 & 6.67 & N/A \\
        & & Max $\abs{T-T^{(0)}}$ & 0.21 & 0.11 & 0.27 & 0.12 & N/A \\
        & & Max $\abs{\Lambda-\Lambda^{(0)}}$ & 0.20 & 0.32 & 0.30 & 0.46 &  \\
        PLASCC$^{\left(M,1\right)}$ & EOM-CCSD & Energy & 6.53 & N/A & N/A & N/A & N/A \\
        & & Max $\abs{T-T^{(0)}}$ & 0.17 & N/A & N/A & N/A & N/A \\
        & & Max $\abs{\Lambda-\Lambda^{(0)}}$ & 0.35 & N/A & N/A & N/A &  \\
        PLASCC$^{\left(M,1\right)}$ & $\omega$B97X-V & Energy & 6.62 & N/A & N/A & N/A & N/A \\
        & & Max $\abs{T-T^{(0)}}$ & 0.32 & N/A & N/A & N/A & N/A \\
        & & Max $\abs{\Lambda-\Lambda^{(0)}}$ & 0.16 & N/A & N/A & N/A &  \\
        PLASCC$^{\left(M,1\right)}$ & ESMF & Energy & 6.58 & 6.69 & N/A & N/A & N/A \\
        & & Max $\abs{T-T^{(0)}}$ & 0.29 & 0.10 & N/A & N/A & N/A \\
        & & Max $\abs{\Lambda-\Lambda^{(0)}}$ & 0.24 & 0.91 & N/A & N/A &  \\
    \end{tabular}
\end{table}

\begin{table}[ht]
    \centering
    \caption{Diagnostics when performing natural orbital refinement on the ammonia-oxygendifluorine $4^1A$ charge transfer excitation. The CIS reference was determined to be 2-CSF. Elsewhere, N/A indicates that convergence could not be reached for the specific calculation or the previous calculation(s).}
    \label{CT_Diagnostics_5}
    \begin{tabular}{lllccccc}
        & & & \multicolumn{5}{c}{Orbital Refinements} \\
        ASCC Variant & Reference & Diagnostic & 0 & 1 & 2 & 3 & 4 \\
        \hline
        ASCC$^{\left(M,1\right)}$ & CIS & Energy & N/A & N/A & N/A & N/A & N/A \\
        & & Max $\abs{T-T^{(0)}}$ & N/A & N/A & N/A & N/A & N/A \\
        & & Max $\abs{\Lambda-\Lambda^{(0)}}$ & N/A & N/A & N/A & N/A &  \\
        ASCC$^{\left(M,1\right)}$ & EOM-CCSD & Energy & 7.08 & 6.99 & 6.98 & 6.97 & 6.97 \\
        & & Max $\abs{T-T^{(0)}}$ & 0.15 & 0.07 & 0.07 & 0.07 & 0.07 \\
        & & Max $\abs{\Lambda-\Lambda^{(0)}}$ & 0.15 & 0.09 & 0.09 & 0.09 &  \\
        ASCC$^{\left(M,1\right)}$ & $\omega$B97X-V & Energy & 6.95 & 6.94 & 6.94 & 6.94 & 6.94 \\
        & & Max $\abs{T-T^{(0)}}$ & 0.10 & 0.07 & 0.07 & 0.07 & 0.07 \\
        & & Max $\abs{\Lambda-\Lambda^{(0)}}$ & 0.17 & 0.10 & 0.09 & 0.09 &  \\
        ASCC$^{\left(M,1\right)}$ & ESMF & Energy & 6.95 & 6.98 & 6.97 & 6.97 & 6.97 \\
        & & Max $\abs{T-T^{(0)}}$ & 0.18 & 0.18 & 0.07 & 0.07 & 0.07 \\
        & & Max $\abs{\Lambda-\Lambda^{(0)}}$ & 0.13 & 0.10 & 0.08 & 0.08 &  \\
        PLASCC$^{\left(M,1\right)}$ & CIS & Energy & N/A & N/A & N/A & N/A & N/A \\
        & & Max $\abs{T-T^{(0)}}$ & N/A & N/A & N/A & N/A & N/A \\
        & & Max $\abs{\Lambda-\Lambda^{(0)}}$ & N/A & N/A & N/A & N/A & N/A \\
        PLASCC$^{\left(M,1\right)}$ & EOM-CCSD & Energy & 6.94 & N/A & N/A & N/A & N/A \\
        & & Max $\abs{T-T^{(0)}}$ & 0.26 & N/A & N/A & N/A & N/A \\
        & & Max $\abs{\Lambda-\Lambda^{(0)}}$ & 1.15 & N/A & N/A & N/A &  \\
        PLASCC$^{\left(M,1\right)}$ & $\omega$B97X-V & Energy & 6.97 & 7.19 & N/A & N/A & N/A \\
        & & Max $\abs{T-T^{(0)}}$ & 0.54 & 1.06 & N/A & N/A & N/A \\
        & & Max $\abs{\Lambda-\Lambda^{(0)}}$ & 0.30 & 0.38 & N/A & N/A &  \\
        PLASCC$^{\left(M,1\right)}$ & ESMF & Energy & 7.01 & 7.06 & N/A & N/A & N/A \\
        & & Max $\abs{T-T^{(0)}}$ & 0.09 & 0.19 & N/A & N/A & N/A \\
        & & Max $\abs{\Lambda-\Lambda^{(0)}}$ & 0.09 & N/A & N/A & N/A &  \\
    \end{tabular}
\end{table}

\begin{table}[ht]
    \centering
    \caption{Diagnostics when performing natural orbital refinement on the tetrafluoroethylene-ethylene $5^1B_1$ charge transfer excitation. N/A indicates that convergence could not be reached for the specific calculation or the previous calculation(s).}
    \label{CT_Diagnostics_6}
    \begin{tabular}{lllccccc}
        & & & \multicolumn{5}{c}{Orbital Refinements} \\
        ASCC Variant & Reference & Diagnostic & 0 & 1 & 2 & 3 & 4 \\
        \hline        
        ASCC$^{\left(M,1\right)}$ & CIS & Energy & 10.55 & N/A & N/A & N/A & N/A \\
        & & Max $\abs{T-T^{(0)}}$ & 0.17 & N/A & N/A & N/A & N/A \\
        & & Max $\abs{\Lambda-\Lambda^{(0)}}$ & 0.22 & N/A & N/A & N/A &  \\
        ASCC$^{\left(M,1\right)}$ & EOM-CCSD & Energy & 10.69 & 10.66 & 10.71 & 10.66 & 10.27 \\
        & & Max $\abs{T-T^{(0)}}$ & 0.10 & 0.30 & 0.35 & 0.34 & 0.25 \\
        & & Max $\abs{\Lambda-\Lambda^{(0)}}$ & 0.16 & 0.19 & 0.25 & 0.58 &  \\
        ASCC$^{\left(M,1\right)}$ & $\omega$B97X-V & Energy & 10.31 & 9.78 & 9.56 & 9.46 & 9.40 \\
        & & Max $\abs{T-T^{(0)}}$ & 0.39 & 0.15 & 0.14 & 0.14 & 0.14 \\
        & & Max $\abs{\Lambda-\Lambda^{(0)}}$ & 0.49 & 0.15 & 0.13 & 0.13 &  \\
        ASCC$^{\left(M,1\right)}$ & ESMF & Energy & 10.39 & 11.18 & -0.02 & -0.03 & -0.03 \\
        & & Max $\abs{T-T^{(0)}}$ & 0.44 & 0.54 & 0.42 & 0.42 & 0.42 \\
        & & Max $\abs{\Lambda-\Lambda^{(0)}}$ & 0.50 & 1.26 & 0.88 & 0.65 &  \\
        PLASCC$^{\left(M,1\right)}$ & CIS & Energy & 10.53 & 10.79 & 10.78 & 10.78 & 10.78 \\
        & & Max $\abs{T-T^{(0)}}$ & 0.10 & 0.03 & 0.03 & 0.03 & 0.03 \\
        & & Max $\abs{\Lambda-\Lambda^{(0)}}$ & 0.12 & 0.09 & 0.09 & 0.09 &  \\
        PLASCC$^{\left(M,1\right)}$ & EOM-CCSD & Energy & 10.68 & 10.78 & 10.78 & 10.78 & 10.78 \\
        & & Max $\abs{T-T^{(0)}}$ & 0.05 & 0.03 & 0.03 & 0.03 & 0.03 \\
        & & Max $\abs{\Lambda-\Lambda^{(0)}}$ & 0.10 & 0.09 & 0.09 & 0.09 &  \\
        PLASCC$^{\left(M,1\right)}$ & $\omega$B97X-V & Energy & 10.57 & 10.75 & 10.78 & 10.78 & 10.78 \\
        & & Max $\abs{T-T^{(0)}}$ & 0.16 & 0.11 & 0.07 & 0.04 & 0.03 \\
        & & Max $\abs{\Lambda-\Lambda^{(0)}}$ & 0.25 & 0.13 & 0.11 & 0.10 &  \\
        PLASCC$^{\left(M,1\right)}$ & ESMF & Energy & 10.58 & 10.78 & 10.78 & 10.78 & 10.78 \\
        & & Max $\abs{T-T^{(0)}}$ & 0.11 & 0.05 & 0.03 & 0.03 & 0.03 \\
        & & Max $\abs{\Lambda-\Lambda^{(0)}}$ & 0.13 & 0.10 & 0.09 & 0.09 &  
    \end{tabular}
\end{table}

\begin{table}[ht]
    \centering
    \caption{Diagnostics when performing natural orbital refinement on the 3,5-difluoro-penta-2,4-dienamine $1^1A_1$ charge transfer excitation. N/A indicates that convergence could not be reached for the specific calculation or the previous calculation(s).}
    \label{CT_Diagnostics_7}
    \begin{tabular}{lllccccc}
        & & & \multicolumn{5}{c}{Orbital Refinements} \\
        ASCC Variant & Reference & Diagnostic & 0 & 1 & 2 & 3 & 4 \\
        \hline
        ASCC$^{\left(M,1\right)}$ & CIS & Energy & 7.07 & 6.96 & 6.76 & 6.34 & 6.24 \\
        & & Max $\abs{T-T^{(0)}}$ & 0.30 & 0.13 & 0.26 & 0.24 & 0.23 \\
        & & Max $\abs{\Lambda-\Lambda^{(0)}}$ & 0.31 & 0.19 & 0.49 & 0.18 &  \\
        ASCC$^{\left(M,1\right)}$ & EOM-CCSD & Energy & 6.96 & 6.94 & 6.85 & 6.45 & 6.26 \\
        & & Max $\abs{T-T^{(0)}}$ & 0.26 & 0.10 & 0.22 & 0.25 & 0.23 \\
        & & Max $\abs{\Lambda-\Lambda^{(0)}}$ & 0.23 & 0.14 & 0.40 & 0.27 &  \\
        ASCC$^{\left(M,1\right)}$ & $\omega$B97X-V & Energy & 6.98 & 6.94 & 6.85 & 6.45 & 6.26 \\
        & & Max $\abs{T-T^{(0)}}$ & 0.30 & 0.10 & 0.22 & 0.25 & 0.24 \\
        & & Max $\abs{\Lambda-\Lambda^{(0)}}$ & 0.27 & 0.13 & 0.40 & 0.24 &  \\
        ASCC$^{\left(M,1\right)}$ & ESMF & Energy & 6.84 & 6.89 & 6.88 & 6.87 & 6.87 \\
        & & Max $\abs{T-T^{(0)}}$ & 0.09 & 0.05 & 0.06 & 0.09 & 0.13 \\
        & & Max $\abs{\Lambda-\Lambda^{(0)}}$ & 0.08 & 0.09 & 0.10 & 0.13 &  \\
        PLASCC$^{\left(M,1\right)}$ & CIS & Energy & N/A & N/A & N/A & N/A & N/A \\
        & & Max $\abs{T-T^{(0)}}$ & N/A & N/A & N/A & N/A & N/A \\
        & & Max $\abs{\Lambda-\Lambda^{(0)}}$ & N/A & N/A & N/A & N/A &  \\
        PLASCC$^{\left(M,1\right)}$ & EOM-CCSD & Energy & 6.74 & 6.67 & N/A & N/A & N/A \\
        & & Max $\abs{T-T^{(0)}}$ & 0.16 & 0.22 & N/A & N/A & N/A \\
        & & Max $\abs{\Lambda-\Lambda^{(0)}}$ & 0.31 & 0.42 & N/A & N/A &  \\
        PLASCC$^{\left(M,1\right)}$ & $\omega$B97X-V & Energy & 6.82 & N/A & N/A & N/A & N/A \\
        & & Max $\abs{T-T^{(0)}}$ & 0.20 & N/A & N/A & N/A & N/A \\
        & & Max $\abs{\Lambda-\Lambda^{(0)}}$ & 0.34 & N/A & N/A & N/A &  \\
        PLASCC$^{\left(M,1\right)}$ & ESMF & Energy & 6.76 & 6.84 & 6.67 & N/A & N/A \\
        & & Max $\abs{T-T^{(0)}}$ & 0.12 & 0.05 & 0.30 & N/A & N/A \\
        & & Max $\abs{\Lambda-\Lambda^{(0)}}$ & 0.17 & 0.11 & 1.00 & N/A &  
    \end{tabular}
\end{table}

\renewcommand{\thesection}{\Roman{section}}
\setcounter{section}{6}

\end{document}